%% file: main.tex
\input{Preamble}

\title{Shortest Paths in Multimode Graphs}
\date{}
\author{Yael Kirkpatrick\thanks{MIT, supported by the National Science Foundation Graduate Research Fellowship under Grant No 2141064.} \and Virginia Vassilevska Williams\thanks{MIT, Supported by NSF Grants CCF-2129139 and CCF-2330048, BSF Grant 2020356 and a Simons Investigator Award.}}

\begin{document}

\maketitle  
\input{0-New_Abstract}

\input{1-New_Introduction}

\input{2-Preliminaries}

\input{4-5-Algorithms}

\input{6-7-LowerBounds}

% \section*{Acknowledgements} 
% The authors would like to thank Asaf Etgar, Jenny Kaufmann and Surya Mathialagan for helpful discussions.

\bibliography{references.bib}

\appendix
\input{3-Exact_Case}

\input{4a-Undir_diam_appendix}

% icalp page limit rearrangments:
% \section{Approximating Undirected $k$-mode Diameter and Radius}\label{section:alg_undir}
% \input{icalp-4-Alg_undir}

% \section*{Acknowledgements} 
% The authors would like to thank Asaf Etgar, Jenny Kaufmann and Surya Mathialagan for helpful discussions.

% \bibliography{references.bib}
% \appendix
% \section{Approximating Directed $k$-mode Diameter and Radius}
% \label{section:alg_dir}\input{5-Alg_Dir}
% \input{6-7-LowerBounds}
% \input{3-Exact_Case}
% \input{icalp-4a-Undir_diam_appendix}
\end{document}

%% file: Preamble.tex
\def\numbered{1}

\documentclass[11pt, final]{article}
\usepackage{color,colortbl,latexsym,amsthm,amsmath,amssymb,path,enumitem,soul,tikz,fullpage,subcaption}
\usetikzlibrary{shapes.geometric}
\usepackage[hidelinks]{hyperref}
\usepackage{algorithm}
\usepackage{algorithmicx}
\usepackage{algpseudocode}
\usepackage{soul}
\usepackage{placeins}
\usepackage{relsize}
\usepackage{diagbox}
\usepackage{comment}
\usepackage{tikz}
\usepackage{verbatim}
\usepackage{mathtools}
\usepackage{trimclip}
\usepackage{amsthm}
\usepackage[font={small,sf}]{caption}
\usepackage[normalem]{ulem}
\useunder{\uline}{\ul}{}
\newcommand{\ignore}[1]{}

\setlist[itemize]{itemsep=-0.1em}
\setlist[enumerate]{itemsep=-0.1em}

\makeatletter
\def\th@plain{%
  \thm@notefont{}% same as heading font
  \itshape % body font
}
\def\th@definition{%
  \thm@notefont{}% same as heading font
  \normalfont % body font
}
\makeatother
\ifnum\numbered=1
    \newtheorem{theorem}{Theorem}[section]
\else
    \newtheorem{theorem}{Theorem}
\fi 
    \newtheorem{lemma}[theorem]{Lemma}
    \newtheorem{corollary}[theorem]{Corollary}

    \newtheorem{hypothesis}[theorem]{Hypothesis}
    
    \newtheorem{claim}[theorem]{Claim}

\usepackage{graphicx}
\usepackage{psfrag}

\usepackage{listings,xcolor}
\usepackage{dsfont}
\newcommand{\NN}{\ensuremath{\mathbb N}}
\newcommand{\ZZ}{\ensuremath{\mathbb Z}}

\newcommand{\Gg}{\ensuremath{\mathcal G}}
\newcommand{\Uu}{\ensuremath{\mathcal U}}
\newcommand{\Tt}{\ensuremath{\mathcal T}}

\renewcommand{\epsilon}{\varepsilon}
\newcommand{\eps}{\varepsilon}

\renewcommand{\subset}{\subseteq}

\newcommand{\pars}[1]{\left( #1 \right)}

\newcommand{\abs}[1]{\left| #1 \right|}
\newcommand{\set}[1]{\left\{ #1 \right\}}

\makeatletter
\newcommand{\thickhline}{%
    \noalign {\ifnum 0=`}\fi \hrule height 1pt
    \futurelet \reserved@a \@xhline
}
\newcolumntype{"}{@{\hskip\tabcolsep\vrule width 1pt\hskip\tabcolsep}}
\makeatother
\newcommand{\mindiam}{\ensuremath{\mathsf{min}\text{-}\mathsf{diam}}}
\newcommand{\gov}{G_{\ensuremath{\mathsf{OV}}}}

\newcommand{\poly}{\textrm{poly}}

%% file: 0-New_Abstract.tex
\begin{abstract}
    In this work we study shortest path problems in multimode graphs, a generalization of the min-distance measure introduced by Abboud, Vassilevska W. and Wang in [SODA'16]. A multimode shortest path is the shortest path using one of multiple `modes' of transportation that cannot be combined. This represents real-world scenarios where different modes are not combinable, such as flights operated by different airline alliances. The problem arises naturally in machine learning in the context of learning with multiple embedding. More precisely, a $k$-multimode graph is a collection of $k$ graphs on the same vertex set and the $k$-mode distance between two vertices is defined as the minimum among the distances computed in each individual graph. 

    We focus on approximating fundamental graph parameters on these graphs, specifically diameter and radius. In undirected multimode graphs we first show an elegant linear time 3-approximation algorithm for 2-mode diameter. We then extend this idea into a general subroutine that can be used as a part of any $\alpha$-approximation, and use it to construct a 2 and 2.5 approximation algorithm for 2-mode diameter. For undirected radius, we introduce a general scheme that can compute a 3-approximation of the $k$-mode radius for any $k$ and runs in near linear time in the case of $k=O(1)$. In the directed case we establish an equivalence between approximating 2-mode diameter on DAGs and approximating the min-diameter, while for general graphs we develop novel techniques and provide a linear time algorithm to determine whether the diameter is finite.

    We also develop many conditional fine-grained lower bounds for various multimode diameter and radius approximation problems. We are able to show that many of our algorithms are tight under popular fine-grained complexity hypotheses, including our linear time 3-approximation for $3$-mode undirected diameter and radius. As part of this effort we propose the first extension to the Hitting Set Hypothesis [SODA'16], which we call the $\ell$-Hitting Set Hypothesis. We use this hypothesis to prove the first parameterized lower bound tradeoff for radius approximation algorithms.

    % First we consider computing all pairwise distances under this measure and ask whether the All-Pairs Shortest Paths (APSP) problem can be computed faster than running APSP on each graph individually (which would run in $kn^{3-o(1)}$ time). We show that under the APSP Hypothesis, APSP in $k$-mode $n$-node graphs with arbitrary edge weights requires $(kn^3)^{1-o(1)}$ time. In contrast, when the weights are small integers, we present an algorithm that performs better than running the current best APSP algorithm $k$ times.
    
    % We then focus on computing and approximating the diameter, radius and eccentricites of a multimode graph. This poses a significant challenge, since the $k$-mode distance lacks many of the nice properties of the traditional distance measure, such as the triangle inequality. Therefore, we need to develop new approaches and techniques in order to approximate distances under this measure. We develop multiple algorithms for approximating these parameters in both directed and undirected multimode graphs. We also prove conditional lower bounds for these problems, showing that in several cases our algorithms are tight. 

\end{abstract}

%% file: 1-New_Introduction.tex
\section{Introduction}
The study of shortest paths algorithms in directed graphs has introduced various distance measures. The most standard notion of distance between two vertices $u$ and $v$ is the minimum length of a path from $u$ to $v$, $d(u,v)$. This asymmetric one-way notion of distance is not always the most natural. In some applications the {\em roundtrip} distance $d(u,v)+d(v,u)$, defined by Cowen and Wagner \cite{CowenW99}, is most useful, since unlike its one-way counterpart, it is a metric. 

In directed acyclic graphs (DAGs), however, both the standard notion of distance and the roundtrip notion produce infinite distances for all or a large fraction of the distance pairs. A more useful notion of distance in this case was defined by Abboud, Vassilevska W. and Wang \cite{radiusdiamapprox2016}.
In this work they introduced the {\em min-distance}, defined as $d_{\textrm{min}}(u,v)=\min \{d(u,v),d(v,u)\}$. Under one-way distance or roundtrip distance, almost all points in a DAG have infinite eccentricity. Introducing min-distance opened an avenue of research which lead to intricate algorithms for computing or approximating min-diameter, radius and eccentricities in DAGs \cite{radiusdiamapprox2016, mindistanceDAG2021} as well as for general graphs \cite{bergermindiam2023, chechiklinearmindiam2022, mindistance2019}.

One real-world motivation for studying the min-distance problem is computing the optimal location for a hospital, as one has the choice of either going to the hospital or having a doctor come to them. The min-distance therefore measures the fastest way to receive care and the point with smallest min-eccentricity in the graph functions as the optimal location for a hospital.
Imagine now that the doctor can take a helicopter, whereas a patient can only take a cab to the doctor. Then the two routes are no longer part of the same graph - the routes by which a helicopter can travel are very different from those available to a taxi. The new distance between the patient and the doctor is now the minimum between the distances in two different graphs.

Consider another common scenario. We want to fly from city $A$ to city $B$. We can take any valid itinerary, except that each itinerary needs to stay within the same airline alliance. Now the fastest flight time is the minimum of the distances over several different ``graphs'', one for each airline alliance. All of these graphs share a vertex set representing the different cities, while each graph has edges representing the flights of a single airline alliance.

Let us formalize a new shortest paths problem that captures both scenarios. Consider a collection of $k$ graphs over the same vertex set $V$. Let their edge sets be $E_1,\ldots,E_k$. Together they form a ``$k$-multimode graph''  $\Gg = (V, E_1, \ldots, E_k)$. One can associate each $E_i$ with a  `mode of transportation', where these modes are separate and cannot be combined.

The $k$-\textit{mode distance} between two vertices $u, v \in V$ in a $k$-multimode graph $\mathcal{G} = (V, E_1, \dots E_k)$ is defined as $$d_\mathcal{G}(u,v) = \min_i d_{G_i}(u,v),$$ where $d_{G_i}$ is the distance in the graph $G_i = (V, E_i)$. Intuitively, $d_{\mathcal{G}}(u,v)$ is the length of the shortest path connecting $u$ and $v$ using a single mode of transportation.

% The $k$-mode distance is also a natural way to formulate the problem of communication routing when there are multiple networks in use. For instance, in space communication, components of the networks can be connected to proprietary networks of various space agencies as well as public networks such as starlink. When two nodes wish to communicate they must choose the network in which they are closest to each other in order to route their communication in the fastest way.

The $k$-mode distance naturally arises in practice in the context of multiple embeddings in machine learning (see \cite{multipleembeddings2019,multipleembeddings2021, multiplieembeddings2014} and many more). In many scenarios, given a high dimensional data set, there is no clear distance function that accurately captures semantic relationships within the data. To address this, deep learning is now commonly used to obtain a data embedding that encodes semantic information. As research in this field advances, we observe a proliferation of embeddings which can vary due to differences in network architecture, learning paradigms or training data, to name a few. This results in a single set of data points with multiple sets of edges representing distances in the different embeddings spaces. For every pair of points we are interested in the embedding in which they are closest together and often care about finding a pair of points which are furthest apart in all embeddings (i.e. computing the $k$-mode diameter) or a point which is closest to all others in some embedding (i.e. computing the $k$-mode radius).\\

% This notion of distance trivially generalizes the traditional (single graph) notion of distance. It also generalizes the {\em min-distance} in directed graphs: the min-distance in $G$ is the same as the $2$-mode distance in the $2$-multimode graph $(V, E(G),E^R(G))$ where $E^R(G)$ are the edges of $G$ in the reverse direction.
% This new notion of distance is more general, as the edges in a multimode graph can now also be undirected, and there can be more than 2 modes. 

% \paragraph{} 
Let us first consider the All-Pairs Shortest Paths (APSP) problem in $k$-multimode graphs on $n$ vertices: compute the $k$-mode distance between all pairs of vertices of $\mathcal{G} = (V, E_1, \dots E_k)$.
One can always run any APSP algorithm on each $(V,E_i)$ separately, resulting in a running time which is $k$ times that of the standard APSP algorithm. This approach will also trivially compute the $k$-mode radius of the multimode graph. In fact, we show that in general no algorithm can have polynomial improvements over this trivial solution.

\begin{theorem}\label{thm:apspk}
    Under the APSP hypothesis \cite{roditty2004dynamic}, no $O((kn^3)^{1-\epsilon})$ time algorithm can solve $k$-mode APSP or $k$-mode radius for any $\epsilon > 0$.
\end{theorem}

% In shortest paths research, there are several distance-based parameters of interest. The eccentricity $\eps(v)$ of a vertex $v$ in a graph $G=(V,E)$ is $\max_{u\in V} d(v,u)$. The {\em diameter} is the maximum eccentricity in the graph, and the {\em radius} is the minimum eccentricity of a graph. 

While it is clear for APSP that one can compute all $k$-multimode distances by solving $k$ instances of APSP separately on each graph, it is no longer obvious how to solve $k$-multimode radius or diameter using algorithms that compute the radius or diameter in a standard graph.

To illustrate why the $k$-mode diameter is not directly related to the standard diameters of the individual graphs $G_1, \ldots, G_k$, denote the diameter of $G_1=(V,E_1)$ by $D_1=D(G_1)$ and let $u_1, v_1\in V$ be such that $d_{G_1}(u_1,v_1)= D_1$. Similarly, denote the diameter of $G_2=(V,E_2)$ by $D_2=D(G_2)$ and let $u_2, v_2\in V$ be such that $d_{G_2}(u_2,v_2) = D_2$. We cannot merely take the minimum of $D_1, D_2$ to get the diameter in the multimode graph $\Gg=(V,E_1,E_2)$. This is because it could be that $d_{G_2}(u_1,v_1)<D_1$ and $d_{G_1}(u_2,v_2)<D_2$ so that the multimode distance between $u_1$ and $v_1$ is not $D_1$ but rather $d_{G_2}(u_1,v_1)$ and similarly for $u_2$ and $v_2$. The multimode diameter can be the distance between some completely different pair of vertices and thus can be arbitrarily smaller than both $D_1$ and $D_2$. It could even be the case that $D_1, D_2$ are infinite, while the multimode diameter $D(\Gg)$ is finite.

However, if we wish to compute the exact $k$-mode diameter or radius of a $k$-multimode undirected graph, we show that we can in fact use any algorithm for computing standard diameter or radius in a blackbox way.

\begin{theorem}\label{thm:exactradiusdiam}
    If there exists a $T(n,m)$ time algorithm for computing the diameter/radius of a weighted, undirected graph with $n$-nodes and $m$-edges, then there exists an algorithm for computing the $k$-mode diameter of a $k$-multimode undirected graph in time $T(O(kn), O(m + kn))$.
\end{theorem}

The proofs of both \autoref{thm:apspk} and \autoref{thm:exactradiusdiam} follows from a straightforward application of standard techniques. We therefore defer their proofs to \autoref{section:exactcase}. We instead turn our focus to approximating these parameters.\\

% \paragraph{} 
There exists a plethora of algorithms and conditional lower bounds for approximating diameter and radius under both the standard notion of (directed or undirected) distance and the min-distance measure \cite{tightdiamapprox2018,radiusdiamapprox2016,chechiklinearmindiam2022,mindistanceDAG2021,DalirrooyfardLW21,DalirrooyfardW019a,ChechikLRSTW14,RodittyW13,CairoGR16,DalirrooyfardW21}. As the $k$-multimode distance generalizes both these distance measures, we would like to develop approximation algorithms and prove tight conditional lower bounds for approximating diameter and radius in the multimode setting as well.

The known fine-grained hardness results for the min-distance and standard diameter, radius and eccentricities would trivially carry over for the multimode distance case (as it generalizes both). We are thus interested in when stronger hardness is possible, and when similar approximation algorithms can be attained.

Devising approximation algorithms for the min-distance measure is particularly difficult  and therefore significantly new techniques had to be developed over the standard diameter ones (see e.g. \cite{mindistance2019,chechiklinearmindiam2022,mindistanceDAG2021}). The main difficulty lies in the fact that the triangle inequality no longer holds: it could be that $d_{min}(u,v)=1$ and $d_{min}(v,w)=1$ but $d_{min}(u,w)=\infty$, as in the graph with vertices $\{u,v,w\}$ and edges $\{(v,u),(v,w)\}$. This difficulty persists and is even more prominent in the multimode distance setting, as now the various graphs that represent modes of transportation may not share any edges. We can therefore expect that the distance parameters of interest can be even more difficult to approximate.

\subsection{Our results}
We develop various new techniques that work in the multimode distance setting, both for directed and undirected graphs, and for fine-grained conditional lower bounds. Table \ref{tab:results} summarizes our results. Here we present some highlights.

\input{1-Results}

Our hardness results are based on the Strong Exponential Time Hypothesis (SETH) \cite{impagliazzopaturi2001,cip10} and the Hitting Set Hypothesis \cite{radiusdiamapprox2016} (see also \cite{vsurvey}). We further extend the Hitting Set Hypothesis to the stronger $\ell$-Hitting Set Hypothesis (Hypothesis \ref{hyp:lhsh} in the text). Based on this new hypothesis we are able to obtain a lower bound tradeoff for approximating the $k$-mode radius. As the introduction of the Hitting Set Lemma in \cite{radiusdiamapprox2016} allowed for the proof of the first lower bounds on radius approximation, we hope this extension will enable future work on lower bound tradeoffs for radius in various distance frameworks.

\begin{theorem} Let $k\geq 3$.
Under SETH, there can be no $O(m^{2-\eps})$ time (for $\eps>0$) algorithm that achieves any finite multiplicative approximation for the $k$-mode diameter in directed unweighted $m$-edge multimode graphs, even if all graphs in the multimode graph are DAGs.
\end{theorem}

The result appears as Theorem \ref{thm:hard3colordir} in the text. It implies that, under SETH, to get any approximation algorithm for directed multimode diameter, even for DAGs, one must focus on $k=1$ or $k=2$. The case $k=1$ is just the standard diameter problem, and so we consider the case where $k=2$ next.

For $2$-mode diameter in directed graphs we provide (1) a near-linear time algorithm that can decide whether the diameter is finite (an $n$-approximation, in Theorem \ref{thm:2colorfindirdiam}), and (2) a near-linear time $2$-approximation algorithm for the case when both graphs in the multimode graph are DAGs (Corollary \ref{cor:dagmindiam}). Both of these results are tight, in the sense that for $k>2$ such results are impossible under SETH. \\

% While directed $2$-mode diameter is a direct generalization of min-diameter, we show that approximating these two parameters is equivalent only in the case of DAGs (Corollary \ref{cor:dagmindiam}). In the general case we develop new techniques for approximating the $k$-mode distances.

% \paragraph{} 
For $k$-mode diameter and radius in undirected graphs we obtain several tight results as well. Here are two highlights:

\begin{theorem}
For every constant $k$, there is an $\tilde{O}(m)$ time\footnote{As is standard, $\Tilde{O}$ hides polylogarithmic factors.} $3$-approximation algorithm for the radius of $m$-edge undirected $k$-multimode graphs. This is tight under the $\ell$-Hitting Set Hypothesis, in that no $3-\eps$-approximation is possible in $\tilde{O}(m)$ time, even for $k=2$.
\end{theorem}

This result is stated in Theorems \ref{thm:radius3approx} and \ref{thm:hard2colorundirradius} and achieves the conditionally best possible result for radius in near-linear time.

\begin{theorem}
Under SETH, there can be no $2-\delta$-approximation algorithm for diameter running in $O(mn^{1-\eps})$ time for $\eps,\delta>0$ for $m$-edge $2$-mode undirected unweighted graphs.
If $\omega=2$,\footnote{Here $\omega$ denotes the fast matrix multiplication exponent, $\omega < 2.371552$ \cite{newomega2024}.} there is an $\tilde{O}(mn^{3/4})$ time algorithm that achieves a $2$-multiplicative, $2$-additive approximation to the diameter in $m$-edge, $n$-node undirected $2$-multimode graphs.
\end{theorem}

This result is stated in Theorems \ref{thm:2approx} and \ref{thm:hardundir2approx}. Up to the small additive error, the approximation algorithm is optimal due to the SETH-based lower bound. The running time of the algorithm is still faster than $mn$, even with the current bound on $\omega$.

% \subsection{Techniques}
% TODO

% \paragraph{Results:} 
% \begin{itemize}
%     \item Exact case
%     \item Linear time 3-approx for undirected 2-mode diam/radius
%     \item Generalization to $<3$ approx for 2-mode diameter and 3-approx for k-mode radius.
%     \item Generalization of HS hypothesis for radius lower bounds.
%     \item Approximating 2 mode diam in DAGs is equivalent to approximating min-diam.
%     \item Fin vs infinite diam/radius algorithm
% \end{itemize}

\subsection{Technical Overview}
In this subsection we describe an outline of some of our results, and highlight their connections to other, well studied problems.

\subsubsection*{Approximating Undirected 2 and 3 Mode Diameter}
Our first result is a simple, linear time 3-approximation for 2-mode diameter. Given a threshold $D$, we would like to either show that the 2-mode diameter is less than $D$ or find a pair of points $u,v$ whose 2-mode distance is greater than $D/3$, meaning $d_{G_1}(u,v) \geq D/3$ and $d_{G_2}(u,v)\geq D/3$. 

To find such a pair, run BFS from an arbitrary point $z$. Let $X$ be the points within distance $D/3$ of $z$ in $G_1$ and let $Y$ be the points within distance $D/3$ of $z$ in $G_2$. 
If any vertex is contained in neither $X$ nor $Y$, then it has multimode distance greater than $D/3$ from $z$ and we are finished. 

Next, run BFS from $X$ (as a set) in $G_1$ to identify the vertices in $Y$ that have distance $\geq D/3$ in $G_1$ from \emph{all} vertices in $X$. Similarly, run BFS from $Y$ in $G_2$ to identify all vertices in $X$ that have distance $\geq D/3$ in $G_2$ from all vertices in $Y$. If there exists a pair of vertices $x\in X,y\in Y$ such that $d_{G_1}(y, X) \geq  D/3$ and $d_{G_2}(x,Y)\geq D/3$ then the multimode distance of $x,y$ is greater than $D/3$ and we are done. Otherwise, we can show that any pair of points has multimode distance $<D$, and thus the 2-mode diameter is $<D$. Indeed, if $x,y\in X$ then $d_{G_1}(x,y) < 2D/3$ using a path through $z$ in $G_1$. Similarly, if $x,y\in Y$ then $d_{G_2}(x,y) < 2D/3$. Otherwise, we can assume w.l.o.g that $x\in X, y\in Y$ and $d_{G_1}(y,X) < D/3$. Then there exists a point $x'\in X$ such that $d_{G_1}(x', y) < D/3$. Now, following a path in $G_1$ from $y$ to $x'$ to $z$ to $x$ we have that $d_{G_1}(x,y) < D/3 + D/3 + D/3 = D$. Thus, for any pair of vertices we have $d_\Gg(x,y) < D$.

This completes our 3-approximation for undirected 2-mode diameter in linear time. Next we generalize this approach to a subroutine that can be used for any $\beta$-approximation for $2\leq \beta \leq 3$. Informally, we take $X$ to be the neighborhood in $G_1$ of some point $x$ and $Y$ to be the neighborhood in $G_2$ of some point $y$. We perform a similar search, running BFS from the set $X$ in $G_1$ and from $Y$ in $G_2$ to find a pair of points in $X\times Y$ that is far apart in both graphs. In order to conclude that the graph has a small diameter in the case that we do not find a pair of large distance, we need to consider all pairs $x,y$ where $x$ is in the $G_1$ neighborhood of a vertex $z$ and $y$ is in the $G_2$ neighborhood of the same vertex $z$. This results in a subroutine that, given a point $z$ with small $G_1$ and $G_2$ neighborhoods, we can compute a $\beta$-approximation to the 2-mode diameter. For details see Lemma \ref{lm:point2smallnbhds}.

We can now use this idea to construct $\beta$-approximation algorithms using a classic large-small neighborhoods tradeoff. We show two ways in which to use this subroutine in a complete algorithm, obtaining a 2-approximation and a 2.5-approximation. 

For our 2-approximation, if we have a point $z$ with small $G_1$ and $G_2$ neighborhoods, we use the aforementioned subroutine. Otherwise, all vertices have a large $G_1$ or $G_2$ neighborhood, including the endpoints of the diameter. We leverage this fact and sample a hitting set, hitting the $G_1$ or $G_2$ neighborhood of a diameter endpoint. For every point $z$ in this hitting set, we compute the set of points $A$ of distance $<D/4$ from $z$  in $G_1$ and the set of points $B$ of distance $>3D/4$ from $z$ in $G_1$. We show that if the 2-mode diameter is $\geq D$ and $z$ hits the $G_1$ neighborhood of a diameter endpoint, then the ST-diameter of the sets $A,B$ in $G_2$ will be $D$. We can therefore use an ST-diameter approximation algorithm in a black box way to obtain our desired approximation.

To obtain our 2.5-approximation we use a similar approach, leveraging a blackbox ST-diameter 2.5-approximation. Using a slightly different method, we apply the simple 3-approximation of ST-diameter to turn our linear time 2-mode diameter approximation into an approximation for 3-mode diameter, incurring only poly-logarithmic factors in the running time.

\subsubsection*{Connections to ST-Diameter}
As hinted to above, the ST-diameter problem turns out to be intricately connected to the problem of computing the multimode diameter. In many cases, one can reduce the problem of computing a multimode diameter to that of computing the ST-diameter of some subsets of the vertices in one of the edge sets. It turns out, this connection between the problems persists in the lower bounds as well. 

Consider a lower bound construction showing that an $\alpha$-approximation of ST-diameter requires $T(n)$ time: a graph $G=(V,E)$ with $S,T\subset V$ such that approximating the ST-diameter of $S,T$ beyond a factor of $\alpha$ takes $\Omega (T(n))$ time. We can now construct a simple lower bound for the problem of approximating 3-mode diameter as follows. Take the 3-multimode graph with $E$ as its first edge set, using a second set of edges connect the vertices of $S$ with the vertices of $V\setminus T$ and using a third set of edges connect $T$ with $V\setminus S$. Now, the 3-mode diameter of this 3-multimode graph will be the ST-diameter of $S$ and $T$ in the original graph. Therefore approximating the 3-mode diameter beyond a factor of $\alpha$  requires $\Omega(T(n))$ time as well.

We use an existing lower bound tradeoff for ST-diameter to establish a lower bound tradeoff for 3-mode diameter approximation. However, for radius approximation no such tradeoff is known. 
We further extend this result to a lower bound tradeoff for 3-mode radius approximation, establishing the first lower bound tradeoff for any form of radius approximation problem. This result also suggests a potential `radius equivalent' of the ST-diameter problem, which could be an avenue for future work.

\subsubsection*{Approximating Directed Multimode Distances}
While the problem of approximating diameter and radius of multimode graphs is closely related to the ST-diameter problem, and often ST-diameter techniques allow to `reduce the number of modes' in the graph, the directed case is very different. In directed graphs we cannot hope for techniques like this to work as we show that any approximation to 3-mode diameter or 2-mode radius requires quadratic time. 

We are able to show, however, that in directed acyclic graphs, the problem of approximating 2-mode diameter is in fact equivalent to the problem of approximating the min-diameter of each of the 2 graphs making up the multimode graph. In the case of a general directed 2-mode graph, we construct a linear time algorithm which can differentiate between finite and infinite 2-mode diameter. It does so by considering the min-distances in the graphs of the strongly connected components of $G_1$ and $G_2$, and the intersections of these two sets of SCCs with each other.

\subsection{Organization}
In \autoref{section:alg_undir} we present approximation algorithms for \textbf{undirected} $k$-mode diameter and radius, with additional proofs deferred to \autoref{section:4a_more_undir_diam}. In \autoref{section:alg_dir} we present approximation algorithms for \textbf{directed} $k$-mode diameter, radius and eccentricities. In \autoref{section:diamLB} we prove lower bounds for approximating $k$-mode \textbf{diameter}. In \autoref{section:radiusLB} we prove lower bounds for approximating $k$-mode \textbf{radius}. We present additional results regarding computing exact $k$-mode distances in \autoref{section:exactcase}.

% \subsection{Open Problems?} \todo{}

% \paragraph{Future work:} Directed algorithms (any constant approximation), undirected algorithms without an additive error, show a lower bound on exact diameter.

% for intro:
% We note that getting an approximate value, or even exact value, of the diameter or radius of each color graph individually ($D(G_1), D(G_2), \ldots$) does not provide much information regarding the $k$-color diameter or radius of $\Gg$. \yael{give examples where G1 G2 diams are infinite and G diam is finite, opposite for radius}

%% file: 1-Results.tex
\begin{table}[]
\centering
\begin{tabular}{|c|c|l|l|c|}

\hline
{$\boldsymbol{k}$}       
& {\textbf{Approx.}}        
& {\textbf{Runtime}}                
& {\textbf{Comments}}                                                 
& \textbf{Reference} \\ 

\hline
\multicolumn{5}{|l|}{\textbf{Undirected $k$-mode Diameter Upper Bounds}}     \\ 

\hline
$2$ & $3$ & $O(m)$ & Weighted & \autoref{thm:2color3approx}     \\ 

\hline
$\boldsymbol{3}$  & $\boldsymbol{3}$ & $\boldsymbol{\Tilde{O}(m)}$ 
& \textbf{Weighted.}                                    
& \textbf{\autoref{thm:3color3approx}}             \\ 

\hline
$2$ & $(2, 2M)$ & $\Tilde{O}(mn^{3/4})^*$     
& \begin{tabular}[c]{@{}l@{}}
    Non negative edge weights $\leq M$.\\ If $\omega > 2$ the runtime is\\  
    $m\cdot \pars{\frac{n^{1.5}}{(m\sqrt{n})^{1/\omega}} + \pars{\frac{m}{n}}^{1/(\omega -      2)}}$.
\end{tabular}
&   \autoref{thm:2approx}                \\ 

\hline
$2$ & $(2.5, 2.5M)$ & $\Tilde{O}(m\sqrt{n})^*$
& \begin{tabular}[c]{@{}l@{}}
    Non negative edge weights $\leq M$.\\ If $\omega > 2$ the runtime  is \\
    $m\cdot \pars{\frac{n}{m^{1/\omega}} + \pars{\frac{m}{n}}^{1/(\omega - 2)}}$.
\end{tabular}            
&    \autoref{thm:2.5approx}                \\ 

\hline
\multicolumn{5}{|l|}{\textbf{Undirected $k$-mode Diameter Lower Bounds (Under SETH)}}   \\

\hline
$2$ & $2 - \delta$ & $\Omega(m^{2-o(1)})$ & Unweighted                                      
&            \autoref{thm:hardundir2approx}        \\ 

\hline
{$3$}                
& {$3 - \frac{2}{\ell} - \delta$} 
& {$\Omega(m^{1 + 1/(\ell - 1) - o(1)})$}  
& {Unweighted.}                                                   
&            \autoref{thm:hard3colorundir}        \\ 

\hline
{$\Omega(\log n)$} & {Any} & {$\Omega(m^{2-o(1)})$} & {Unweighted.}               
&            \autoref{thm:hardlogcolordiam}        \\ 

\hline
\multicolumn{5}{|l|}{\textbf{Directed $k$-mode Diameter Upper Bounds}} \\ 

\hline
$2$ & $n$ & $O(m)$ & Weighted. 
&         \autoref{thm:2colorfindirdiam}           \\ 

\hline
$2$ & $2$ & {$\Tilde{O}(m)$} & {Weighted DAG. }
&         \cite{radiusdiamapprox2016} + Cor. \ref{cor:dagmindiam}           \\ 

\hline
{$2$}                
& {$\pars{\frac{3}{2},1}$
}                             
& {$O(m^{0.414}n^{1.522}+ n^{2+o(1)})$}                            
& {Unweighted DAG. }
&         \cite{mindistanceDAG2021} + Cor. \ref{cor:dagmindiam}          \\ 

\hline
\multicolumn{5}{|l|}{\textbf{Directed $k$-mode Diameter Lower Bounds (Under SETH)}}      \\ 

\hline
{$2$}                
& {$2 - \delta$}                  
& {$\Omega(m^{2-o(1)})$}                   
& {Unweighted.}                                                         &        \autoref{thm:hardundir2approx}            \\ 

\hline
{$3$}                
& {Any}                           
& {$\Omega(m^{2-o(1)})$}                   
& {Unweighted DAG.}                                                    
&         \autoref{thm:hard3colordir}           \\ 

% \hline

% \end{tabular}   
% \captionsetup{justification=centering}
% \caption{Diameter Results. Algorithms that match their lower bound are in bold. \\The runtimes marked with $*$ assume $\omega = 2$.}
% \end{table}

% \begin{table}[]
% \begin{tabular}{|c|c|l|l|c|}
% \hline
% {\textbf{k}}       
% & {\textbf{Approx.}}        
% & {\textbf{Runtime}}                
% & {\textbf{Comments}}                                                   & \textbf{Reference} \\ 

\hline
\multicolumn{5}{|l|}{\textbf{Undirected $k$-mode Radius Upper Bounds}}    \\ 

\hline
{$\boldsymbol{k}$}              
& {$\boldsymbol{3}$}                             
& {$\boldsymbol{\Tilde{O}(mk!)}$}                
& 
{\begin{tabular}
    [c]{@{}l@{}}\textbf{Weighted.} \\\textbf{Tight for constant } $k\geq 3$.
\end{tabular}}                                                                           
&           \textbf{\autoref{thm:radius3approx}}         \\ 

\hline
\multicolumn{5}{|l|}{\textbf{Undirected $k$-mode Radius Lower Bounds (Under the HS Hypothesis)}} \\ 

\hline
{$2$}                
& {$2 - \delta$}                  
& {$\Omega(m^{2-o(1)})$}                   
& {Unweighted.}
&          \autoref{thm:hard2colorundirradius}          \\ 

\hline
{$3$}                
& {$3 - \frac{2}{\ell} - \delta$}          
& {$\Omega(m^{1 + 1/(3\ell - 5) - o(1)})$} 
& {\begin{tabular}
    [c]{@{}l@{}}Weighted. \\Conditioned on the new $\ell$-HSC.
\end{tabular}} 
&      \autoref{thm:hard3colorundirradius}              \\ 

\hline
{$\Omega(\log n)$} 
& {Any}                           
& {$\Omega(m^{2-o(1)})$}                   
& {Unweighted.}
&         \autoref{thm:hardlogcolorradius}           \\ 

\hline
\multicolumn{5}{|l|}{\textbf{Directed $k$-mode Radius Upper Bounds for DAGs}}\\ 

\hline
{$2$}                
& {$n$}                           
& {$O(m)$}                   
& {Weighted DAG.}
&            \autoref{thm:dagfinecc}       \\ 

% \hline
% {2}                
% & {$3$}                           
% & {$O(m\sqrt{n}\log (Mn))$}                   
% & {Weighted DAG.}
% &            TODO       \\ 

\hline

\multicolumn{5}{|l|}{\textbf{Directed $k$-mode Radius Lower Bounds (Under the HS Hypothesis)}}\\ 

\hline
{$2$}                
& {Any}                           
& {$\Omega(m^{2-o(1)})$}                   
& {Unweighted.}
&            \autoref{thm:hard2colordirradius}        \\ 

\hline
{$2$}                
& {$2-\delta$}                           
& {$\Omega(m^{2-o(1)})$}                   
& {Unweighted DAG.}
&            \autoref{thm:hard2colordagradius}        \\ 

\hline
{$3$}                
& {Any}                           
& {$\Omega(m^{2-o(1)})$}                   
& {Unweighted DAG.}
&            \autoref{thm:hard3colordagradius}       \\ 

\hline

\end{tabular}
\captionsetup{justification=centering}
% \caption{Radius Results. Algorithms that match their lower bound are in bold. }
\caption{Our Results. Algorithms that match their lower bound are in bold. \\The runtimes marked with $*$ assume $\omega = 2$.}
\label{tab:results}
\end{table}

%% file: 2-Preliminaries.tex
% \yael{Hitting set lemma, SETH and HSC?}

\section{Preliminaries}
Let $G = (V,E)$ be a graph with $n = |V|$ vertices and $m = |E|$ edges. For a $k$-multimode graph $\Gg=(V, E_1, \ldots, E_k)$, denote $n =|V|$ and $m = |E_1| + \ldots, + |E_k|$, we will also use $e(\Gg)$ to denote the number of edges in $\Gg$.
Define $G_i = (V, E_i)$ and for any $u,v\in V$ let $d_{G_i}(u,v)$ be the length of the shortest path from $u$ to $v$ in the graph $G_i$. When $\Gg$ is clear from context, we denote this distance by $d_i(u,v)$. The $k$-mode distance of $u,v$ is defined as $d_{\Gg}(u,v) = \min_{i\in [k]} d_i(u,v)$.

Throughout this paper, we will often associate each `mode' with a color. We can assign a color to each set of edges $E_i$ and think of $\Gg$ as the graph $G = (V, E = \bigcup_i E_i)$ where every edge is assigned at least one color. A path that uses a single mode corresponds to a monochromatic path under this coloring.

%Throughout this paper we use colors to represent the different edge sets in a multimode graph $E_1, E_2, \ldots$.
Consider the special case of $2$-multimode graphs with edge sets $E_1,E_2$. We use `red distance' to refer to $d_{G_1}$, meaning paths using only edges in $E_1$ (`red edges'). Likewise, we use `blue distance' to refer to $d_{G_2}$. In $3$-multimode graphs, $\Gg = (V, E_1, E_2, E_3)$, we use `green distance' to refer to $d_{G_3}$. 

In this work we present many approximation algorithms to $k$-mode diameter, radius and eccentricities of $k$-multimode graphs. We define an $(\alpha, \beta)$-approximation algorithm for a value $D$ to be such that outputs $\Tilde{D}$ satisfying $D \leq \Tilde{D}\leq \alpha D + \beta$. When $\beta = 0$ we have a multiplicative approximation and call it an $\alpha$-approximation.

We call a directed $k$-multimode graph $\Gg = (V, E_1, \ldots, E_k)$ a $k$-multimode DAG (Directed Acyclic Graph) if $G_1, \ldots, G_k$ are all DAGs.

Denote the ball around a vertex $v$ in a graph $G$, or the neighborhood of $v$, by $B_G(v, r)\coloneqq \set{u\in V : d_G(u,v) < r}$. In a $k$-multimode graph, we will use $B_i(v,r)$ to denote $B_{G_i}(v,r)$.

Given a vertex subset $A\subset V$, denote by $\Gg[A]$ the induced subgraph defined by $\Gg[A]\coloneqq (A, E_1\cap A\times A, \ldots, E_k\cap A\times A)$.

The eccentricity of a vertex $v$ in a graph $G$ is defined as $ecc_G(u)\coloneqq\max_{v\in V}d_G(u,v)$. The diameter of $G$, or $D(G)$, is defined as the largest eccentricity in the graph and the radius, or $R(G)$, is defined as the smallest eccentricity. We naturally extend these definitions to $k$-multimode graphs using the $k$-mode distance. We refer to points $u,v$ such that $d_G(u,v) = D(G)$ as `diameter endpoints' and say that these points `achieve the diameter'. We refer to a point $c$ such that $ecc_G(c) = R(G)$ as a `center' of $G$.

The $k$-mode distance is a generalization of the well studied min-distance. In a directed graph, the min-distance between two vertices $u,v$ is defined as $d_{\min}(u,v) \coloneqq\min\set{d(u,v), d(v,u)}$. Approximating the diameter of a graph under the min-distance (or \mindiam) has been studied in \cite{radiusdiamapprox2016}, \cite{chechiklinearmindiam2022}, \cite{mindistanceDAG2021},  \cite{mindistance2019}. In our approximation algorithms we will use the following lemma from \cite{radiusdiamapprox2016}:
\begin{lemma}[\cite{radiusdiamapprox2016}, Lemma C.2]\label{lm:finecc}
    There is a $O(m+n)$-time algorithm that determines which vertices in a directed graph G have finite min-eccentricity.
\end{lemma}

% ST diam stuff
A variant of the diameter that we use throughout this paper is the $ST$-diameter. Given two sets $S,T\subset V$, the $ST$-diameter is defined as $D_{S,T}\coloneqq \max_{s\in S, t\in T}d(s,t)$. Backurs, Roditty, Segal, Vassilevska W. and Wein explored this variant in \cite{tightdiamapprox2018}. We will use many of their results throughout this paper. We will use the following algorithms in approximating the undirected $k$-mode diameter.

\begin{lemma}[\cite{tightdiamapprox2018}, Claim 24, Theorem 25]\label{lm:stdiamalg}
    Given $G = (V,E)$ and $S,T\subset V$, there exists an $\Tilde{O}(m\sqrt{n})$ time algorithm that computes a $2$-approximation for the $ST$-diameter and a $O(m)$ time algorithm that computes a $3$-approximation for the $ST$-diameter.
\end{lemma}

Backurs et al. \cite{tightdiamapprox2018} also proved conditional lower bounds surrounding the hardness of approximating the $ST$-diameter.

\begin{lemma}[\cite{tightdiamapprox2018}, Theorem 7]\label{lm:stdiamlb}
    Under the Strong Exponential Time Hypothesis (SETH), for every integer $\ell\geq 2$, any $3 - \frac{2}{\ell}-\delta$-approximation algorithm for $ST$-diameter requires $m^{1 + 1/(\ell - 1) - o(1)}$ time for any $\delta > 0$.
\end{lemma}

%% file: 4-5-Algorithms.tex
\section{Approximating Undirected $k$-mode Diameter and Radius}\label{section:alg_undir}
\input{4-Alg_Undir}

\section{Approximating Directed $k$-mode Diameter and Radius}
\label{section:alg_dir}\input{5-Alg_Dir}

%% file: 4-Alg_Undir.tex
In the following two sections we focus on $k$-multimode graphs with small values of $k$, $k \leq \poly\log(n)$. Under SETH and the Hitting Set Hypothesis, computing the exact $k$-mode diameter or radius requires $\Omega((nm)^{1-o(1)})$ time, as it is at least as hard as computing these values in ordinary graphs \cite{radiusdiamapprox2016}. Therefore, we have little hope of computing the exact diameter or radius polynomially faster than the trivial $\tilde{O}(mn)$ time algorithm. Instead, we focus on approximating these parameters. Our results are summarized in table \ref{tab:undirAlg}.

% Computing the exact values of the $k$-mode diameter or radius of a $k$-multimode graph is at least as hard as computing these values in ordinary graphs. Therefore, under SETH and the Hitting Set Conjecture, we cannot hope to compute these values faster than $O(nm)$ time \cite{radiusdiamapprox2016}. We now focus our attention on approximating the $k$-mode diameter and radius for small values of $k$, beginning with undirected graphs. Our results are summarized in table \ref{tab:undirAlg}.

\begin{table}[] 
\center
\begin{tabular}{|c|c|c|l|l|c|}

\hline
\textbf{}
& {$\boldsymbol{k}$}       
& {\textbf{Approx.}}        
& {\textbf{Runtime}}                
& {\textbf{Comments}}                                                 
& \textbf{Reference} \\ 

\hline
Diameter & $2$ & $3$ & $O(m)$ & Weighted & \autoref{thm:2color3approx}     \\

\hline
Diameter & $2$ & $(2, 2M)$ & $\Tilde{O}(mn^{3/4})^*$     
& \begin{tabular}[c]{@{}l@{}}
    Non negative edge weights $\leq M$.\\ If $\omega > 2$ the runtime is\\  
    $m\cdot \pars{\frac{n^{1.5}}{(m\sqrt{n})^{1/\omega}} + \pars{\frac{m}{n}}^{1/(\omega -      2)}}$.
\end{tabular}
&   \autoref{thm:2approx}                \\ 

\hline
Diameter & $2$ & $(2.5, 2.5M)$ & $\Tilde{O}(m\sqrt{n})^*$
& \begin{tabular}[c]{@{}l@{}}
    Non negative edge weights $\leq M$.\\ If $\omega > 2$ the runtime  is \\
    $m\cdot \pars{\frac{n}{m^{1/\omega}} + \pars{\frac{m}{n}}^{1/(\omega - 2)}}$.
\end{tabular}            
&    \autoref{thm:2.5approx}                \\

\hline
\textbf{Diameter} & $\boldsymbol{3}$  & $\boldsymbol{3}$ & $\boldsymbol{\Tilde{O}(m)}$ 
& \textbf{Weighted.}                                    
& \textbf{\autoref{thm:3color3approx}}             \\ 

\hline
\textbf{Radius} & {$\boldsymbol{k}$}              
& {$\boldsymbol{3}$}                             
& {$\boldsymbol{\Tilde{O}(mk!)}$}                
& 
{\begin{tabular}
    [c]{@{}l@{}}\textbf{Weighted.} \\\textbf{Tight for constant } $k\geq 3$.
\end{tabular}}                                                                           
&           \textbf{\autoref{thm:radius3approx}}         \\

\hline
\end{tabular}
\captionsetup{justification=centering}
\caption{Result Summary: Undirected Algorithms. \\Algorithms that match their lower bounds are in bold.}\label{tab:undirAlg}
\end{table}

We begin with undirected graphs. First, we present a simple algorithm running in linear time that provides a $3$-approximation to the $k$-mode diameter when $k=2$. We then show how to extend the idea of this algorithm to use in the context of a general $\alpha$-approximation. We use this extended idea to obtain a near $2$-approximation and a near $2.5$-approximation algorithm for $2$-mode diameter.

Next, we extend the $3$-approximation algorithm for $2$-mode diameter to the case of $k=3$ and obtain a near linear\footnote{Running in $\Tilde{O}(m)$ time.} time $3$-approximation algorithm for the $3$-mode diameter. 

Finally, we turn our attention to approximating the $k$-mode radius of a $k$-multimode graph. Here we are able to show a $3$-approximation algorithm for the $k$-mode radius for any $k$, with a running time of $O(mk!)$. 

In \autoref{section:diamLB} we prove that for $k\geq 3$, any $3-\delta$ approximation algorithm for $k$-mode diameter must run in super linear time, showing that our near linear time $3$-approximation algorithm for $3$-mode diameter is in fact tight. Similarly, in \autoref{section:radiusLB} we show that for $k\geq 3$, any $3-\delta$ approximation algorithm for $k$-mode radius must run in super linear time. Therefore, for any constant $k\geq 3$ our $3$-approximation algorithm for $k$-mode radius runs in near linear time and is tight.

\input{4-Alg_Undir_Diam}

% \input{4-icalp_Alg_UndirDiam}
\input{4-Alg_Undir_Radius}

%% file: 4-Alg_Undir_Diam.tex
\newcommand{\sptwoapprox}{\ensuremath{\mathsf{SP}~2\text{-}\mathsf{MODE}\text{-}\mathsf{DIAM}~2\text{-}\mathsf{APPROX}}}
\newcommand{\sptwohalfapprox}{\ensuremath{\mathsf{SP}~2\text{-}\mathsf{MODE}\text{-}\mathsf{DIAM}~2.5\text{-}\mathsf{APPROX}}}
\newcommand{\stdiamtwoapprox}{\ensuremath{\mathsf{ST}\text{-}\mathsf{DIAM}~2\text{-}\mathsf{APPROX}}}

\subsection{Linear Time 3-Approximation for 2-mode Diameter}

In this section we show our first approximation algorithm for undirected $2$-mode diameter, running in linear time and providing a multiplicative $3$-approximation. We prove this result for unweighted graphs but note that by replacing BFS with Dijkstra's algorithm we obtain the same result for weighted graphs while only adding a $\log n$ factor to the running time.

\begin{theorem}\label{thm:2color3approx}
There exists an $O(m)$ time algorithm that computes a $3$-approximation of the $2$-mode diameter of an unweighted, undirected $2$-multimode graph. 
\end{theorem}

\begin{proof}
Given a $2$-multimode graph $\Gg = (V, E_1, E_2)$ with $2$-mode diameter $D$, our algorithm will output a pair of points $a,b$ such that $\frac{D}{3}\leq d_{\Gg}(a,b) \leq D$.

Start by running BFS in the red graph $G_1$ and the blue graph $G_2$ from an arbitrary node $z\in V$. Let $X = \set{v\in V \mid d_1(v,z) < d_2(v,z)}$, the points closer to $z$ in red than in blue. Let $Y = V\setminus X$. Denote by $\alpha_0 \coloneqq \max_{v\in X}d_1(z,v)$, the largest distance between $z$ and a node in $X$. Likewise define $\beta_0\coloneqq \max_{v\in Y}d_2(z,v)$.

Run BFS in $G_1$ from the set $X$ by contracting the set $X$ into a single vertex and running BFS from it. Let $y\in Y$ be the point furthest away from $X$ in red. Similarly, run BFS in $G_2$ from the set $Y$ and find the point $x\in X$ furthest away from $Y$ in blue. 

Return $\Tilde{D} = \max(d_{\mathcal{G}}(x,y), \alpha_0, \beta_0)$. We claim that $D/3 \leq \Tilde{D} \leq D$. \\

Clearly $\Tilde{D} \leq D$. We are left to show $\Tilde{D} \geq D/3$.
If $\alpha_0 \geq  D/3$ or $\beta_0 \geq D/3$, we are finished. Otherwise, consider a pair of diameter endpoints $s,t$. Since any two points in $X$ are within distance $\leq 2\alpha_0 < D$, the points $s,t$ cannot both be in $X$. Likewise, they cannot both be in $Y$. Thus, we can assume without loss of generality $s\in Y, t\in X$. 

Denote by $\alpha_1\coloneqq d_1(y, X)$. By our choice of $y$, we have $d_1(s, X) \leq \alpha_1$. Therefore, there is some point $x'\in X$ such that $d_1(s,x') \leq \alpha_1$. Thus,
\[
D = d_1(s,t) \leq d_1(s,x') + d_1(x', z) + d_1(z, t) \leq \alpha_1 + 2 \alpha_0.
\]
Since $\alpha_0 < D/3$ we conclude $\alpha_1 \geq D/3$. By a similar argument, if we define $\beta_1 = d_2(x, Y)$, we conclude that $\beta_1 \geq D/3$. Therefore, $d_1(x,y) \geq d_1(y, X) = \alpha_1\geq D/3$ and $d_2(x,y) \geq d_2(x,Y) = \beta_1 \geq D/3$. The pair of points $x,y$ have both red and blue distance greater than $D/3$ and so their $2$-multimode distance is  $d_\mathcal{G}(x,y) \geq D/3$.

We conclude that $\Tilde{D}\geq D/3$.

\end{proof}

\subsection{Subquadratic Time $<3$-Approximation Technique}
Still considering a $2$-multimode graph, our next goal is to obtain a better-than-$3$ approximation for the $2$-mode diameter.
The main tool we will use in this section is a generalization of our linear time $3$-approximation algorithm. For the following algorithms we will be given a value of $D$ and determine whether $D(\Gg) < D$ or $D(\Gg) \geq \alpha D$. Thus, binary searching over $D$ will give a $\frac{1}{\alpha}$-approximation to the diameter. For simplicity, we will assume $\frac{1-\alpha}{2} D$ is an integer and $\Gg$ is unweighted. If that is not the case, we get an additional additive error to our approximation, which we address in \autoref{section:4a_more_undir_diam}.

Given $D$, we can rephrase the above 3-approximation algorithm as follows: pick an arbitrary node $z$ and run BFS from it to compute its `red neighborhood' $X_1\coloneqq B_1\pars{z, \frac{D}{3}}$ and its `blue neighborhood' $Y_1\coloneqq B_2\pars{z, \frac{D}{3}}$. Find the points in the blue neighborhood $Y_1$ that are within red distance $<\frac{D}{3}$  of some point in the red neighborhood, call these points $X_2$. Similarly, take the points in $X_1$ that are within blue distance $<\frac{D}{3}$ of some point in the blue neighborhood and call them $Y_2$.  If we take a pair of points $a\in X_1 \setminus Y_2, b\in Y_1\setminus X_2$ we know that their distance from each other in both red and blue is greater than $\frac{D}{3}$ and thus the pair $(a,b)$ provides a $3$-approximation to the diameter. Furthermore, we know any point in $X_2$ is within red distance $<D$ of all points in $X_1$ and within blue distance $<\frac{2D}{3}$ of all points in $Y_1$. Thus, if the diameter is $\geq D$, no diameter endpoint can be in $X_2$ and similarly no diameter endpoint can be in $Y_2$. We conclude that if $X_1\setminus Y_2$ and $Y_1\setminus X_2$ are not empty we can find points $a,b$ that provide a $3$-approximation to the $2$-mode diameter, and otherwise $D(\Gg) < D$.

\begin{figure}[ht]
    \centering
    \includegraphics[width=0.5\textwidth]{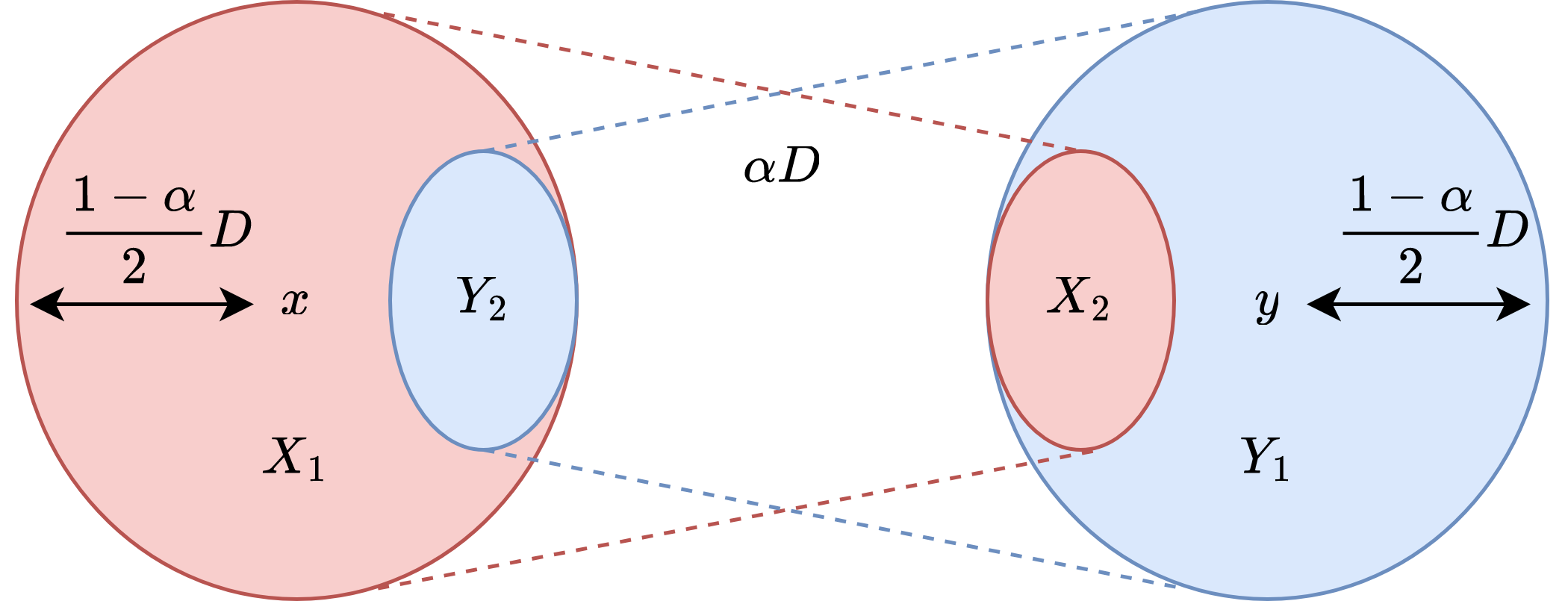}
    \caption{Defining the sets $X_1, X_2, Y_1, Y_2$.}
    \label{fig:pair_alpha_approx}
\end{figure}

For a general approximation factor $\alpha$, instead of considering the red and the blue neighborhoods of a single point $z$, consider a pair of points $x,y$ and take the red neighborhood of one and the blue neighborhood of the other.
Define the red neighborhood of $x$, $X_1 = B_1(x, \frac{1-\alpha}{2}D)$ and the blue neighborhood of $y$, $Y_1 = B_2(y, \frac{1-\alpha}{2}D)$. Consider the points in $Y_1$ that are within red distance $\alpha D$ of some point in $X_1$, $X_2 = B_1(x, \frac{1+\alpha}{2}D)\cap Y_1$. Similarly, define $Y_2 = B_1(y, \frac{1+\alpha}{2}D)\cap X_1$. The sets are illustrated in figure \ref{fig:pair_alpha_approx}, where $d_1$ distance is illustrated in red and $d_2$ distance is illustrated in blue.

As with the 3-approximation, these sets have two useful properties.

\begin{claim}\label{clm:pairdist}
If $X_1\setminus Y_2 \neq \emptyset$ and $Y_1 \setminus X_2 \neq \emptyset$ then we can find a pair of points $a,b$ with $d_\Gg(a,b) \geq \alpha D$.
\end{claim}

\begin{proof}
Take an arbitrary $a\in X_1 \setminus Y_2$ and $b\in Y_1 \setminus X_2$. If $d_1(a,b) < \alpha D$, then 
$$d_1(x,b) \leq d_1(x,a) + d_1(a,b) < \alpha D + \frac{1-\alpha}{2}D = \frac{1+\alpha}{2}D$$
and so $b\in X_2$, a contradiction. Thus $d_1(a,b)\geq \alpha D$ and similarly $d_2(a,b)\geq \alpha D$. Therefore $d_\Gg(a,b) \geq \alpha D$. 
\end{proof}

\begin{claim}\label{clm:stpair}
Let $s,t$ have distance $d_\Gg(s,t)\geq D$.
If $s \in X_1$ and $t\in Y_1$, then $X_1\setminus Y_2 \neq \emptyset$ and $Y_1 \setminus X_2 \neq \emptyset$.
\end{claim}

\begin{proof}
If $X_1 = Y_2$, then $s\in X_1 = Y_2$ and therefore $d_2(s,y) < \frac{1 + \alpha}{2}D$. Thus, $d_2(s,t) \leq d_2(s,y) + d_2(y,t) < \frac{1 + \alpha}{2}D + \frac{1 - \alpha}{2}D = D$, in contradiction to $d_\Gg(s,t) \geq D$. We arrive at a similar contradiction if $Y_1 = X_2$.
\end{proof}

Therefore, our goal is now to find a pair of points $x,y$ such that $s\in X_1$ and $t\in Y_1$. Using the above claims, this will allow us to find a pair of points with $d_\Gg(a,b)\geq \alpha D$ in linear time.\\

To do so, consider running BFS from an arbitrary point $z$ in both the red and the blue graphs $G_1, G_2$. If the algorithm found no point $p$ such that $d_\Gg(z,p) \geq \alpha D$, then $z$ is close to $s$ and $t$ in either red or blue. Since $\alpha \leq \frac{1}{2}$, we can assume without loss of generality that $d_1(z,s) < \alpha D$ and $d_2(z,t) < \alpha D$. Now, if $d_1(z,a) \geq \frac{1-\alpha}{2}D$, since we assumed $\frac{1-\alpha}{2}D\in \NN$, there is some point $x$ on the shortest red path between $z$ and $s$ such that  $d_1(x,s) = \frac{1-\alpha}{2}D$, and so \[d_1(z,x)  = d_1(z,s) - d_1(x,s) <  \pars{\alpha - \frac{1-\alpha}{2}}D = \frac{3\alpha - 1}{2}D.\] 

If $d_1(z,a) < \frac{1-\alpha}{2}D$ then there exists a point $x$ on this shortest path such that $d_1(x,s)\leq \frac{1-\alpha}{2}D$ and $d_1(z,x) < \frac{3\alpha -1}{2}D$.
Similarly, there exists a point $y$ such that $d_2(y,t) \leq \frac{1-\alpha}{2}D$ and $d_2(z,y) < \frac{3\alpha - 1}{2}D$.

Therefore, by considering all pairs of points $x\in B_1(z, \frac{3\alpha -1}{2}D ), y\in B_2(z, \frac{3\alpha -1}{2}D)$ we can find a pair that satisfy the conditions of claim \ref{clm:stpair} and thus also the conditions of claim \ref{clm:pairdist}. Now, by running a linear time algorithm for every pair of points $x\in B_1(z, \frac{3\alpha -1}{2}D), y\in B_2(z, \frac{3\alpha -1}{2}D)$ we can find a pair of points of distance $\geq \alpha D$. In fact, we don't need to run the above algorithm for every pair of potential points $x,y$ independently. We can improve our running time using fast rectangular matrix multiplication. We use the standard notation $M(J,K,L)$ to indicate the time it takes to multiply a $J\times K$ matrix by a $K\times L$ matrix using fast rectangular matrix multiplication.

We are now ready to claim the following generalization to our $2$-approximation algorithm.

\begin{lemma}\label{lm:point2smallnbhds}
Let $\Gg$ be a $2$-multimode graph  with $2$-mode diameter $\geq D$. Given  $\frac{1}{3}< \alpha \leq \frac{1}{2}$ and a point $z$ with $|B_1(z, \frac{3\alpha - 1}{2} D)|\leq n^\delta$ and $|B_2(z, \frac{3\alpha - 1}{2} D)| \leq n^\delta$, we can find a pair of points $a,b$ with $d_{\Gg}(a,b) \geq \alpha D$ in time $O(n^\delta m) + M(n^\delta, n, n^\delta)$.
\end{lemma}

\begin{proof}
Define $X = B_1\pars{z, \frac{3\alpha - 1}{2} D}$ and $Y = B_2\pars{z, \frac{3\alpha - 1}{2}D}$. From every $x\in X$ run BFS in $d_1$ to compute $X_1 \coloneqq B_1(x, \frac{1-\alpha}{2}D)$ and $X_2 \coloneqq B_1(x, \frac{1+\alpha}{2}D)$ as illustrated in figure \ref{fig:pair_alpha_approx} (without taking an intersection with $Y_1$). Let $M_X^1$ be a $n^\delta \times n$ matrix with rows indexed by $X$ and columns indexed by $V$ and let $M_X^2$ be a $n\times n^\delta$ matrix with rows indexed by $V$ and columns indexed by $X$. Define $M_X^1$ and $M_X^2$ as follows,
\begin{align*}
M_X^1[x,u] &= 
\begin{cases}
    1 & \text{if } u\in X_1,\\
    0 & \text{otherwise}.
\end{cases}
\\
M_X^2[u, x] &= 
\begin{cases}
    1 & \text{if } u\notin X_2,\\
    0 & \text{otherwise}.
\end{cases}
\end{align*}

Similarly, run BFS from every $y\in Y$ in $d_2$ and define $M_Y^1, M_Y^2$ by,

\begin{align*}
M_Y^1[y,u] &= 
\begin{cases}
    1 & \text{if } u\in Y_1\coloneqq B_2(y, \frac{1-\alpha}{2}D),\\
    0 & \text{otherwise}.
\end{cases}
\\
M_Y^2[u, y] &= 
\begin{cases}
    1 & \text{if } u\notin Y_2\coloneqq B_2(y, \frac{1+\alpha}{2}D),\\
    0 & \text{otherwise}.
\end{cases}
\end{align*}

Finally, compute the boolean matrix product
\[
Z = M_X^1 \cdot M_Y^2 \land M_Y^1 \cdot M_X^2.
\]

For any $x\in X, y\in Y$, we have $Z[x,y] =1 \iff X_1\setminus Y_2 \neq \emptyset$ and $Y_1 \setminus X_2 \neq \emptyset$. 
Therefore, by claim \ref{clm:pairdist}, if $Z[x,y] = 1$ we can find a pair of points with $d_\Gg(a,b) \geq \alpha D$. If $D(\Gg)\geq D$, there exist some $x,y$ for which $s\in X_1$ and $t\in Y_1$, and so by claim \ref{clm:stpair} at least one pair will have $Z[x,y] = 1$.\\

To summarize, in order to find a pair of points with $d_\Gg(a,b)\geq \alpha D$ we first run BFS from every point in $X$ and $Y$. We construct the matrices $M_X^1, M_X^2, M_Y^1, M_Y^2$ and use them to compute $Z$. We then find a pair $x,y$ such that $Z[x,y] = 1$. For this pair, we compute the sets $X_1, X_2, Y_1, Y_2$ and return $a\in X_1\setminus Y_2$ and $b\in Y_1 \setminus X_2$. These points satisfy $d_\Gg(a,b)\geq \alpha D$. If we are unable to find such points we conclude that $D(\Gg) < D$.

The final running time is dominated by $O(n^\delta m)$ for the BFS searches and $M(n^\delta, n, n^\delta)$ to preform the matrix multiplications.

\end{proof}

\subsection{$2$-Approximation for $2$-mode Diameter}

Using lemma \ref{lm:point2smallnbhds} we can state our first subquadratic $<3$ approximation for 2-mode diameter. For the sake of clarity, we maintain the simplifying assumptions we used for lemma \ref{lm:point2smallnbhds}. In \autoref{section:4a_more_undir_diam} we show that removing these assumptions adds an additive error to our approximation and results in a $(2,2)$-approximation, rather than $2$-approximation. However, here we will ignore the additive error and focus on the algorithm.

We will use lemma \ref{lm:point2smallnbhds} for our desired 2-approximation using $\alpha = \frac{1}{2}$. Denote this algorithm by $\sptwoapprox(\Gg, z)$. 

\begin{corollary}[Lemma \ref{lm:point2smallnbhds}, $\alpha = \frac{1}{2}$]\label{cor:pair2approx}
Given a point $z$ with $|B_1\pars{z, \frac{D}{4}}|\leq n^\delta$ and $|B_2\pars{z, \frac{D}{4}}|\leq n^\delta$, if $D(\Gg)\geq D$ then we can find a pair of points with $d_\Gg(a,b) \geq \frac{D}{2}$ in time $O(n^\delta m) + M(n^\delta, n, n^\delta)$.
\end{corollary}

Using this, we can construct an algorithm that computes a $2$-approximation of the 2-mode diameter. We will consider two cases, the first is that some diameter endpoint has a large $\frac{D}{4}$ neighborhood in either blue or red. In this case we can sample a hitting set and find a point that is within distance $<\frac{D}{4}$ of a diameter endpoint. We will show how to use this point to find a pair of points of distance $\geq \frac{D}{2}$. Otherwise, we will have a point with small red and blue neighborhoods and we can apply corollary \ref{cor:pair2approx}.

\begin{theorem} \label{thm:2approx}
Given a $2$-multimode undirected, unweighted graph $\Gg$ and an integer $D$, if $\omega = 2$, there exists an $O(m\cdot n^{3/4})$ time algorithm that does one of the following with high probability:
\begin{enumerate}
    \item Finds a pair of points $a,b$ with $d_\Gg(a,b) \geq \frac{D}{2}$.
    \item Determines that the 2-mode diameter of $\Gg$ is $< D$.
\end{enumerate}
If $\omega > 2$ the runtime of the algorithm is $m\cdot \pars{\frac{n^{1.5}}{(m\sqrt{n})^{1/\omega}} + \pars{\frac{m}{n}}^{1/(\omega - 2)}}$.
\end{theorem}

\begin{proof}
Fix a pair of diameter endpoints $s,t$ and assume $d_\Gg(s,t)\geq D$. In this case our algorithm will output a pair of points with distance $\geq \frac{D}{2}$. 

Consider a point $z$ of distance $<\frac{D}{4}$ from a diameter endpoint, w.l.o.g $d_1(s,z) < \frac{D}{4}$. Run BFS from $z$ to compute $A = B_1\pars{z, \frac{D}{4}}$ and $B = V\setminus B_1\pars{z, \frac{3D}{4}}$. By our assumption, $s\in A$. Any point $x\in V\setminus B$ is within distance $<D$ of $s$ since $d_1(s,x) \leq d_1(s,z) + d_1(z, x) < \frac{D}{4} + \frac{3D}{4} = D$. Therefore, $t\in B$. Furthermore, any pair of points $a\in A, b\in B$ have $d_1(a,b) \geq \frac{D}{2}$.

In \cite{tightdiamapprox2018}, Backurs, Roditty, Segal, Vassilevska W. and Wein provide an algorithm that computes a $2$-approximation to the $ST$-diameter of two sets in an undirected graph running in time $\Tilde{O}(m\sqrt{n})$ (Lemma \ref{lm:stdiamalg}). Denote their algorithm for approximating the $ST$-diameter in graph $G$ between sets $S,T$ by $\stdiamtwoapprox(G, S, T)$.

In $G_2$, the $ST$-diameter of $A,B$ is $\geq D$ since $d_2(s,t)\geq D$. Thus, running $\stdiamtwoapprox(G_2,A,B)$ will return a pair of points $a\in A, b\in B$ with $d_2(a,b) \geq \frac{D}{2}$ and hence $d_\Gg(a,b) \geq \frac{D}{2}$. 

Therefore, if we are able to find such a point $z$ of distance $<\frac{D}{4}$ from a diameter endpoint, then we are able to find $a,b$ with $d_\Gg(a,b)\geq \frac{D}{2}$ in time $\Tilde{O}(m\sqrt{n})$.\\

Let $\delta$ be a parameter to be set later. We claim that algorithm \ref{alg:diam2approx} does in fact output a pair of points $a,b$ with $d_\Gg(a,b) \geq \frac{D}{2}$ if $D(\Gg)\geq D$. For simplicity, we assume that when running BFS from some vertex $x$ in the algorithm we never find a point $y$ with $2$-mode distance $d_\Gg(x,y) \geq \frac{D}{2}$, otherwise we would output the pair $x,y$ and finish.

 \begin{algorithm}
     \caption{2-Approximation Algorithm for $2$-mode Undirected Diameter.}\label{alg:diam2approx}
     \begin{algorithmic}[1]
        \item \textbf{Input:} 2-multimode graph $\Gg = (V, E_1, E_2), D$.
        \item \textbf{Output:} Nodes $a,b$ with $d_\Gg(a,b) \geq \frac{D}{2}$ or determine $D(\Gg) < D$.
        \State Randomly sample a subset $S\subset V$ of size $|S| = n^{1-\delta}\log n$.
        \State $W\leftarrow \emptyset$
        \For{$x\in S$}
            \State $W \leftarrow W \cup B_1\pars{x, \frac{D}{4}}\cup B_2\pars{x, \frac{D}{4}}$
            \For{$i = 1,2$}
                \State $A\leftarrow B_i\pars{x, \frac{D}{4}}, B \leftarrow V\setminus B_i\pars{x, \frac{3D}{4}}$
                \State $(a, b) \leftarrow \stdiamtwoapprox(G_{3-i},A,B)$
                \If{$d_\Gg(a,b) \geq \frac{D}{2}$} \Return $(a,b)$ \EndIf
            \EndFor
        \EndFor
        \If{$W\neq V$}
            \State Select $z\notin W$
            \State $(a,b) \leftarrow \sptwoapprox(\Gg, z)$
            \If{$d_\Gg(a,b) \geq \frac{D}{2}$} \Return $(a,b)$ \EndIf
        \EndIf\\
        \Return $D(\Gg) < D$.
        \end{algorithmic}
\end{algorithm}

\paragraph{Correctness:} For every node $x\in V$, call the ball of radius $\frac{D}{4}$ around it in $G_1$ its red neighborhood and the ball of radius $\frac{D}{4}$ in $G_2$ its blue neighborhood. By the hitting set lemma, with high probability the set $S$ will hit the red neighborhood of any node with red neighborhood larger than $n^\delta$ and likewise for its blue neighborhood. Thus, after considering all the blue and red neighborhoods of points in $W$, if we have not covered all points, then the remaining uncovered vertices will have small red and blue neighborhoods. W.h.p the vertex $z$ selected in line 15 will have small neighborhoods. Thus, $\sptwoapprox(\Gg, z)$ will run in time $O(n^\delta m) + M(n^\delta, n, n^\delta)$ and obtain a $2$-approximation to the diameter.

If no point remains after covering all neighborhoods ($W=V$ in line 14), then $s$ must have been covered at some point. This means that one of the iterations of $\stdiamtwoapprox(G_{3-i},A,B)$ (line 9) was run from a point within distance $<\frac{D}{4}$ of $s$ in the appropriate color. Therefore, this run will find a pair of points with $d_\Gg(a,b)\geq \frac{D}{2}$.

\paragraph{Runtime:} The running time of this algorithm is dominated by line 16, which is run once and takes $O(n^\delta m) + M(n^\delta, n, n^\delta)$, and line 9 which is run $\Tilde{O}(n^{1-\delta})$ time and take $\Tilde{O}(m\sqrt{n})$. The total running time is therefore,
\[
n^\delta \cdot m + M(n^\delta, n, n^\delta) + n^{1.5 - \delta}\cdot m.
\]

If the fast matrix multiplication exponent $\omega = 2$, then $M(n^\delta, n , n^\delta) = O(n^{1+\delta})$, which is dominated by $m\cdot n^\delta$. Thus, our final running time is $O(m\cdot (n^\delta + n^{1.5 - \delta})$. Setting $\delta = \frac{3}{4}$ gives us the desired running time of $O(m\cdot n^{3/4})$.\\

If $\omega > 2$, we will use the bound $M(n^\delta, n , n^\delta) \leq O(n^{1-\delta + \delta \omega})$. Denote $m = n^{1+\mu}$ and consider two cases. First, if $\delta \leq \frac{3}{4}$ the running time of the algorithm is $O(n^{2.5 + \mu -\delta} + n^{1-\delta + \omega \delta})$, setting $\delta = \frac{1.5 + \mu}{\omega}$ gives us a running time of $m\cdot \frac{n^{1.5}}{(m\sqrt{n})^{1/\omega}}$. In this case $\delta = \frac{1.5 + \mu}{\omega} \leq \frac{3}{4}$ when $\mu \leq \frac{3}{4}(\omega - 2)$.

On the other hand, if $\delta > \frac{3}{4}$, the running time of the algorithm is $O(n^{1+\mu + \delta} + n^{1-\delta + \omega \delta})$. Setting $\delta = \frac{\mu}{\omega - 2}$ gives a running time of $m\cdot \pars{\frac{m}{n}}^{1/(\omega - 2)}$ and in this case $\delta > \frac{3}{4}$ when $\mu > \frac{3}{4}(\omega - 2)$. 

Note that the term $m\cdot \frac{n^{1.5}}{(m\sqrt{n})^{1/\omega}}$ dominates the term $m\cdot \pars{\frac{m}{n}}^{1/(\omega - 2)}$ if and only if $\mu \geq \frac{3}{4}(\omega - 2)$. Therefore, the final running time of the algorithm comes out to
\[
m\cdot \pars{\frac{n^{1.5}}{(m\sqrt{n})^{1/\omega}} + \pars{\frac{m}{n}}^{1/(\omega - 2)}}.
\]

We can improve upon this running time by using fast rectangular matrix multiplication. For instance, if $m=n$ we can obtain a running time of $n^{1.846}$.

\end{proof}

% By running a binary search over the possible values of $D$ we conclude the following corollary.

% \begin{corollary}
% If $\omega =2$, there exists an $\Tilde{O}(m\cdot n^{3/4})$ time algorithm to compute a $2$-approximation of the $2$-mode diameter of an unweighted, undirected $2$-multimode graph.
% \end{corollary}

% In this algorithm we used lemma \ref{lm:point2smallnbhds} with the simplifying assumption that $\Gg$ is unweighted and $\frac{1-\alpha}{2}D\in \NN$. Removing these assumptions produces an additive error of twice the maximum edge weight in $\Gg$, which we address in section \ref{section:4afixadditiveerror} of the appendix.

\subsection{2.5-Approximation for 2-mode Diameter}

In this section we present a faster algorithm than the previous $2$-approximation, that achieves a slightly weaker approximation factor of 2.5. Again we use a small vs. large neighborhood argument combined with lemma \ref{lm:point2smallnbhds}, now for $\alpha = \frac{2}{5}$. We maintain the simplifying assumptions we used in lemma \ref{lm:point2smallnbhds} and theorem \ref{thm:2approx}. Removing them results in an additional additive error, which we compute in \autoref{section:4a_more_undir_diam}.

In this section we will make use of lemma \ref{lm:point2smallnbhds} again, this time using $\alpha = \frac{1}{2.5} = \frac{2}{5}$. Denote this algorithm by \sptwohalfapprox$(\Gg,z)$.

\begin{corollary}[Lemma \ref{lm:point2smallnbhds}, $\alpha = \frac{2}{5}$] \label{cor:pair2.5approx}
Given a point $z$ with $|B_1\pars{z, \frac{D}{10}}|\leq n^\delta$ and $|B_2\pars{z, \frac{D}{10}}|\leq n^\delta$, if $D(\Gg) \geq D$ then we can find a pair of points with $d_\Gg(a,b) \geq \frac{2D}{5}$ in time $O(n^\delta m) + M(n^\delta, n, n^\delta)$.
\end{corollary}

We can now construct a 2.5-approximation algorithm using corollary \ref{cor:pair2.5approx}. If we are able to find a point with small $\frac{D}{10}$ neighborhoods in both red and blue, we can run \sptwohalfapprox. Otherwise, we can sample and hit the $\frac{D}{10}$ neighborhood of some diameter endpoint $s$. We can then use this point to find in linear time another point $y$ that is far from both endpoints of the diameter and show that this is sufficient for achieving our desired approximation. 

\begin{theorem} \label{thm:2.5approx}
Given a $2$-multimode undirected, unweighted graph $\Gg$ and integer $D$, if $\omega = 2$ there exists an $O(m\cdot \sqrt{n})$ time algorithm that does one of the following:
\begin{enumerate}
    \item Finds a pair of points $a,b$ with $d_\Gg(a,b) \geq \frac{2D}{5}$.
    \item Determines that the 2-mode diameter of $G$ is $< D$.
\end{enumerate}
If $\omega > 2$ the runtime of the algorithm is $m\cdot \pars{\frac{n}{m^{1/\omega}} + \pars{\frac{m}{n}}^{1/(\omega - 2)}}$.
\end{theorem}
% m\cdot \pars{\frac{n}{m^{1/\omega}} + \pars{\frac{m}{n}}^{1/(\omega - 2)}}.

As was the case with the $2$-approximation, we can obtain the following corollary.

\begin{corollary}
If $\omega = 2$, there exists an $\Tilde{O}(m\sqrt{n})$ time algorithm to compute a $5/2$ approximation of the $2$-mode diameter of an unweighted, undirected $2$-multimode graph.
\end{corollary}

\begin{proof} [Proof of theorem \ref{thm:2.5approx}]
Let $\delta$ be a parameter to be set later. We claim that algorithm \ref{alg:diam2.5approx} outputs a pair of points $a,b$ with $d_\Gg(a,b) \geq \frac{2D}{5}$ if $D(\Gg)\geq D$. Again we assume that if while running BFS from $x$ (whether from a point in $S$, from $y$ in line 8 or within $\sptwohalfapprox$), if we find a point $y$ with $2$-mode distance $d_\Gg(x,y) \geq \frac{2D}{5}$ we return the pair and finish.

 \begin{algorithm}
     \caption{2.5-Approximation Algorithm for $2$-mode Undirected Diameter.}\label{alg:diam2.5approx}
     \begin{algorithmic}[1]
        \item \textbf{Input:} 2-multimode graph $\Gg = (V, E_1, E_2), D$.
        \item \textbf{Output:} Nodes $a,b$ with $d_\Gg(a,b) \geq \frac{2D}{5}$ or determine $D(\Gg) < D$.
        \State Randomly sample a subset $S\subset V$ of size $|S| = n^{1-\delta}\log n$.
        \State $W\leftarrow \emptyset$
        \For{$x\in S$}
            \State $W \leftarrow W \cup B_1\pars{x, \frac{D}{10}}\cup B_2\pars{x, \frac{D}{10}}$\label{line:setw}
            \For{$i = 1,2$}
                \State Find $y$ such that $d_i(x,y) = \frac{D}{2}$. Run BFS from $y$. \label{line:findy}
            \EndFor
        \EndFor
        \If{$W\neq V$} \label{line:wneqv}
            \State Select $z\notin W$
            \State $(a,b) \leftarrow \sptwohalfapprox(\Gg, z)$ \label{line:calltwohalfapprox}
            \If{$d_\Gg(a,b) \geq \frac{D}{2}$} \Return $(a,b)$ \EndIf
        \EndIf\\
        \Return $D(\Gg) < D$.
        \end{algorithmic}
\end{algorithm}

\paragraph{Correctness:} As in the proof of theorem \ref{thm:2approx}, if $W\neq V$ in line \ref{line:wneqv} then we run the algorithm from corollary \ref{cor:pair2.5approx} on a point with small red and blue neighborhoods. Thus $\sptwoapprox(\Gg, z)$ will run in $O(n^\delta m) + M(n^\delta, n, n^\delta)$ time and obtain a $2.5$-approximation to the diameter.

If no point remains after covering all neighborhoods ($W=V$ in line \ref{line:wneqv}), then line \ref{line:findy} will have run at least once from a point $x$ such that $d_i(x,s) < \frac{D}{10}$. In this case, $d_i(x,t) > \frac{D}{2}$ so we will be able to find a $y$ with $d_i(x,y) = \frac{D}{2}$. Now we claim that running BFS from $y$ will find a point $z$ with $d_\Gg(y,z) \geq \frac{2D}{5}$. 

First, $y$ cannot have $d_i(s,y)< \frac{2D}{5}$ as that would imply $d_i(x,y) \leq d_i(x,s) + d_i(s,y) < \frac{D}{10} + \frac{2D}{5} = \frac{D}{2}$. On the other hand, $d_i(t,y)\geq \frac{2D}{5}$ since otherwise $d_i(s,t) \leq d_i(s,x) + d_i(x,y) + d_i(y,t) < \frac{D}{10} + \frac{D}{2} + \frac{2D}{5} = D$. Thus, since some $c\in \set{s,t}$ must have $d_{3-i}(c,y)\geq \frac{2D}{5}$, it must have $d_\Gg(c,y)\geq \frac{2D}{5}$.

\paragraph{Runtime:} The running time of this algorithm is dominated by line \ref{line:calltwohalfapprox}, which takes $O(n^\delta m) + M(n^\delta, n, n^\delta)$, and lines \ref{line:setw}-\ref{line:findy}, in which we run a constant number of BFS searches for every $s\in S$. The total running time is therefore,
\[
n^\delta \cdot m + M(n^\delta, n, n^\delta) + n^{1 - \delta}\cdot m.
\]

If the fast matrix multiplication exponent $\omega = 2$, then $M(n^\delta, n , n^\delta) = O(n^{1+\delta})$, and thus our final running time is $O(m\cdot (n^\delta + n^{1 - \delta})$. Setting $\delta = \frac{1}{2}$ gives us a running time of $O(m\cdot \sqrt{n})$.\\

If $\omega > 2$, we will again use the bound $M(n^\delta, n , n^\delta) \leq O(n^{1-\delta + \delta \omega})$ and denote $m = n^{1+\mu}$. If $\delta \leq \frac{1}{2}$ the running time of the algorithm is $O(n^{2 + \mu -\delta} + n^{1-\delta + \omega \delta})$, setting $\delta = \frac{1 + \mu}{\omega}$ gives us a running time of $m\cdot \frac{n}{m^{1/\omega}}$. In this case $\delta\leq \frac{1}{2}$ when $\mu \leq \frac{\omega - 2}{2}$.

On the other hand, if $\delta > \frac{1}{2}$, the running time of the algorithm is $O(n^{1+\mu + \delta} + n^{1-\delta + \omega \delta})$. Setting $\delta = \frac{\mu}{\omega - 2}$ gives a running time of $m\cdot \pars{\frac{m}{n}}^{1/(\omega - 2)}$ and  $\delta > \frac{1}{2}$ when $\mu > \frac{\omega - 2}{2}$. 

The term $m\cdot \frac{n}{m^{1/\omega}}$ dominates the term $m\cdot \pars{\frac{m}{n}}^{1/(\omega - 2)}$ if and only if $\mu \geq \frac{\omega - 2}{2}$. Therefore, the final running time of the algorithm comes out to
\[
m\cdot \pars{\frac{n}{m^{1/\omega}} + \pars{\frac{m}{n}}^{1/(\omega - 2)}}.
\]

We can improve upon this running time by using fast rectangular matrix multiplication. For instance, if $m=n$ we can obtain a running time of $n^{1.553}$.
\end{proof}

\subsection{Linear Time $3$-Approximation for $3$-mode Diameter}

In this section we consider a $3$-multimode graph $\Gg = (V, E_1, E_2, E_3)$ and attempt to approximate its $3$-mode diameter. We extend our result for $2$-multimode graphs and provide a near linear time $3$-approximation algorithm for the $3$-mode diameter.

To find a $3$-approximation to the $3$-mode diameter we will look for a pair of points with $d_\Gg(a,b)\geq \frac{D}{3}$.
We will use a similar approach to the 2-mode case, finding two sets in which all points are far apart in both the red and blue graphs ($G_1, G_2$). We can then compute the ST-diameter of the two sets in green ($G_3$), finding two points with large $d_3$ distance as well. As in previous sections, we provide an algorithm for unweighted graphs and note that by replacing BFS with Dijkstra's algorithm the algorithm works for weighted graphs, while adding a $\log n$ factor to the running time.

\begin{theorem} \label{thm:3color3approx}
Given a $3$-multimode undirected, unweighted graph $\Gg$ and integer $D$, there exists an $O(m)$ time algorithm that does one of the following with high probability:
\begin{enumerate}
    \item Finds a pair of points $a,b$ with $d_\Gg(a,b) \geq \frac{D}{3}$.
    \item Determines that the 3-mode diameter of $G$ is $< D$.
\end{enumerate}
\end{theorem}

By performing a binary search over the possible values of $D$ we obtain our desired result.
% \begin{corollary}
% There exists an $\Tilde{O}(m)$ time algorithm for computing a $3$-approximation of the colorful diameter of an unweighted, undirected $3$-multimode graph.
% \end{corollary}

\begin{proof}
% [Proof of theorem \ref{thm:3color3approx}]
% Given a $3$-multimode graph $\Gg$, assume that $D(\Gg)\geq D$. 
% We will show that in this case we are able find a pair of points $a,b$ with $d_\Gg(a,b)\geq \frac{D}{3}$ and thus if we fail to find such a pair, we output $D(\Gg)< D$. Denote the diameter endpoints by $s,t$, $d_\Gg(s,t)\geq D$.
Choose an arbitrary vertex $p\in V$ and run BFS from it in all three graphs, $G_1, G_2, G_3$. Let $X = B_1(p, D/3), Y = B_2(p, D/3) \setminus  X$ and $Z = V \setminus (X\cup Y)$. If there exists a vertex $z\in Z$ such that $d_3(p,z) \geq D/3$, then $z\notin \bigcup_{i=1}^3 B_i(p, D/3)$ and so $d_{\mathcal{G}}(p,z) \geq D/3$. We can therefore return $(p,z)$ and finish. Otherwise $Z \subset B_3(p,D/3)$.

As was the case for 2-mode diameter, we note that for any pair of points $u,u'\in X$, $d_1(u,u') \leq d_1(u,p) + d_1 (p,u') < 2D/3$ and so $d_{\mathcal{G}}(u,u') < 2D/3$. Likewise, for any pair $v,v'\in Y$ or $w,w'\in Z$ we have $d_{\mathcal{G}}(v,v')<2D/3$ and $d_{\mathcal{G}}(w,w') < 2D/3$.

Thus, any pair of points within one of the sets $X,Y,Z$ has $3$-mode distance $<D$. Next, consider pairs of points $x\in X$ and $y\in Y$. Run BFS in red  from the set $X$ and run BFS in blue from the set $Y$. Let $X_0 = Y \cap B_1(X, D/3)$ be the points in $Y$ of red distance $<D/3$ from the set $X$. We claim that any pair of points $x'\in X, y'\in X_0$ have 3-mode distance $<D$.
For any point $y\in X_0$ there exists $x'\in X$ with $d_1(x',y) < D/3$. Therefore,  $d_1(x,y) \leq d_1(x,x') + d_1(x',y) < 2D/3 + D/3 = D$. 

Now define $Y_0 = X \cap B_2(Y, D/3)$. Using a symmetric argument we can show every pair of points $y\in Y, x\in Y_0$ have $d_2$ distance $<D$. Therefore, if $X=Y_0$ or $Y= X_0$, then every pair of points $x\in X, y\in Y$ have distance $d_\Gg(x,y)<D$.

Denote $X_1 = X \setminus X_0$ and $Y_1 = Y \setminus Y_0$. We are left to handle the case when both $X_1, Y_1$ are non-empty. Every pair of points $x\in X_1, y\in Y_1$ has $d_1(x,y)\geq D/3$ and $d_2(x,y) \geq D/3$ so in order to determine if $d_{\mathcal{G}}(x,y)\geq D/3$ we only need to consider $d_3(x,y)$. 

To do so, preform a simple approximation to the ST-diameter of the sets $X_1, Y_1$ in $G_3$ as follows.
Pick two arbitrary nodes $\hat{x}\in X_1, \hat{y}\in Y_1$ and run BFS from each one of them in $G_3$. If $Y_1\not \subset B_3(\hat{x}, D/3)$  then we will have found a point $y\in Y_1$ such that $d_3(\hat{x}, y) \geq D/3$ and so $d_{\mathcal{G}}(\hat{x}, y) \geq D/3$. Similarly, if $X_1\not \subset B_3(\hat{y}, D/3)$, we will have found a point $x\in X$ such that $d_\mathcal{G}(x,\hat{y}) \geq D/3$. 

Otherwise, $X_1 \subset B_3(\hat{y}, D/3)$ and $Y_1 \subset B_3(\hat{x}, D/3)$. Thus, for every pair of points $x\in X_1, y\in Y_1$, 
\[
d_\mathcal{G}(x,y)\leq d_3(x,y) \leq d_3(x,\hat{y}) + d_3(\hat{y}, \hat{x}) + d_3(\hat{x}, y) < D/3 + D/3 + D/3 = D.
\]

And so we conclude that any pair of points $x\in X, y\in Y$ have $d_\mathcal{G}(x,y) < D$.\\

We can now repeat this algorithm for the pairs of sets $X,Z$ and $Y,Z$ and their respective pairs of colors. If the algorithms finds no pair of points with $d_\mathcal{G}(u,v)\geq D/3$, we conclude that the 3-mode diameter of $\mathcal{G}$ is $< D$.

\end{proof}

% Picture:
% \includegraphics[width=0.5\textwidth]{3color-3approx.png}
% \yael{TODO: make picture?}

%% file: 4-Alg_Undir_Radius.tex
\subsection{Linear Time 3-Approximation for $k$-mode Radius}
\newcommand{\radiusapprox}{\ensuremath{\mathsf{RADIUS}\text{-}3\text{-}\mathsf{APPROX}}}

In the following section we consider the problem of approximating the $k$-mode radius of a $k$-multimode graph. Unlike the algorithms we constructed for approximating the $k$-mode diameter, in this case we provide a general algorithm for all $k$, with running time parameterized by $k$ and $m$. Our algorithm provides a $3$-approximation for the $k$-mode radius and runs in time $\Tilde{O}(mk!)$.
Thus, for a constant $k$ we have a near linear time algorithm. We again present our algorithm for unweighted graphs and note that by replacing BFS with Dijkstra's algorithm we can extend the result to weighted graphs while adding $\log n$ to the running time.

\begin{theorem}\label{thm:radius3approx}
There exists an $\Tilde{O}(m k!)$ time algorithm that computes a $3$-approximation of the $k$-mode radius of an unweighted, undirected $k$-multimode graph.
\end{theorem}

\begin{proof}
Our algorithm receives a threshold $R$ and either finds a vertex $v$ with $ecc(v) \leq 3R$ or determines that $R(\Gg)> R$. Performing a binary search over the values of $R$ will give the desired approximation. 

First we consider the 2-mode radius ($k=2$) and then we will show how to generalize this algorithm to any $k$. Assuming $R(\Gg)\leq R$, our approach will be to find a point that is close to the center $c$ of the graph in both colors. If we find a vertex $x$ with $d_1(c,x)\leq 2R$ and $d_2(x,c) \leq 2R$, then for any vertex $v$ we have $d_i(c,v)\leq R$ for some $i\in \set{1,2}$ and so $d_i(x,v)\leq 3R$.

To find such a point, run BFS from an arbitrary node $z$. If $ecc(z)\leq 3R$ we are finished, otherwise we have a point $w$ with $d_{\mathcal{G}}(z,w) > 3R$. The points $z,w$ are both within distance $R$ of the center $c$ in either red or blue, if $z,w$ were both within distance $R$ of the center in the same color we would have a contradiction. Thus, if $d_{\mathcal{G}}(z,c) = d_1(z,c)\leq R$, then me must have $d_{\mathcal{G}}(w,c) = d_2(w,c)\leq R$ and if $d_{\mathcal{G}}(z,c) = d_2(z,c)\leq R$, then me must have $d_{\mathcal{G}}(w,c) = d_1(w,c)\leq R$. 

Let $X: = B_1(z,R) \cap B_2(w, R)$ and $Y:= B_2(z,R) \cap B_1(w,R)$. By the above observation, $c$ is either in $X$ or in $Y$. If $c\in X$, then any point $x\in X$ will have $d_1(x,c) \leq d_1(x,z) + d_1(z,c) \leq 2R$ and $d_2(x,c) \leq d_2(x,w) + d_2(w,c) \leq 2R$. This would imply $ecc(x) \leq 3R$. Similarly, if $c\in Y$ then any point $y\in Y$ will have $ecc(y) \leq 3R$. Therefore, by choosing arbitrary $x\in X, y\in Y$ and running BFS from each, we will have found a vertex of eccentricity $\leq 3R$.

This algorithm runs a constant number of BFSs and if $R(\Gg)\leq R$ it finds a point with $ecc(v)\leq 3R$. Otherwise, we can conclude $R(\Gg) > R$. Therefore, in $\Tilde{O}(m)$ time our algorithm provides a $3$-approximation the the $2$-mode radius. We can now extend this idea to any value of $k$.\\

For a general $k$ we can define a recursive algorithm, where at each node we `guess' the color (or mode) in which it achieves its distance to the center (in fact we try all possible colors). We run algorithm~\ref{alg:radius3approx}, \radiusapprox, on input $\Gg,R$, a subset of colors $C$ and a subset of vertices $W\subset V$. We would like $\Gg,R,C,W$ to have the following property:
\begin{enumerate}[label=(P\arabic*)]
    \item \label{radiusprop}If $R(\Gg)\leq R$, then $c\in W$ and $d_i(w, c) \leq 2R$ for $w\in W, i\in C$. 
\end{enumerate}

 \begin{algorithm}
     \caption{\radiusapprox}\label{alg:radius3approx}
     \begin{algorithmic}[1]
        \item \textbf{Input:} Graph $G = (V, E)$, threshold $R$, color subset $C \subseteq [k]$ and vertex subset $W\subseteq V$.
        \item \textbf{Output:} Node $x$ with $ecc(x) \leq 3R$ or determine $R(G) > R$.
        \State Select arbitrary $x \in W$. \label{line:3radiuschoosex}
        \If {$ecc(x) \leq 3R$} \Return $x$.\label{line:3radiusreturn}
        \EndIf
        % \If {$C = [k]$ and $W \neq \emptyset$} Continue.
        % \EndIf
        \State Let $y$ be a point with $d_\mathcal{G}(x, y) > 3R$.\label{line:3radiuschoosey}
        \For{$i\in [k] \setminus C$}\label{line:3radiusforloop}
        \State $C'\leftarrow C \cup \set{i}, W' \leftarrow W \cap B_i(y, R)$.
        \State \radiusapprox$(G, R, C', W')$.\label{line:3radiusrecusivecall}
        \EndFor
        \end{algorithmic}
\end{algorithm}

\paragraph{Correctness:} We run algorithm \ref{alg:radius3approx} on $W=V$ and $C=\emptyset$ at first, at which point \ref{radiusprop}~trivially holds. Subsequently, we claim that if $C,W$ satisfy \ref{radiusprop}, then in at least one of the recursive calls in line \ref{line:3radiusrecusivecall}, the sets $C', W'$ satisfy \ref{radiusprop}.

Let $x\in W$ and $y$ be such that $d_\Gg(x,y)> 3R$ (as chosen in lines \ref{line:3radiuschoosex},\ref{line:3radiuschoosey}). Since $d_j(x,c)\leq 2R$ for any $j\in C$, we must have $d_j(c,y)> R$. Thus, there exists $i\in [k]\setminus C$ such that $d_i(c,y)\leq R$. For this iteration of the for the loop in line \ref{line:3radiusforloop}, we have $c\in B_i(y,R)$ and so any point  $z\in B_i(y,R)$ has $d_i(z,c)\leq 2R$. Hence $C',W'$ satisfy property \ref{radiusprop}.

Therefore, there exists at least one path of length $k$ in the recursion tree for which \ref{radiusprop}~is maintained at all levels of the recursion. Since each call adds a color to $C$, after $k$ calls we will have $C = [k]$. In this case, $W$ is a nonempty set which has $d_i(c,x)\leq 2R$ for any $x\in W, i\in [k]$. Thus, any $x\in W$ will have $ecc(x)\leq 3R$ and will be returned in line \ref{line:3radiusreturn}. If no $x$ is returned by the algorithm we can conclude that $R(\Gg) > R$.

\paragraph{Runtime:} Every call to $\radiusapprox(G,R,C,W)$ preforms a constant number of BFS searches and $k - |C|$ calls to $\radiusapprox$ with $|C'| = |C| + 1$. Therefore, the total running time of the algorithm is $O(mk!)$.
\end{proof}

% non recursive pseudo code

% The algorithm is described in the following pseudo-code:
%  \begin{algorithm}\label{alg:linearradius3approx}
%      \caption{3-Approximation Algorithm for $k$-mode Undirected Radius.}
%      \begin{algorithmic}
%         \item \textbf{Input:} Graph $G = (V, E), R$.
%         \item \textbf{Output:} Node $x$ with $ecc(x) \leq 3R$ or determine $R(G) > R$.
%         \State $W \leftarrow V$, $C \leftarrow \emptyset$.
%         \State Pick arbitrary $x_1 \in V$.
%         \If {$ecc(x_1) \leq 3R$} \Return $x_1$
%         \Else ~~ Let $y_1$ be a point with $d_\mathcal{G}(x_1, y_1) > 3R$.
%         \For {$c_1 \neq c_2 \in [k]$}
%             \State $ W \leftarrow B_{c_1}(x_1, R) \cap B_{c_2}(y_1, R)$.
%             \For {$i = 2, \ldots, k$}
%                 \State Pick arbitrary $x_i \in W$.
%                 \If {$ecc(x_i) \leq 3R$} \Return $x_i$
%                 \Else ~~ Let $y_i$ be a point with $d_\mathcal{G}(x_i, y_i) > 3R$.
%                 \For{$c_i \notin C$}
%                     \State $W \leftarrow W \cap B_{c_i}(y_i, R)$.
%                     \State $C \leftarrow C \cup \set{c_i}$.
%                 \EndFor
%                 \EndIf        
%             \EndFor
%         \EndFor
%         \EndIf
%         \end{algorithmic}
%  \end{algorithm}
 
%

%% file: 5-Alg_Dir.tex
In this section we consider the problem of approximating the $k$-mode diameter and radius of a directed $k$-multimode graph for $k=2$. This is a direct generalization of the \textit{min-distance} problem.

As we show later in our lower bounds, approximating directed $k$-mode diameter or radius faster than the trivial $O(knm)$ time algorithm is significantly harder than the undirected case. We show that under popular conjectures, no subquadratic algorithm exists for approximating the $k$-mode diameter for $k\geq 3$ or the $k$-mode radius for $k\geq 2$ for general graphs. We are therefore left to focus on approximating directed 2-mode diameter for general graphs or limiting ourselves to a subset of directed graphs, the most natural choice being directed acyclic graphs (DAGs). In this case, we show that the problem of approximating the 2-mode diameter of a DAG is equivalent to the problem of approximating the min-diameter of a DAG. For a general 2-multimode directed graph, we show that one can differentiate between finite and infinite 2-mode diameter in linear time. We then consider the problem of approximating the 2-mode radius of a DAG and show that in this case one can also differentiate between finite and infinite 2-mode radius in linear time. Our results are summarized in table \ref{tab:dirAlg}.

Most of the algorithms we discuss in this section address reachability and thus we are not concerned with edge weights.

\begin{table}[] 
\center
\begin{tabular}{|c|c|c|l|l|c|}

\hline
\textbf{}
& {$\boldsymbol{k}$}       
& {\textbf{Approx.}}        
& {\textbf{Runtime}}                
& {\textbf{Comments}}                                                 
& \textbf{Reference} \\

\hline
Diameter & $2$ & $2$ & {$\Tilde{O}(m)$} & {Weighted DAG. }
&         \cite{radiusdiamapprox2016} + Cor. \ref{cor:dagmindiam}           \\ 

\hline
Diameter & {$2$}                
& {$\pars{\frac{3}{2},1}$}                             
& {$O(m^{0.414}n^{1.522}+ n^{2+o(1)})$}                            
& {Unweighted DAG. }
&         \cite{mindistanceDAG2021} + Cor. \ref{cor:dagmindiam}          \\

\hline
Diameter & $2$ & $n$ & $O(m)$ & Weighted. 
&         \autoref{thm:2colorfindirdiam}           \\ 

\hline
Radius & {$2$}                
& {$n$}                           
& {$O(m)$}                   
& {Weighted DAG.}
&            \autoref{thm:dagfinecc}       \\

\hline
\end{tabular}
\captionsetup{justification=centering}
\caption{Result Summary: Directed Algorithms. }\label{tab:dirAlg}
\end{table}

\subsection{Approximating the 2-mode Diameter of a DAG}
First we consider the case where $\Gg = (V,E_1,E_2)$ is a DAG, i.e. $G_1$ and $G_2$ are acyclic. In this case the problem of approximating the 2-mode diameter is in fact equivalent to a different, well studied problem, of approximating the min-diameter of a DAG.  Much work has gone into approximating the min-diameter of general and acyclic graphs \cite{radiusdiamapprox2016, chechiklinearmindiam2022, mindistanceDAG2021, mindistance2019}. In the case of DAGs, Abboud, Vassilevska W. and Wang \cite{radiusdiamapprox2016} introduced a simple linear time algorithm for computing a 2-approximation of the min-diameter. In later work, Dalirrooyfard and Kaufmann \cite{mindistanceDAG2021} developed an algorithm running in $O(m^{0.414}n^{1.522}+n^{2+o(1)})$ time that gives an $\pars{\frac{3}{2},1}$ approximation for the min-diameter of DAGs. We show how to extend these results to approximating the 2-mode diameter of a DAG.

Set a topological ordering $<_1$ of $G_1$ and a topological order $<_2$ of $G_2$. We call a 2-multimode DAG $\Gg$ \textit{aligned} if $G_1, G_2$ have opposite topological order, meaning there exists an ordering of $V$, $v_1, \ldots, v_n$, such that $v_1 <_1 \ldots <_1 v_n$ and $v_1 >_2 \ldots >_2 v_n$. Note that if $\Gg$ is not aligned then it has infinite 2-mode diameter. If there exist $u,v\in V$ such that $u<_1 v$ and $u<_2 v$, then $d_1(v,u) = d_2(v,u) = \infty$ and so $d_\Gg(v,u) = \infty$. 

The following claim relates the 2-mode diameter of an aligned 2-multimode DAG $\Gg$ to the min-diameters of $G_1, G_2$.

\begin{claim}\label{clm:dagmindiam}
If $\Gg = (V,E_1, E_2)$ is an aligned 2-multimode DAG, then $$D(\Gg) = \max(\mindiam(G_1), \mindiam(G_2)).$$
\end{claim}

\begin{proof}
Denote by $v_1, \ldots, v_n$ the vertices of $\Gg$ in $<_1$-order and reverse $<_2$-order. For any $i<j$, $d_2(v_i, v_j) = \infty$ and so $d_\Gg(v_i, v_j) = d_1(v_i, v_j) = d_{\min, G_1}(v_i, v_j)$. Similarly, $d_\Gg(v_j, v_i) = d_2(v_j, v_i) = d_{\min, G_2}(v_j, v_i)$. Therefore,
\begin{align*}
D(\Gg) &= \max_{i, j\in [n],i\neq j}\set{d_\Gg(v_i, v_j)} = 
\max\pars{\max_{i<j}\set{d_\Gg(v_i, v_j)},\max_{i>j}\set{d_\Gg(v_i, v_j)}}\\
&= \max\pars{\max_{i<j}\set{d_{\text{min}, G_1}(v_i, v_j)},\max_{i>j}\set{d_{\text{min}, G_2}(v_i, v_j)}}\\
&= \max(\mindiam(G_1), \mindiam(G_2)).
\end{align*}
\end{proof}

We can check in $O(m)$ time if a 2-multimode graph $\Gg$ is aligned. If it isn't, we determine $D(\Gg) = \infty$ and otherwise, using claim \ref{clm:dagmindiam}, we can use any algorithm for approximating the min-diameter of DAGs to approximate the 2-mode diameter of a 2-multimode DAG.

\begin{corollary} \label{cor:dagmindiam}
If there exists a $T(n,m)$ time algorithm computing an $\alpha$-approximation of the min-diameter of a DAG, then there exists a $T(n,m) + O(m)$ time algorithm computing an $\alpha$-approximation on the 2-mode diameter of a 2-multimode DAG.
\end{corollary}

Using the $2$-approximation for $\mindiam$ of \cite{radiusdiamapprox2016} running in $\Tilde{O}(m)$ time, or the $\pars{\frac{3}{2}, 1}$-approximation of \cite{mindistanceDAG2021} running in $O(m^{0.414}n^{1.522} + n^{2 +o(1)})$ time, we obtain the same results for approximating $2$-mode diameter of a DAG.

\subsection{Approximating the 2-mode Diameter of a General Directed Graph}
Now consider the case of general directed graphs. In this case, the problem of computing the 2-mode diameter is no longer equivalent to that of computing the min-diameter of $G_1, G_2$ individually. The problem seems to be harder than the min-diameter problem and it remains an open question whether one can compute a constant approximation for this value in linear or even subquadratic time. In the following theorem we are able to show that in near linear time we can determine whether a 2-multimode graph has finite or infinite diameter.

\begin{theorem}\label{thm:2colorfindirdiam}
    There exists an $\Tilde{O}(m)$ time algorithm that determines whether a directed 2-multimode graph $\Gg = (V, E_1, E_2)$ has finite 2-mode diameter.
\end{theorem}

To prove theorem \ref{thm:2colorfindirdiam}, consider the graph of strongly connected components (SCCs) of the red graph $G_1$. Denote this graph by $H_1$ and note that it is a DAG and so has a topological order. Call this the `red topological order', referring to an ordering of the red SCCs, or $G_1$-SCCs. For each node in $H_1$ (red SCC) consider its min-eccentricity, i.e. its eccentricity under min-distance in the DAG $H_1$. Using lemma \ref{lm:finecc}, we can determine in linear time which red SCCs have finite eccentricity, meaning every SCC behind them in the topological ordering can reach them and they can reach any SCC ahead of them. Call red SCCs with finite min-eccentricity in $H_1$ `finite' SCCs and otherwise we call them `infinite'.

Similarly, let $H_2$ be the DAG of the strongly connected components of the blue graph $G_2$. We can in linear time compute a topological ordering of $H_2$, the `blue topological order' and determine which blue SCCs are finite. 

Finally, we introduce one additional set of notations. Given a red  SCC $S$, denote the vertices behind $S$ in the red topological order by $A_1(S)$ and the vertices ahead of it in the red topological order by $B_1(S)$. Similarly, define $A_2(S), B_2(S)$ for a blue SCC $S$. When $S$ and $i=1/2$ are clear from context we refer to these sets as $A$ and $B$.

We define an algorithm to determine whether the 2-mode diameter of $\Gg$ is finite using the following three claims.

\begin{claim}\label{clm:infiniteintersecting}
    If some infinite red SCC intersects an infinite blue SCC, then $D(\Gg)=\infty$.
\end{claim}

\begin{proof} 
    Let $S_1$ be an infinite red SCC intersecting an infinite blue SCC $T_1$. Since $S_1$ has infinite eccentricity, there exists some other red SCC $S_2\neq S_1$ such that in the red graph $S_1$ can't reach $S_2$ and $S_2$ can't reach $S_1$. Therefore, if $\Gg$ had finite 2-mode diameter, $S_1$ and $S_2$ would both have to be able to reach each other in blue and thus be contained in the same blue SCC. Since $S_1$ intersects $T_1$ we conclude $S_1\subseteq T_1$ and $S_2\subseteq T_1$. However, by the same argument we can conclude that $T_1\subseteq S_1$ and so $S_1 = T_1$, meaning $S_2\subseteq S_1$, contradiction.
\end{proof}

Now consider a finite SCC $S$ in graph $G_i$ ($i = 1/2$). Since $S$ has finite min-eccentricity, we know all vertices contained in SCCs behind $S$, or $A_i(S)$, can reach $S$ in $G_i$ and all vertices in $S$ can reach all vertices in $B_i(S)$. In the next claim, we show that verifying whether the reverse is true can be done in linear time. We defer the proof of this and the following claim to the end of this section.

\begin{claim}\label{clm:BreachSreachA}
    Given a finite $G_i$-SCC $S$,  we can determine in $O(m)$ time whether $B_i(S)$ can reach $S$ in $\Gg$ and whether $S$ can reach $A_i(S)$ in $\Gg$. 

    Furthermore, we can add a vertex and edges to $\Gg[A_i(S)]$ to define $\Gg_A = (V_A, E^A_1, E^A_2)$ such that all the vertices in $A_i(S)$ can reach each other in $\Gg$ if and only if $D(\Gg_A) < \infty$. The graph $\Gg_A$ satisfies  $|V_A| \leq  |A| + 1$ and $e(\Gg_A) \leq e(\Gg[A]) + |A|$. Likewise, we can define $\Gg_B$.
\end{claim}

We would like use claim \ref{clm:BreachSreachA} to recurse on the sets $A$ and $B$. The following claim guarantees that we can always find an $S$ such that this recursive step cuts down the size of the sets by a constant factor, thus bounding the depth of our recursion. 

\begin{claim}\label{clm:boundrecursion}
    If no infinite red SCC intersects an infinite blue SCC, then there exists a finite $G_i$-SCC $S$  such that the sets $A_i(S),B_i(S)$  satisfy $|A_i(S)|\leq \frac{7}{8}\cdot n,|B_i(S)|\leq \frac{7}{8}\cdot n$.
\end{claim}

We can now prove theorem \ref{thm:2colorfindirdiam}.

\begin{proof}[Proof of theorem \ref{thm:2colorfindirdiam}]
    Combining claims \ref{clm:infiniteintersecting}, \ref{clm:BreachSreachA} and \ref{clm:boundrecursion}, we claim that Algorithm \ref{alg:dirdiamapprox} runs in $\Tilde{O}(m)$ time and determines whether the 2-mode diameter of $\Gg$ is finite.

    \begin{algorithm}
     \caption{$n$-Approximation Algorithm for $2$-mode Directed Diameter.}\label{alg:dirdiamapprox}
     \begin{algorithmic}[1]
        \item \textbf{Input:} 2-multimode graph $\Gg = (V, E_1, E_2)$.
        \item \textbf{Output:}  $D(\Gg)=\infty$ or $D(\Gg) < \infty$.
        \If{$V = \emptyset$}
            \State \Return $D(\Gg) < \infty$ 
        \EndIf
        \State Compute the SCCs of $G_1, G_2$ and determine which are finite.
        \If{An infinite red SCC intersects an infinite blue SCC}
            \State \Return $D(\Gg) = \infty$  \label{line:infiniteintersecting}
        \EndIf
        \State $S\leftarrow$ finite $G_i$-SCC such that $|A_i(S)|\leq \frac{7}{8}n, |B_i(S)|\leq \frac{7}{8}n$, for $i\in \set{1,2}$.
        \State Determine whether $B_i(S)$ can reach $S$ and $S$ can reach $A_i(S)$ and define $\Gg_A, \Gg_B$ as in claim \ref{clm:BreachSreachA}.
        \If{$B_i(S)$ can't reach $S$ or $S$ can't reach $A_i(S)$}
            \State \Return $D(\Gg) = \infty$ 
        \EndIf
        \If{$D(\Gg_A)=\infty$ or $D(\Gg_B) = \infty$} \label{line:recursivestep}
            \State \Return $D(\Gg) = \infty$ 
            \Else \State \Return $D(\Gg) < \infty$
        \EndIf
        \end{algorithmic}
\end{algorithm}

    \paragraph{Correctness:}Claim \ref{clm:infiniteintersecting} shows that if in fact two infinite SCCs intersect then $D(\Gg)=\infty$ and so the value returned in line \ref{line:infiniteintersecting} is correct. Otherwise, by claim \ref{clm:boundrecursion}, there exists a finite $G_i$-SCC $S$ such that $|A_i(S)|\leq \frac{7}{8}n, |B_i(S)|\leq \frac{7}{8}n$. If $B$ can't reach $S$ or $S$ can't reach $A$ then clearly $D(\Gg) = \infty$. Otherwise, $\Gg$ has finite 2-mode diameter if and only if all the vertices in $A$ can reach each other and all the vertices in $B$ can reach each other. By claim \ref{clm:BreachSreachA}, this is true if and only if $\Gg_A$ and $\Gg_B$ have finite 2-mode diameter.

    \paragraph{Runtime:}Other than the recursive step in line \ref{line:recursivestep}, all steps of the algorithm are linear in the size of the input. By claim \ref{clm:BreachSreachA}, $|V(\Gg_A)| \leq |A| + 1 \leq \frac{7}{8}n + 1$ and similarly $|V(\Gg_B)| \leq \frac{7}{8}n + 1$. Therefore, we can bound the depth of the recursion by $O(\log n)$. 
    At each step of the recursion the total number of vertices grows by at most 2, and so at level $d$ of the recursion the total number of new vertices added is $O(d)$. Thus, at level $d$ of the recursion, the total number of vertices is $n + O(d^2)$. Furthermore, at each level of recursion, the total number of edges grows by at most $n$. Therefore, at the last level of recursion the total number of edges is $O(m + n\log^2 n) = \Tilde{O}(m)$ and so the total runtime is $\Tilde{O}(m)$.
\end{proof}

To complete the proof we are left to prove claims \ref{clm:BreachSreachA} and \ref{clm:boundrecursion}.

\begin{proof}[Proof of claim \ref{clm:BreachSreachA}]
    Let $S$ be a finite red SCC (the blue case is identical) and let $T_1, \ldots, T_\ell$ be the blue SCCs that intersect $S$, labeled such that $T_1 < \ldots < T_\ell$ in the blue topological order. Define $A \coloneqq A_1(S)$ and $B \coloneqq B_1(S)$.
    
    Let $U$ be a blue SCC such that $T_1 < U < T_\ell$ in the blue topological order and consider points $x\in U$, $y\in T_1\cap S$ and $y'\in T_\ell \cap S$. The point $x$ is ahead of $y$ in the blue topological order and thus can't reach it in blue. Therefore if $D(\Gg) < \infty$, $x$ must be able to reach $y$ in red. Similarly, $y'$ must be able to reach $x$ in red. Thus, in the red graph $x$ must be able to reach a point in the SCC $S$ and be reached by a point in $S$ and so $x\in S$. We conclude that any blue SCC between $T_1$ and $T_\ell$ must be contained in $S$, otherwise we can determine that $D(\Gg) = \infty$. 
    
    Furthermore, this implies that $T_1,\ldots, T_\ell$ must form a continuous range of SCCs in the red graph $H_1$. Denote the set of SCCs behind this range by $C\coloneqq A_2(T_1)$ and the set of SCCs ahead of this range by $D\coloneqq B_2(T_\ell)$. The sets are depicted in figure \ref{fig:linear_dir_diam}.

    \begin{figure}[ht]
        \centering
        \includegraphics[width=0.5\textwidth]{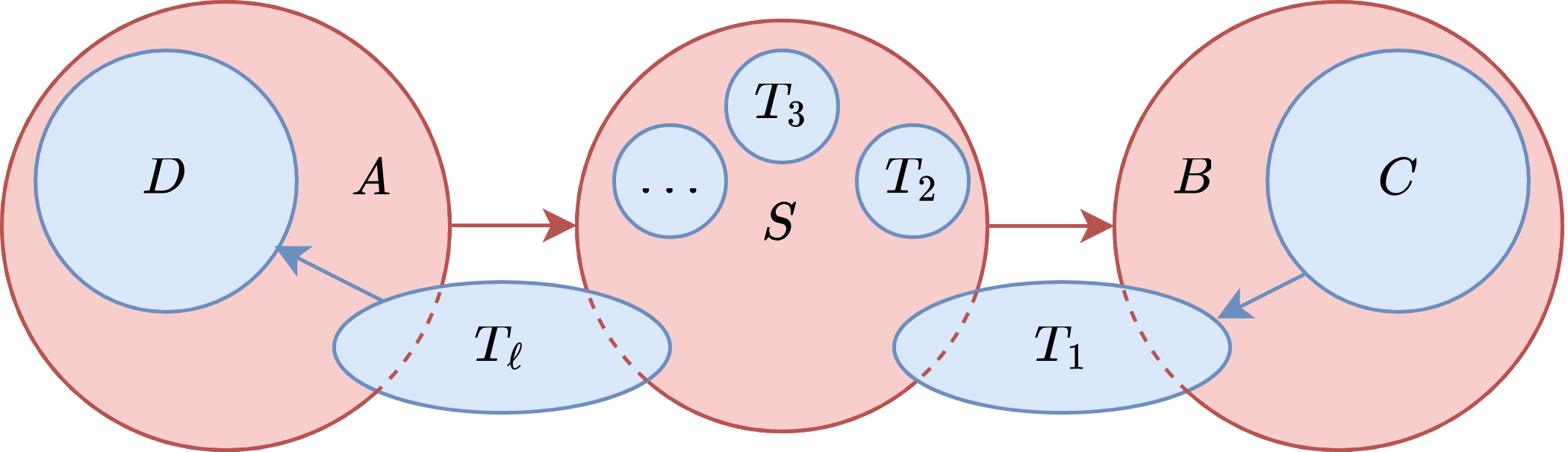}
        \caption{Defining the sets $A,B,C,D,S,T_1,\ldots, T_\ell$.}
        \label{fig:linear_dir_diam}
    \end{figure}

    Note that if $B\not \subseteq C\cup T_1$, then any point $x\in B \setminus (C\cup T_1)$ can't reach $S$ in neither red nor blue. Thus, if $B\not \subseteq C\cup T_1$ we conclude that $D(\Gg) = \infty$, and likewise if $A\not \subseteq D\cup T_\ell$. We can therefore assume that $B\subseteq C\cup T_1$ and $A\subseteq D\cup T_\ell$ (otherwise determine $D(\Gg) = \infty$).
    \\

    If $T_1$ is a finite blue SCC, then all the points in $C$ can reach $T_1$ in blue and $T_1$ can reach all the sets $T_2,\ldots, T_\ell, D$ in blue. Since $B\subseteq C \cup T_1$ we conclude that $B$ can reach $S$. Now consider paths between two points in $B$. If this path is red it must be contained in $\Gg[B]$, since no red path goes from a point in $B$ to a point in $A\cup S$. If this path is blue, it must be contained in $C\cup T_1$. 
    
    Define the 2-multimode graph $\Gg_B$ by adding an additional vertex to $\Gg[B]$ and connecting it with a blue edge to all points in $T_1\cap B$. This graph preserves the connectivity of $B$, since all points in $T_1\cap B$ could already reach each other in blue. Furthermore, any pair of points in $B$ that could reach each other in $\Gg$ can reach each other in $\Gg_B$. Therefore, all points in $B$ can reach each other in $\Gg$ if and only if $D(\Gg_B) < \infty$.

    Conversely, if $T_1$ is an infinite blue SCC, then it must be contained in a single red SCC, otherwise $D(\Gg) = \infty$. Hence, we can assume $T_1 \subseteq S$. Now  $B = C$, so any path between two vertices in $B$, in either red or blue, must be contained in $B$ and so we can define $\Gg_B = \Gg[B]$. We are left to determine if $B$ can reach $S$. 

    Let $U_1, \ldots, U_r$ be the blue SCCs making up $C$. We would like to determine whether all of $U_1, \ldots, U_r$ can reach all of $T_1, \ldots, T_\ell$. If there exists an edge from $U_i$ to $U_j$ for some $i,j\in [r]$, then $U_i$ and $U_j$ can reach $T_1, \ldots, T_\ell$ if and only if $U_j$ can reach all of $T_1,\ldots, T_\ell$. Thus, we can iteratively delete any $U_i$ that has an outgoing edge to another blue SCC contained in $C$. Similarly, we can iteratively delete any $T_i$ that has an incoming edge from another $T_j$. We are left with a subset of SCCs $\Uu \subset \set{U_i}_{i=1}^r, \Tt\subset \set{T_j}_{j=1}^\ell$ such that no edges exists between two SCCs in $\Uu$ or two SCCs in $\Tt$. To check whether all the SCCs in $\Uu$ can reach all the SCCs in $\Tt$ all we have to do is check whether the induced graph of these SCC is a complete bipartite graph.

    Similarly, we can determine whether $A$ can be reached by $S$ in blue and define $\Gg_A$ such that all points in $A$ can reach each other if and only if $D(\Gg_A) < \infty$.

    We note that all the steps above can be preformed in $O(m)$ time and that in the graphs we defined $|V(\Gg_A)|\leq |A| + 1, e(\Gg_A) = e(\Gg[A]) + |A\cap T_\ell|\leq e(\Gg[A]) + |A|$, which completes the proof.
    
\end{proof}

\begin{proof}[Proof of claim \ref{clm:boundrecursion}]
    Let $\Gg$ be a 2-multimode graph in which no infinite red SCC intersects an infinite blue SCC. If there exists some finite $G_i$-SCC of size $|S|\geq \frac{n}{8}$, then  $|A_i(S)|\leq \frac{7}{8}\cdot n,|B_i(S)|\leq \frac{7}{8}\cdot n$ since they are disjoint from $S$. So we can assume all SCCs are smaller than $\frac{n}{8}$.

    Assume no finite red SCC $S$ has $|A_1(S)|,|B_1(S)|\leq \frac{7}{8}$.
    Let $S_1$ be the last finite red SCC (last in red topological order) such that $|A_1(S_1)| < \frac{n}{8}$. If no such SCC exists, let $S_1$ be an auxiliary empty SCC that precedes all others in the red topological order. Similarly, let $S_2$ be the first finite red SCC such that $|B_1(S_2)| < \frac{n}{8}$ or, if none exist, an empty SCC that follows all others in the red topological order.

    Note that any SCC $S$ that falls between $S_1$ and $S_2$ in the red topological order must be infinite. Otherwise, $A_1(S) \geq \frac{n}{8}$ and $B_1(S)\geq \frac{n}{8}$, in contradiction to our assumption. Since the sets $|A_1(S_1)|, |S_1|,|B_1(S_2)|, |S_2|$ all have size $ < \frac{n}{8}$, this gives us a range of $>\frac{n}{2}$ vertices all contained in infinite red SCCs.

    Similarly, if no finite blue SCC $S$ has $|A_2(S)|,|B_2(S)|\leq \frac{7}{8}$ we conclude that $>\frac{n}{2}$ nodes are contained in infinite blue SCCs. Therefore, some node must be contained in both an infinite red SCC and an infinite blue SCC, in contradiction to our assumption that no infinite red SCC intersects an infinite blue SCC.
\end{proof}

\subsection{Approximating the 2-mode Radius of a DAG}
Next, we consider the problem of approximating the $2$-mode radius of a 2-multimode DAG. In the following theorem we show that in linear time, we can identify all vertices of finite eccentricity and thus determine if the 2-mode radius of $\Gg$ is finite or infinite.

\begin{theorem}\label{thm:dagfinecc}
    There exists an $O(m)$ time algorithm that determines which vertices in a 2-multimode DAG $\Gg = (V, E_1, E_2)$ have finite 2-mode eccentricity.
\end{theorem}

\begin{proof}
    Our algorithms consists of three steps. First we sort the vertex set $V$ into 'good' and 'bad' vertices such that all 'bad' vertices have infinite eccentricity and the 'good' vertices are in reverse topological order (meaning that there exists an ordering of $V$, $v_1, \ldots, v_n$, a topological order $<_1$ of $G_1$ and a topological order $<_2$ of $G_2$ such that $v_1 <_1 \ldots <_1 v_n$ and $v_1 > _2 \ldots >_2 v_n$). Next, we determine which good vertices can reach all other good vertices using a similar technique to lemma \ref{lm:finecc}. We move any good vertex that can't reach all other good vertices to the bad set. In the final step, we determine for every bad vertex the set of good vertices that cannot reach it. Since all good vertices can reach each other, these sets will be continuous intervals under the topological orders of $G_1$ and $G_2$. We then determine for every good vertex if it is contained in any one of these intervals, in which case it has infinite eccentricity and otherwise it has finite eccentricity.

    \paragraph{Step 1:}Begin by performing a topological sort of the vertex set in the red graph $G_1$ and the blue graph $G_2$ to obtain the orders $<_1$ and $<_2$ respectively.  Consider the last vertex in the red ordering $<_1$. This vertex cannot reach any other vertex in red and so if any vertex comes before it in the blue topological order it is unable to reach it in $\Gg$. Thus, if this vertex is not first in the blue topological order, it has infinite eccentricity and so we move it to the 'bad' vertex set. If it is first in the blue order, we move the vertex to the 'good' vertex set. We now remove the vertex from the ordering and repeat the process with the next vertex that is now last under $<_1$. After we have sorted all vertices we note that all bad vertices have infinite eccentricity. On the other hand, the good vertices have reverse topological order, since whenever a vertex was moved to the `good' set it was last in the red order and first in the blue order among all vertices yet to be sorted. 

    We can perform this step in $O(m)$ time as follows. Maintain two lists pointing to the same set of vertices, one sorted by $<_1$ and the other by $<_2$. Keep a pointer $p_1$ for the top of the red list and a pointer $p_2$ for the bottom of the blue list. At each step, check if the node that $p_1$ and $p_2$ point to are the same. If they differ, move the node that $p_1$ points at to the bad set, mark it as visited and move $p_1$ down by 1. If they point at the same node, move it to the good set, move $p_1$ down by one and $p_2$ up by one. Keep moving $p_2$ until it is pointing at an unvisited node. This process goes over the list of nodes a constant number of times and thus runs in $O(m)$ time.

    \paragraph{Step 2:} Label the good vertices $v_1, \ldots, v_k$ such that  $v_1 <_1 \ldots <_1 v_k$ and  $v_1 >_2 \ldots>_2 v_k$. Since the vertices have reverse topological order, to have finite eccentricity vertex $v_i$ must be able to reach $v_j$ in red if $j>i$ and in blue if $j <i$.
    
    Using dynamic programming, we determine for each vertex $v$ (good or bad, ordered by red topological order) the last good vertex that can reach it in red, i.e. the maximum $i$ such that $v_i$ can reach $v$. We do this by considering all incoming red edges to the vertex, $(u,v)$, and taking the maximum $i$ such that $u = v_i$ or $v_i$ can reach $u$. Denote this value by $\ell(v)$. This computation takes $O(m)$ time as we consider every edge once.
    
    Now consider a good vertex $v_i$. If for every $j>i$ we have $\ell(v_j)\geq i$, then every good node ahead of $v_i$ can be reached by $v_i$ or some good node ahead of $v_i$. Thus, we can keep taking paths that do not take us past $v_i$ and conclude that $v_i$ can reach $v_j$. To check for which good vertices this property holds we count $s(i) \coloneqq \abs{\set{j>i : \ell(v_j)\geq i}}$. We can compute this value for all $i$ recursively in linear time. First we compute $\abs{\set{j>i : \ell(v_j) = i}}$ for every $i$ by running over the values of $\ell(v_j)$ once and updating relevant counters. For the final good node we have $s(k) = 0$, and for every $i<k$ we have $s(i) = s(i+1) + \abs{\set{j>i : \ell(v_j) = i}}$. 

    We repeat the process for the blue graph and can now determine which good nodes can reach all other good nodes. Any node that cannot has infinite eccentricity and so we move it to the bad vertex set.

    \paragraph{Step 3:} Using the same dynamic programming as in step 2, we now determine for every bad vertex $v$ the largest $i$ such that $v_i$ that can reach it in red and the smallest $j$ such that $v_j$ can reach $v$ in blue. Since for any $i'<i$ we know $v_{i'}$ can reach $v_i$ in red and for any $j'>j$ we know $v_{j'}$ can reach $v_j$ in blue, we conclude that $v_{i'}$ and $v_{j'}$ can reach $v$. If $j > i + 1$, then for any $r\in [i + 1, j - 1]$ the vertex $v_r$ can't reach $v$ and has infinite eccentricity. Thus, every bad vertex gives us an interval of good vertices that cannot reach it. 

    We now have $O(n)$ intervals, relabel them as $\set{[a_j, b_j]}_{j=1}^t$. We are left to check which good vertices don't fall into any of these intervals. To do so in linear time, we iterate over $j\in [t]$, adding $+1$ to $v_{a_j}$ and $-1$ to $v_{b_j + 1}$. We then compute a running sum of the values assigned to each $v_i: v_1 + v_2 + \ldots + v_i$. We claim that a good vertex $v_i$ has finite eccentricity if and only if this sum is zero. If $v_i$ has finite eccentricity, then no interval $[a_j, b_j]$ contains it and so either $a_j, b_j < i$ or $a_j, b_j > i$. In both cases, the interval contributes 0 to the sum up to $v_i$. On the other hand, if $v_i$ has infinite eccentricity then it falls into some interval, $a_j \leq v_i \leq b_j$. In this case the $+1$ added to $v_{a_j}$ contributes to the sum while the $-1$ added to $v_{b_j + 1}$ doesn't. Since no interval contributes negatively to the sum, the sum at $v_i$ will be strictly positive. 
    
    Therefore, any vertex with a positive sum has infinite eccentricity and any vertex with a zero sum has finite eccentricity. This step also runs in $O(m)$ time.

    We return that any good vertex with running sum 0 has finite eccentricity, and all other vertices have infinite eccentricity. All steps of the algorithm run in linear time, which completes our proof.
\end{proof}

%% file: 6-7-LowerBounds.tex
\section{Lower Bounds for Approximating $k$-mode Diameter}\label{section:diamLB}
\input{6-LB_diam}

\section{Lower Bounds for Approximating $k$-mode Radius}\label{section:radiusLB}
\input{7-LB_radius}

%% file: 6-LB_diam.tex
In this section we present our conditional lower bounds for approximating $k$-mode diameter. 
First we show that under the Strong Exponential Time Hypothesis (SETH)\cite{impagliazzopaturi2001, impagliazzo2001problems}, no subquadratic algorithm for approximating either directed or undirected 2-mode diameter can achieve a $2-\delta$ approximation. For $3$-multimode graphs we can show stronger results: In the directed case, we show that no subquadratic algorithm can achieve any approximation guarantees. In the undirected case, we show a lower bounds tradeoff, where any $3 - \frac{2}{\ell}-\delta$ approximation for $3$-mode diameter requires $\Omega(m^{1 + 1/(\ell - 1)-o(1)})$ time. Finally, we show that for large enough $k$, $k = \Omega( \log n)$, even in the undirected case no algorithm for $k$-mode diameter can achieve any approximation guarantees. The results are summarized in table \ref{tab:diamLB}.

\begin{table}[] 
\center
\begin{tabular}{|c|c|c|l|l|c|}

\hline
\textbf{Directed?}
& {$\boldsymbol{k}$}       
& {\textbf{Approx.}}        
& {\textbf{Runtime}}                
& {\textbf{Comments}}                                                 
& \textbf{Reference} \\ 

\hline
Yes and No & $2$ & $2 - \delta$ & $\Omega(m^{2-o(1)})$ & Unweighted                                      
&            \autoref{thm:hardundir2approx}        \\

%\hline
%Yes & {$2$}                
%& {$2 - \delta$}                  
%& {$\Omega(m^2)$}                   
%& {Unweighted.}                                                         &        \autoref{thm:hardundir2approx}     %       \\ 

\hline
Yes & {$3$}                
& {Any}                           
& {$\Omega(m^{2-o(1)})$}                   
& {Unweighted DAG.}                                                    
&         \autoref{thm:hard3colordir}           \\ 

\hline
No & {$3$}                
& {$3 - \frac{2}{\ell} - \delta$} 
& {$\Omega(m^{1 + 1/(\ell - 1)-o(1)})$}  
& {Unweighted.}                                                   
&            \autoref{thm:hard3colorundir}        \\ 

\hline
No & {$\Omega(\log n)$} & {Any} & {$\Omega(m^{2-o(1)})$} & {Unweighted.}               
&            \autoref{thm:hardlogcolordiam}        \\

\hline
\end{tabular}
\caption{Result Summary: Diameter Lower Bounds.}\label{tab:diamLB}
\end{table}

Our lower bounds for $k$-mode diameter approximation follow from the Orthogonal Vector (OV) Hypothesis. The OV-problem for $n,d$ is defined as follows - given two lists of $n$ boolean vectors $A,B\subset \set{0,1}^d$, determine if there exists an orthogonal pair $a\in A,b\in B$. This problem can be extended to the $\ell$-OV-problem for $n,d$ - given $\ell$ lists of $n$ boolean vectors in $\set{0,1}^d$ $A_1, \ldots, A_\ell$, determine whether there exists a $\ell$-tuple $a_1\in A_1, \ldots, a_\ell\in A_\ell$ such that $a_1\cdot a_2 \cdot \ldots a_\ell = \sum_{i = 1}^n a_1[i]\cdot a_2[i]\cdot \ldots \cdot a_\ell[i] = 0$. Williams \cite{williams2005} showed the following hypothesis is implied by SETH \cite{impagliazzopaturi2001},\cite{impagliazzo2001problems}. 

\begin{hypothesis}[Orthogonal Vectors Hypothesis]
There is no $\epsilon > 0$ such that for any $c\geq 1$ there exists an algorithm that can solve the $\ell$-OV problem on $n, d = c\log n$ in $O(n^{\ell - \epsilon})$ time.
\end{hypothesis}

The proofs of our results follow the following format - we begin with the OV problem $A,B$ and construct a graph where if the answer to the OV question is NO (i.e. any pair of vectors $a\in A, b\in B$ is non orthogonal), then the $k$-mode diameter is $\leq \alpha$ and otherwise (there exist $a\in A, b\in B$ such that $a\cdot b = 0$), the $k$-mode diameter is $\geq \beta$. Thus, any algorithm with approximation factor $\frac{\beta}{\alpha} - \delta$, for some $\delta > 0$, is able to determine the answer to the OV problem. Under SETH, this implies that this algorithm must run in time quadratic in the size of the graph.

Our reductions will use the following representation of the 2-OV problem. Given two sets of $n$ vectors $A,B\subset \set{0,1}^d$, define $U =[d]$ and define the following graph $\gov$. The vertex set of $\gov$ is $A\cup B\cup U$ and we will add an edge between $v\in A\cup B$ and $i\in U$ if $v[i] = 1$. Note that in this graph, if the answer to the OV-problem is NO, then any $a\in A$ can reach any $b\in B$ in two steps, through any coordinate in $U$ in which they both have a 1. \label{govconstruction}

\begin{theorem}\label{thm:hardundir2approx}
If one can distinguish between 2-mode diameter 2 and 4 in an undirected $2$-multimode graph $\Gg$ in $O(m^{2-\epsilon})$ time for some constant $\epsilon >0$, then SETH is false.
\end{theorem}

\begin{proof}
Consider the base OV-construction $\gov=(A\cup B\cup U, E)$. Define the following $2$-multimode graph $\Gg = (V, E_1, E_2)$. Let $V = A\cup B \cup U \cup \set{x,y}$ where $x,y$ are two new vertices not in $A,B,U$. The red edge set $E_1$ consists of the OV-construction edges $E$, along with all the edges between $A$ and $y$ and between $x$ and $B$, $E_1 = E \cup (\set{y}\times A)\cup(\set{x}\times B)$. The blue edge set $E_2$ consists of all the edges between $A\cup U$ and $x$ and between $U\cup B$ and $y$, $E_2 = (\set{x}\times (A \cup U ))\cup (\set{y}\times (U\cup B))$.

First we claim that in the NO case (any $a\in A, b\in B$ are not orthogonal), $D(\Gg) = 2$. Any point $a\in A, b\in B$ have some index $i\in [d]$ such that $a[i] = b[i] = 1$, therefore there exists a red path $a \to i \to b$ and $d_\Gg(a,b) = 2$. We can easily check that in the blue graph any pair of points within $A\times U, U\times U$ or $U\times B$ have distance $2$. Finally, the points $x,y$ are within blue distance $\leq 2$ of all point in the graph. Thus, in the NO case the 2-mode diameter of $\Gg$ is 2.

In the YES case, there exists an orthogonal pair $a\in A, b\in B$. Therefore, there cannot exist a red path $a\to i \to b$ using some $i\in U$, as this would imply $a[i] = b[i] = 1$ and $a\cdot b \neq 0$. Any other red or blue path in the graph between $A$ and $B$ is of length at least 4, thus $d_\Gg(a,b)\geq 4$. We conclude that in the YES case, $D(\Gg)\geq 4$.

Therefore, an algorithm that can distinguish between $2$-mode diameter $2$ and $\geq 4$ in an undirected $2$-multimode graph can determine the answer to the $OV$ problem. Under SETH, it requires $\Omega(n^{2-o(1)}) = \Omega(m^{2-o(1)})$ time.
\end{proof}

As a corollary of theorem \ref{thm:hardundir2approx} we can conclude that in directed graphs no subquadratic algorithm can achieve a $2-\delta$ approximation for 2-mode diameter. If we allow a third mode (or color), we can now show that no subquadratic algorithm can achieve any approximation guarantee for directed $k$-mode diameter. In fact, this claim holds even if we limit the graphs to DAGs (i.e. a $k$-multimode graph $\Gg = (V, E_1, \ldots, E_k)$ such that every $G_i = (V, E_i)$ is a directed acyclic graph).

\begin{theorem}\label{thm:hard3colordir}
If one can distinguish between $3$-mode diameter 2 and $\infty$ in a directed acyclic $3$-multimode graph $\Gg$ in $O(m^{2-\epsilon})$ time for some constant $\epsilon >0$, then SETH is false.
\end{theorem}

Our proof will use the following gadget, introduced by Abboud, Vassilevska W. and Wang \cite{radiusdiamapprox2016}.

\begin{lemma}[\cite{radiusdiamapprox2016}]\label{lm:daggadget}
    Given an ordered list of nodes $V = \set{v_1, \ldots, v_n}$, one can construct an $O(n)$ node, $O(n\log n)$ edge DAG $DG(V)$ containing $V$ in its vertex set, such that in the topological ordering of $DG(V)$, $v_i < v_{i+1}$ and for any two nodes $x,y$ in $DG(V)$ where $x<y$ in the topological order, $d(x,y) \leq 2$. 
    % Furthermore, for any $i<j$, in $DG(V)$ we have $d(v_i, v_j) = 2$.
\end{lemma}

\begin{proof}[Proof of theorem \ref{thm:hard3colordir}]
Let $\gov = (A\cup B \cup U , E)$ be the base OV-construction with an arbitrary order of the vertices, $A = \set{a_1, \ldots, a_n}, B = \set{b_1, \ldots, b_n}, U = \set{1, \ldots, d}$. Denote by $(\hat{A}, E_A) = DG(A)$ the gadget obtained from lemma \ref{lm:daggadget} and let $E_A^R$ be the set of edges $E_A$ in reverse direction. Similarly, define $\hat{B}, E_B, E_B^R, \hat{U}, E_U, E_U^R$.

Define the $3$-mode DAG $\Gg = (V, E_1, E_2, E_3)$ as follows. The vertex set $V$ consists of $\hat{A},\hat{B},\hat{U}$ and an additional vertex $x$. The red edge set $E_1$ consists of the edges $E$ of $\gov$ directed from $A$ to $U$ and $U$ to $B$, in addition to all the edges from $\hat{A}\setminus A$ to $x$ and all edges from $x$ to $\hat{B}$.

Next, take the blue edge set $E_2$ to be the edge sets $E_A, E_B, E_U$ along with all the edges from $\hat{A}\cup \hat{B}$ to $\hat{U}$, from $\hat{B}$ to $x$ and from $x$ to $\hat{A}$. Finally, take the green edge set $E_3$ to be the edge sets $E_A^R, E_B^R, E_U^R$ along with all the edges from $\hat{U}$ to $\hat{A}
\cup \hat{B}$, from $\hat{A}$ to $x$ and from $\hat{A}$ to $\hat{B}\setminus B$.

In the NO case, any $a\in A, b\in B$ have $d_1(a,b) = 2$ from the OV-construction. For any pair $a,a'\in \hat{A}$, if $a<a'$ in the topological ordering of $DG(A)$, then $d_2(a,a')\leq 2$ and otherwise $d_3(a,a')\leq 2$. Similarly, any $b,b'\in \hat{B}$ and $u,u'\in \hat{U}$ have $d_\Gg(b,b')\leq 2$ and $d_\Gg(u,u')\leq 2$. The remaining distances are covered in table \ref{tbl:3dagdist}. We conclude that in the NO case $D(\Gg) = 2$.

\begin{table}[]
\begin{tabular}{|l"l|l|l|l|l|l|}
\hline
\textbf{}                           & \textbf{$a\in A$} & \textbf{$a'\in \hat{A}\setminus A$} & \textbf{$b\in B$} & \textbf{$b'\in \hat{B}\setminus B$} & \textbf{$u\in \hat U$} & \textbf{$x$}      \\ \thickhline
\textbf{$a\in A$}                   &                   &                                     &                   & $d_3(a,b') \leq 2$                  & $d_2(a,u) \leq 2$      & $d_3(a,x)\leq2$   \\ \hline
\textbf{$a'\in \hat{A}\setminus A$} &                   &                                     & $d_3(a',b)\leq 2$ & $d_3(a',b') \leq 2$                  & $d_2(a',u) \leq 2$     & $d_3(a',x)\leq2$  \\ \hline
\textbf{$b\in B$}                   & $d_2(b,a)\leq 2$  & $d_2(b,a')\leq 2$                   &                   &                                     & $d_2(b,u)\leq 2$       & $d_2(b,x)\leq 2$  \\ \hline
\textbf{$b'\in \hat{B}\setminus B$} & $d_2(b',a)\leq 2$ & $d_2(b',a')\leq 2$                  &                   &                                     & $d_2(b',u)\leq 2$      & $d_2(b',x)\leq 2$ \\ \hline
\textbf{$u\in \hat U$}              & $d_3(u,a)\leq 2$  & $d_3(u,a')\leq 2$                   & $d_3(u,b)\leq 2$  & $d_3(u,b')\leq 2$                   &                        & $d_3(u,x)\leq 2$  \\ \hline
\textbf{$x$}                        & $d_2(x,a)\leq 2$  & $d_2(x,a')\leq 2$                   & $d_1(x,b)\leq 2$  & $d_1(x,b')\leq 2$                   & $d_2(x,u)\leq 2$       &                   \\ \hline
\end{tabular}\caption{Distances in $\Gg$.}\label{tbl:3dagdist}
\end{table}

In the YES case, there exists an orthogonal pair $a\in A, b\in B$. For this pair no path exists between $a$ and $b$ in $G_1$, since all such paths must go through $i\in U$ such that $a[i] = b[i] = 1$. Furthermore, no path of any length exists from $A$ to $B$ in $G_2$ or $G_3$. Therefore $d_\Gg(a,b) = \infty$ and we conclude $D(\Gg) = \infty$.

Therefore, an algorithm that can distinguish between $3$-mode diameter $2$ and $\infty$ in a $3$-multimode DAG can answer the $OV$ problem. Thus, under SETH it requires $\Omega(n^{2-o(1)}) = \Omega(m^{2-o(1)})$ time.

\end{proof}

In \textit{undirected} $3$-multimode graphs, a subquadratic approximation is possible to achieve, as we showed in the algorithms section of this paper. In fact, the following lower bounds shows that our linear time $3$-approximation algorithm (\autoref{thm:3color3approx}) is tight. In essence, our proof shows that any lower bound for approximating ST-diameter can be converted into a lower bound for approximating 3-mode diameter.

\begin{theorem}\label{thm:hard3colorundir}
If one can distinguish between $3$-mode diameter $\ell$ and $3\ell - 2$ in an undirected $3$-multimode graph in $O(m^{1+\frac{1}{\ell - 1} - \epsilon})$ time for some constant $\epsilon >0$, then SETH is false.
\end{theorem}

Our proof relies on a construction introduced by Backurs et. al in \cite{tightdiamapprox2018} to show hardness of approximation of the ST-diameter $D_{S,T}$ (Lemma \ref{lm:stdiamlb}).
For any $\ell\geq 2$ their construction produces a layered graph with $\ell + 1$ layers, the first being $S$ and the last being $T$. We will use this construction as a black box, disregarding the differentiation between intermediary layers.

\begin{lemma}[Theorem 7, \cite{tightdiamapprox2018}]\label{lm:stapprox}
Let $\ell \geq 2$. There exists an $O(\ell n^{\ell -1}d^{\ell -1})$ time reduction that transforms any instance of $\ell-OV$ on $\ell$ sets of $n$ $d$ dimensional vectors, $A_1, \ldots, A_\ell$ into a graph on $O(n^{\ell -1})$ nodes and $m=O(\ell n^{\ell - 1})$ edges. The graph contains two disjoint subset $S,T$ of $n^{\ell - 1}$ nodes each, such that if the $\ell$-OV instance has a solution then $D_{S,T}\geq 3\ell - 2$ and if it does not, $D_{S,T}\leq \ell$.
\end{lemma}

% \begin{lemma}[Theorem 7, \cite{tightdiamapprox2018}]\label{lm:stapprox}
% Let $\ell \geq 2$. There exists an $O(\ell n^{\ell -1}d^{\ell -1})$ time reduction that transforms any instance of $\ell-OV$ on $\ell$ sets of $n$ $d$ dimensional vectors, $A_1, \ldots, A_\ell$ into a graph on $O(n^{\ell -1})$ nodes and $O(\ell n^{\ell - 1})$ edges. The graph contains two disjoint subset $S,T$ of $n^{\ell - 1}$ nodes each, $S = A_1\times \ldots \times A_{\ell-1}, T = A_2\times \ldots \times A_\ell$ such that 

% In particular, if if the $\ell$-OV instance has a solution then $D_{S,T}\geq 3\ell - 2$ and if it does not, $D_{S,T}\leq \ell$.
% \yael{TODO - set it up to use in the radius proof as well, need to allow the sets to have different size}
% \end{lemma}

\begin{proof}[Proof of theorem \ref{thm:hard3colorundir}]
Let $G = (\hat{V},E)$ be the graph obtained from lemma \ref{lm:stapprox} with subsets $S,T\subset \hat{V}$. Define the following $3$-multimode graph $\Gg = (V, E_1, E_2, E_3)$. For the vertex set take the vertices of $G$ with one additional vertex, $V = \hat{V}\cup \set{x}$. The red edge set $E_1$ is identical to the edge set of $G$, $E_1 = E$. The blue edge set $E_2$ consists of all the edges from $x$ to nodes outside of $T$, $E_2 = \set{x}\times (\hat{V}\setminus T)$. The green edge set $E_3$ consists of all edges from $x$ to nodes outside of $S$, $E_3 = \set{x}\times (\hat{V}\setminus S)$.

First note that for any $u,v\notin T$ we have $d_\Gg(u,v)\leq d_2(u,v) \leq 2 \leq \ell$. Similarly for any $u,v\notin S$ we have $d_\Gg(u,v) \leq \ell$. We are left to consider pairs of points $u\in S, v\in T$. By our construction $d_2(u,v) = d_3(u,v) = \infty$ so $d_\Gg(u,v) = d_1(u,v)$. 

We conclude that the $3$-mode diameter of $\Gg$ is equal to the $ST$-diameter of $G$. Thus, by lemma \ref{lm:stapprox}, distinguishing between $3$-mode diameter $\ell$ and $3\ell - 2$ determines the answer to the $\ell$-OV problem. Under SETH, this requires $\Omega(n^{\ell-o(1)}) = \Omega(m^{1 + \frac{1}{\ell - 1}-o(1)})$ time.
\end{proof}

Theorems \ref{thm:hard3colordir} and \ref{thm:hard3colorundir} show that approximating the directed $k$-mode diameter is significantly harder than approximating its undirected counterpart. The following theorem shows that for large enough $k$ we cannot approximate the undirected $k$-diameter either.

\begin{theorem}\label{thm:hardlogcolordiam}
If one can distinguish between $k$-mode diameter $2$ and $\infty$ for $k = \Omega(\log n)$ in $O(m^{2-\epsilon})$ time for some constant $\epsilon >0$, then SETH is false.
\end{theorem}

\begin{proof}
Consider the base OV-construction $\gov = (A\cup B \cup U, E)$ on a set of $n$ $d$ dimensional vectors where $d = \Theta(\log n)$. Let $k = d+2$ and define the following $k$-multimode graph $\Gg = (V, E_1, \ldots, E_{d+2})$. Let $V = A\cup B \cup U$. For $i\in [d]$, the edge set $E_i$ consists of the edges of $E$ incident to to the vertex $i\in [d] = U$. The edge set $E_{d+1}$ consists of all edges between $A$ and $U$, $E_{d+1} = A\times U$ and similarly $E_{d+2} = U\times B$.

In the NO case, any $a\in A, b\in B$ have some $i\in [d]$ such that $a[i] = b[i] = 1$ and so $d_i(a,b) = 2$. Any pair of points $u,v\notin B$ have $d_{d+1}(u,v) = 1$ and for any pair $u,v\notin A$ we have $d_{d+2}(u,v) = 1$. Thus, in the NO case $D(\Gg) = 2$.

In the YES case, there exists an orthogonal pair $a\in A, b\in B$. For this pair, any $i\in [d]$ has either $a[i] = 0$ or $b[i]=0$ and so no path exists between $a$ and $b$ in $G_i$. Furthermore, $d_{d+1}(a,b) = d_{d+2}(a,b) = \infty$. Therefore $d_{\Gg}(a,b) = \infty$ and $D(\Gg) = \infty$.

Therefore, an algorithm that can distinguish between $k$-mode diameter $2$ and $\infty$ in an undirected $k$-multimode graph can determine the answer to the $OV$ problem and so, under SETH it requires $\Omega(n^{2-o(1)} = \Omega(m^{2-o(1)})$ time.
\end{proof}

%% file: 7-LB_radius.tex
We now present our conditional lower bounds for approximating $k$-mode radius. To prove one of lower bounds, we introduce in this section a new hypothesis, the $\ell$-Hitting Set Hypothesis, which we hope can aid in proving finer lower bounds to radius approximation problems.

First we show that in the directed case there is not much hope of achieving any approximation, as any algorithm that can distinguish between a finite and infinite 2-mode radius in subquadratic time would refute the Hitting Set Hypothesis (HSH). If we require that our directed graph is acyclic, we show that any $2 - \delta$ approximation for the 2-mode radius of a DAG and any approximation for the $3$-mode radius of a DAG need quadratic time under HSH. For the undirected case, we show that for 2-mode radius no subquadratic algorithm can beat a 2-approximation under HSH. For $3$-mode undirected radius, we show a similar tradeoff to the case of approximating the diameter, where any algorithm that can distinguish between radius $\ell$ and $3\ell - 2$ faster than $\Omega(m^{1 + 1/(3\ell - 5)-o(1)})$ would refute a generalization of the Hitting Set Hypothesis, which we introduce as the $\ell$-HSH. Finally, we show that under HSH, for $k = \Omega(\log n)$, no subquadratic algorithm can achieve any approximation guarantees for undirected $k$-mode radius. Our results are summarized in table \ref{tab:radiusLB}.

\begin{table}[] 
\center
\begin{tabular}{|c|c|c|l|l|c|}

\hline
\textbf{Directed?}
& {$\boldsymbol{k}$}       
& {\textbf{Approx.}}        
& {\textbf{Runtime}}                
& {\textbf{Comments}}                                                 
& \textbf{Reference} \\

\hline
Yes & {$2$}                
& {Any}                           
& {$\Omega(m^{2-o(1)})$}                   
& {Unweighted.}
&            \autoref{thm:hard2colordirradius}        \\

\hline
Yes & {$3$}                
& {Any}                           
& {$\Omega(m^{2-o(1)})$}                   
& {Unweighted DAG.}
&            \autoref{thm:hard3colordagradius}       \\

\hline
Yes & {$2$}                
& {$2-\delta$}                           
& {$\Omega(m^{2-o(1)})$}                   
& {Unweighted DAG.}
&            \autoref{thm:hard2colordagradius}        \\ 

\hline
No & {$2$}                
& {$2 - \delta$}                  
& {$\Omega(m^{2-o(1)})$}                   
& {Unweighted.}
&          \autoref{thm:hard2colorundirradius}          \\ 

\hline
No & {$3$}                
& {$3 - \frac{2}{\ell} - \delta$}          
& {$\Omega(m^{1 + 1/(3\ell - 5)-o(1)})$} 
& \begin{tabular}[c]{@{}l@{}}
    Weighted.\\  
    Under the new $\ell$-HSH.
\end{tabular}
&      \autoref{thm:hard3colorundirradius}              \\ 

\hline
No & {$\Omega(\log n)$} 
& {Any}                           
& {$\Omega(m^{2-o(1)})$}                   
& {Unweighted.}
&         \autoref{thm:hardlogcolorradius}           \\

\hline
\end{tabular}
\caption{Result Summary: Radius Lower Bounds.}\label{tab:radiusLB}
\end{table}

As mentioned, our lower bounds for radius approximation follow from the Hitting Set Hypothesis (HSH). This hypothesis was introduced by Abboud, Vassilevska W. and Wang \cite{radiusdiamapprox2016} with a hope to aid in proving lower bounds for radius problems, where reductions from OV are elusive. The Hitting Set Existence (HSE) problem for $n,d$ is defined as follows - given two lists of $n$ boolean vectors $A,B\subset \set{0,1}^d$, determine if there exists a vector $a\in A$ that is non-orthogonal to all $b\in B$ (i.e. $a$ is a \textit{hitting set} for $B$). Note that this is the same problem as the OV-problem with a change of quantifiers, it asks whether $\exists a\in A, \forall b\in B ~ a\cdot b \neq 0$ instead of whether $\forall a\in A, \forall b\in B ~ a\cdot b \neq 0$. Abboud, Vassilevska W. and Wang \cite{radiusdiamapprox2016} proposed the following:

\begin{hypothesis}[Hitting Set Hypothesis]
There is no $\epsilon > 0$ such that for any $c\geq 1$ there exists an algorithm that can solve the Hitting Set problem on $n, d=c\log n$ in $O(n^{2-\epsilon})$ time.
\end{hypothesis}

As with the OV-problem, we wish to  extend the HSE-problem to $\ell$ sets. The $\ell$-OV problem has been used to prove many lower bound tradeoffs for approximating various notions of diameter: directed \cite{DalirrooyfardW21}, undirected \cite{tightdiamapprox2018, DalirrooyfardLW21}, ST-diameter \cite{tightdiamapprox2018} and min-diameter \cite{radiusdiamapprox2016}. The Hitting Set Hypothesis was introduced in \cite{radiusdiamapprox2016} to help prove radius-approximation lower bounds, but has only been used to prove quadratic lower bounds. We propose one way to generalize the Hitting Set Hypothesis to allow for finer lower bound tradeoffs.

Define the $\ell$-HSE problem for $n,d$ as follows. Given $\ell$ lists of $n$ boolean vectors $A_1, \ldots, A_\ell \subset \set{0,1}^\ell$, determine whether there exist $a_1\in A_1,\ldots, a_{\ell-1}\in A_{\ell - 1}$ such that for any $a_\ell\in A_\ell$ we have $a_1\cdot a_2\cdot\ldots\cdot a_\ell \neq 0$. If we think of the vectors as indicator for subsets of $[d]$, this asks if there exist sets $a_1\in A_1,\ldots, a_{\ell-1}\in A_{\ell - 1}$ such that their intersection $a_1\cap\ldots\cap a_{\ell-1}$ hits (i.e. intersects) any $a_\ell\in A_\ell$. We raise the following hypothesis:

\begin{hypothesis}[$\ell$-Hitting Set Hypothesis]\label{hyp:lhsh}
There is no $\epsilon > 0$ such that for any $c\geq 1$ there exists an algorithm that can solve the $\ell$-Hitting Set Existence problem on $n, d=c\log n$ in $O(n^{\ell-\epsilon})$ time.
\end{hypothesis}

% As with the $\ell$-OV problem, we conjecture that the $\ell$-HSE problem requires $\Omega(n^{\ell-o(1)})$ time, a stronger conjecture that implies the Hitting Set Conjecture. 
This hypothesis is in fact equivalent to the following unbalanced version, where the sets $A_1, \ldots, A_\ell$ can have different sizes, which we will need later in our proof of theorem \ref{thm:hard3colorundirradius}. 

\begin{hypothesis}[Unbalanced $\ell$-Hitting Set Hypothesis]
There is no $\epsilon > 0$ such that for any $c\geq 1$ there exists an algorithm that can solve the $\ell$-Hitting Set problem on sets $A_1, \ldots, A_\ell\subset \set{0,1}^d$ of sizes $|A_i| = n^{\lambda_i}$, with $d=c\log n$ in $O(n^{\lambda_1 + \ldots + \lambda_\ell-\epsilon})$ time.
\end{hypothesis}

% For any $\ell$ the $\ell$-HSC implies HSC.

We can now proceed to proving our lower bounds. Our radius lower bounds will follow the same format as our diameter lower bounds. We begin with the HSE problem and construct a graph where if the answer to the HSE problem is YES, then the $k$-mode radius is $\leq \alpha$ and otherwise, the $k$-mode radius is $\geq \beta$. Thus, any algorithm with approximation factor $\frac{\beta}{\alpha} - \delta$, for some $\delta > 0$, is able to the answer the HSE problem and under the HS-hypothesis  must require quadratic time.
In our reduction we will use the graph $\gov$ as defined in section \ref{govconstruction}.

\begin{theorem}\label{thm:hard2colordirradius}
If one can distinguish between 2-mode radius $2$ and $\infty$ in a directed, unweighted 2-multimode graph in $O(m^{2-\epsilon})$ time for some constant $\epsilon >0$, then HSH is false.
\end{theorem}

\begin{proof}
Consider the OV-construction $\gov = (A\cup B \cup U, E)$ and define the following 2-multimode graph $\Gg = (V,E_1, E_2)$. Take $V = A\cup B \cup U\cup \set{x}$ where $x$ is a new vertex not in $A,B,U$. The red edge set $E_1$ consist of the edges of $E$ directed from $A$ to $U$ and from $U$ to $B$. The blue edge set $E_2$ consists of all edges from $A$ to $U$, along with all bidirectional edges between $A$ and $x$ - $E_2 = A\times (E\cup \set{x})\cup \set{x}\times A$. 

In the YES case of the HSE problem, there exists a point $a\in A$ that is non-orthogonal to any point in $B$ and so $d_1(a,b) = 2$ for any $b\in B$. Furthermore, for any $v\in A\cup U \cup \set{x}$ we have $d_2(a,v)\leq 2$. Therefore, $ecc_\Gg(a) = 2$ and so $R(\Gg) \leq 2$.

On the other hand, in the NO case, any point $a\in A$ has a point $b\in B$ for which $a\cdot b = 0$. This implies $d_1(a,b) = \infty$. Since $d_2(a,b) = \infty$ for any $a\in A, b\in B$ we conclude $d_\Gg(a,b) = \infty$ and $ecc(a) = \infty$ for any $a\in A$. In our construction any point $v\in U\cup B$ cannot reach $x$ and $x$ cannot reach any point in $B$. Thus, any $v\in V\setminus A$ has $ecc(v) = \infty$ and so in the NO case we have $R(\Gg) = \infty$.

Hence, an algorithm that can distinguish between $2$-mode diameter $2$ and $\infty$ in a directed $2$-multimode graph can determine the answer to the HSE problem. Therefore, under HSH it requires $\Omega(n^{2-o(1)}) = \Omega(m^{2-o(1)})$ time.
\end{proof}

By adding an additional mode, the claim above holds for acyclic graphs as well:
\begin{theorem}\label{thm:hard3colordagradius}
If one can distinguish between 2-mode radius $2$ and $\infty$ in an unweighted $3$-mode DAG in $O(m^{2-\epsilon})$ time for some constant $\epsilon >0$, then HSH is false.
\end{theorem}

\begin{proof}
Take $\gov = (A\cup B \cup U, E)$ and pick an arbitrary ordering of the vertices of $A$, $A = \set{a_1, \ldots, a_n}$. Recall the gadget introduced in lemma \ref{lm:daggadget}, an $O(n)$ node, $O(n\log n)$ edge DAG $DG(A)=(\hat{A}, E_A)$ containing $A$ as a subset of its vertices. The construction satisfies that $a_i < a_{i+1}$ for $i<n$ in the topological ordering of $DG(A)$ and for any two vertices $x,y\in \hat{A}$ such that $x<y$ in the topological ordering, $d(x,y)\leq 2$. 

Construct two copies of this gadget, $(\hat{A},E_A)$ and $(\hat{A}', E_A')$, such that $\hat{A}, \hat{A}'$  overlap on the vertices of $A$.
Define the following $3$-multimode graph $\Gg = (V, E_1, E_2, E_3)$. Take $V = \hat{A}\cup \hat{A}'\cup B \cup U$. The red edge set $E_1$ consists of the edges of $E$ directed from $A$ to $U$ and from $U$ to $B$. The blue edge set $E_2$ set consists of the edges  $E_A,E_A'$ along with all the edges from $A$ to $U$. Finally, the green edge set $E_3$ consists of the edges of $E_A$ and $E_A'$ in the reverse direction.

In the YES case of the HSE-problem, we have $a\in A$ which can reach any point in $U\cup B$ in two red steps as before. For any $x\in \hat{A}\cup \hat{A}'$, if $a<x$ in the topological ordering, then $d_2(a,x)\leq 2$ and otherwise $d_3(a,x)\leq 2$. Thus $ecc(a)\leq 2$ and so $R(\Gg)\leq 2$.

In the NO case of the HSE-problem, first we note that any $x\in U\cup B$ has $ecc(x)= \infty$ as it cannot reach any vertex in $A$. For any $x\in \hat{A}\setminus A$, consider its copy $x'\in \hat{A}'\setminus A$. Any path from $x$ to $x'$ would have to go through some node $a\in A$. If $x<a$ in the topological order on $\hat{A}$, then the only path from $x$ to $a$ is red (using the edges of $E_2$). However, in this case there exists no red path from $a$ to $x'$ since  $a > x'$ in the topological order on $\hat{A}'$, and so a path from $a$ to $x'$ exists only in the green edge edge set $E_3$. Thus, $d_2(x,x')=\infty$ and similarly $d_3(x,x') = \infty$. Since $d_1(x,x')=\infty$ we conclude $ecc(x) \geq d_\Gg(x,x') = \infty$. Likewise, any $x'\in \hat{A}'\setminus A$ has infinite eccentricity. Finally, any $a\in A$ has some orthogonal $b\in B$ for which $d_1(a,b) = \infty$. Since $d_2(a,b) = d_3(a,b)=\infty$ we conclude $ecc(a)=\infty$ as well and so $R(\Gg) = \infty$.

Therefore, an algorithm that can distinguish between $2$-mode diameter $2$ and $\infty$ in a $3$-multimode DAG can determine the answer to the HSE problem and thus requires $\Omega(n^{2-o(1)}) = \Omega(m^{2-o(1)})$ time under HSH.
\end{proof}

In the 2-mode case for DAGs, it is in fact possible to guarantee some approximation in subquadratic time, as we see in theorem \ref{thm:dagfinecc}. In this case, we show that in subquadratic time we cannot achieve an approximation better than $2$.

\begin{theorem}\label{thm:hard2colordagradius}
If one can distinguish between 2-mode radius $2$ and $4$ in an unweighted 2-mode DAG in $O(m^{2-\epsilon})$ time for some constant $\epsilon >0$, then HSH is false.
\end{theorem}

\begin{proof}
    Let $\gov = (A\cup B \cup U , E)$ be the base OV-construction with an arbitrary order of the vertices, $A = \set{a_1, \ldots, a_n}$. Denote by $DG(A) = (\hat{A}, E_A)$ the gadget obtained from lemma \ref{lm:daggadget} and $(\hat{A}', E_A')$ be an additional copy of it, overlapping on the vertices of $A$. 

    Construct the following $2$-multimode graph $\Gg = (V, E_1, E_2)$. Define $V = \hat{A}\cup \hat{A}'\cup B \cup U$. The red edge set $E_1$ consists of the edges of $\gov$ directed from $A$ to $U$ and from $U$ to $B$, in addition to the edges $E_A, E_A'$. Define the blue edge set $E_2$ to be the edges of $E_A,E_A'$ in reverse direction, in addition to all the edges from $A$ to $U$.  

    In the YES case, there exists a node $a\in A$ that hits $B$ and so has $d_1(a,b)\leq 2$ for any $b\in B$. For any $a'\in \hat{A}\cup \hat{A}'$, if $a<a'$ in the topological order of $\hat{A}$ or $\hat{A}'$ then $d_1(a,a')\leq 2$ and otherwise $d_2(a,a')\leq 2$. Finally, for any $u\in U$ we have $d_2(a,u) \leq 2$. Therefore $ecc(a)\leq 2$ and in the YES case $R(\Gg)\leq 2$.

    In the NO case, any $a\in A$ has a point $b\in B$ with $a\cdot b = 0$. Therefore, any path from $a$ to $b$ in $G_1$ will have to go through some other $a'\in A$. Since $d_1(a,a')=2$, this path will have length $\geq 4$. In $G_2$, $d_2(a,b)=\infty$, so $d_\Gg(a,b)\geq 4$ and so $ecc(a)\geq 4$. As we showed in the proof of \autoref{thm:hard3colordagradius}, any node $x\notin A$ has infinite eccentricity and so $R(\Gg)\geq 4$.

    We conclude that an algorithm that can distinguish between $2$-mode diameter $2$ and $4$ in a $2$-multimode DAG can determine the answer to the HSE problem and thus requires $\Omega(n^{2-o(1)} = \Omega(m^{2-o(1)})$ time under HSH.
\end{proof}

Now consider the case of approximating the $k$-mode radius of an \textit{undirected} $k$-multimode graph. In the following theorem we show a lower bound similar to that of a $2$-multimode DAG, that any $2-\delta$ approximation to the $2$-mode radius requires quadratic time.

\begin{theorem}\label{thm:hard2colorundirradius}
If one can distinguish between 2-mode radius $2$ and $4$ in an undirected, unweighted 2-multimode graph in $O(m^{2-\epsilon})$ time for some constant $\epsilon >0$, then HSH is false.
\end{theorem}

\begin{proof}
Consider the OV-construction $\gov = (A\cup B \cup U, E)$. Define the $2$-multimode graph in which $G_1 = \gov$ and $G_2$ contains all the edges between $A$ and $U$, i.e.  $\Gg = (V, E_1, E_2)$ where $V = A\cup B \cup U,E_1 = E, E_2 = A\times U$. 

As before, in the YES case of the HSE problem, there exists a point $a\in A$ with $ecc(a) = 2$ and so $R(\Gg)\leq 2$. In the NO case, for every point $a\in A$ there is at least one point  $b\in B$ that it cannot reach in 2 steps. Since no edges exists within $A,B$ or $U$, this means $d_\Gg(a,b)\geq 4$, implying $ecc(a)\geq 4$ and thus $R(\Gg)\geq 4$. Under HSH, any algorithm that can distinguish between $2$-mode radius $2$ and $4$ requires $\Omega(n^{2-o(1)}) = \Omega(m^{2-o(1)})$ time.
\end{proof}

Now consider approximating the $k$-mode radius of an undirected $k$-multimode graph for $k=3$. Recall theorem \ref{thm:hard3colorundir}, showing that a $3-\frac{2}{\ell}$ approximation for undirected $3$-mode diameter requires a runtime of $\Omega(m^{1 + 1/(\ell-1)-o(1)})$. We show a similar tradeoff for approximating undirected $3$-mode radius.

\begin{theorem}\label{thm:hard3colorundirradius}
If one can distinguish between $3$-mode radius $\ell$ and $3 \ell - 2$ in an undirected, weighted $3$-multimode graph in $O(m^{1 + 1/(3\ell - 5) - \epsilon})$ time for some constant $\epsilon >0$, then the $\ell$-HSH is false.
\end{theorem}

As in the proof of \autoref{thm:hard3colorundir}, we will use the construction introduced in \cite{tightdiamapprox2018} as part of proving lower bounds to approximating ST-diameter (referenced in Lemma \ref{lm:stapprox}). However, we will use the proof of the lemma and not its conclusion, as our proof will require more details of the construction.

\begin{lemma}[Theorem 9, \cite{tightdiamapprox2018}]\label{lm:stapproxdetailed}
Let $\ell \geq 2$ and let $A_1, \ldots, A_\ell$ be an instance of $\ell$-HSE of size $|A_i| = n^{\lambda_i}$. Denote $\lambda \coloneqq \sum_{i=1}^\ell = \lambda_i$ and $\mu \coloneqq \min_i \lambda_i$.

There exists an $O(\ell n^{\ell -1}d^{\ell -1})$ time reduction that transforms  $A_1, \ldots, A_\ell$ , into an unweighted graph on $O(n^{\lambda - \mu})$ nodes and $O(\ell n^{\lambda - \mu})$ edges with the following properties.

\begin{itemize}
\item The graph contains two disjoint subsets $S,T$ where $S = A_1\times \ldots \times A_{\ell-1}, T = A_2\times \ldots \times A_\ell$.
\item For any $(a_1, \ldots, a_\ell)\in A_1\times \ldots \times A_\ell$, denote $s = (a_1, \ldots, a_{\ell - 1})\in S$ and $t = (a_2, \ldots, a_\ell)\in T$. If $a_1\cdot \ldots a_\ell \neq 0$, then $d(s,t) = \ell$ and otherwise $d(s,t)\geq 3\ell - 2$.
\end{itemize}
\end{lemma}

\begin{proof}[Proof of \autoref{thm:hard3colorundirradius}]
Let let $A_1, \ldots, A_\ell$ be an instance of $\ell$-HSE of size $|A_i| = n^{\lambda_i}$ with parameter $\lambda_i$ we will set at a later point.
Consider the graph $G = (\hat{V}, E)$ obtained from lemma \ref{lm:stapproxdetailed}. Define the $3$-multimode weighted graph $\Gg = (V,E_1, E_2,E_3)$ as follows. For the vertex set $V$ take $\hat{V}$ along with $W\coloneqq A_2\times A_3\times \ldots \times A_{\ell - 1}$ and two additional vertices, $V = \hat{V}\cup W \cup \set{x,y}$.

The red edge set $E_1$ consists of the construction edge set $E$ along with edges between any $(a_1, a_2, \ldots, a_{\ell - 1})\in S$ and $(a_2, \ldots, a_{\ell - 1})\in W$. We will assign edges from $E$ weight $1$ and edges between $S$ and $W$ weight $\ell - 1$.

The blue edge set $E_2$ consists of all edges between $S$ and $x$ with weight $\ell - 1$ and all edges between $x$ and $\hat{V}\setminus (S\cup T)$ with weight 1. Additionally, take all edges between $(a_1, \ldots, a_{\ell - 1})\in S$ and $(a_2', \ldots, a_{\ell -1}')\in W$ such that $(a_2, \ldots, a_{\ell - 1}) \neq (a_2', \ldots, a_{\ell -1}')$ and assign them weight $\ell - 1$. Finally, add an edge of weight 1 between any $(a_2, \ldots, a_{\ell - 1}, a_\ell)\in T$ and $(a_2, \ldots, a_{\ell - 1})\in W$.

For the final edge set $E_3$, take all the edges between $y$ and $S$ with weight $1$.\\

We claim that if $A_1, \ldots, A_\ell$ is a YES instance of the $\ell$-HSE problem, then $R(\Gg) \leq \ell$ and otherwise $R(\Gg)\geq 3\ell - 2$. In the YES case, there exists an $(\ell - 1)$-tuple $c \coloneqq (a_1, \ldots, a_{\ell -1})\in S$ such that for any $a_\ell \in A_\ell$, $a_1\cdot \ldots \cdot a_{\ell - 1}\cdot a_\ell \neq 0$. Therefore, by lemma \ref{lm:stapproxdetailed}, for any $a_\ell \in A_\ell$, $d_1(c, (a_2, \ldots, a_\ell)) = \ell$. For any $(a_2', \ldots, a_{\ell -1}')\neq (a_2, \ldots, a_{\ell - 1)}$, we have $d_2(c, (a_2', \ldots, a_{\ell -1}', a_\ell)) = \ell$.  So for any $t\in T$ we have $d_\Gg(c,t) = \ell$. For any $w\in W$, if $w = (a'_2, \ldots, a_{\ell - 1})$ then $d_1(c,w) = 1$, otherwise $d_2(c,w) = \ell - 1$. Finally, for any $s\in S$, $d_3(c,s) = 2$, $d_3(c,y) = 1$ and $d_2(c,x) = \ell - 1$. We conclude $ecc(c)= \ell$ and so $R(\Gg)\leq \ell$.

In the NO case, first we note that any $c\notin S$ has $ecc(c) = \infty$ since $y$ can't reach any node outside of $S$. Now consider $c \coloneqq (a_1, \ldots, a_{\ell - 1})\in S$. Since $c$ is not a hitting set for $A_\ell$, there exists $a_\ell$ such that $a_1\cdot \ldots \cdot a_{\ell - 1}\cdot a_\ell = 0$. Denote $t\coloneqq (a_2, \ldots, a_{\ell - 1},a_\ell)$. By lemma \ref{lm:stapproxdetailed}, this means $d_1(c, t) \geq 3\ell - 2$. Any path in $G_2$ between $c$ and $t$ can only go through $W$. Since $c$ and $t$ agree on $a_2, \ldots, a_{\ell - 1}$, this path has to go back to into $S$ at least once, as edges between $W$ and $T$ don't change the values of $a_2, \ldots, a_{\ell - 1}$. Therefore, $d_2(c,t)\geq 3\cdot (\ell - 1) + 1 = 3\ell - 2$. Finally, $d_3(c, t) = \infty$, thus $d_\Gg(c,t) \geq 3\ell - 2$ and $ecc(c) \geq 3\ell - 2$. We conclude that in the NO case, $R(\Gg) \geq 3\ell - 2$.\\

Set $\lambda_1 = \lambda_\ell = 1$ and $\lambda_2 = \ldots = \lambda_{\ell - 1} =\frac{1}{\ell - 1}$, so $\lambda = \frac{3\ell - 4}{\ell - 1}$ and $\mu = \frac{1}{\ell - 1}$. For a constant $\ell$, $\Gg$ has $O(n^{\lambda - \mu}) = O(n^{\frac{3\ell - 5}{\ell - 1}})$ vertices. The number of edges in $\Gg$ is dominated by the blue edges between $(a_1, \ldots, a_{\ell-1})\in S$ and $(a_2', \ldots, a_{\ell - 1}')\in W$ and between $(a_2, \ldots, a_{\ell})\in T$ and $(a_2', \ldots, a_{\ell - 1}')\in W$. So the number of edges in $\Gg$ is 
\[
M = O((|S| + |T|)\cdot |W|) = O\pars{n\cdot n^{(\ell - 2)\cdot \frac{1}{\ell - 2}} \cdot n^{(\ell - 2)\cdot \frac{1}{\ell - 1}}} = O\pars{n^{\frac{3\ell - 5}{\ell - 1}}}.
\]

Any algorithm that can distinguish between radius $\leq \ell$ and $\geq 3\ell - 2$ in $\Gg$ can solve the HSE-problem on $A_1, \ldots, A_\ell$. By the unbalanced $\ell$-HS Hypothesis, this algorithm must run in time $\Omega(n^{\lambda-o(1)}) = \Omega(n^{\frac{3\ell - 4}{\ell - 1}-o(1)}) = \Omega(M^{1 + \frac{1}{3\ell - 5} - o(1)})$. 
\end{proof} 

Finally, we show a similar result to Theorem \ref{thm:hardlogcolordiam}, where in the case of $k = \Omega(\log n)$ no subquadratic algorithm can achieve any approximation guarantees.

\begin{theorem}\label{thm:hardlogcolorradius}
If one can distinguish between $k$-mode radius $2$ and $\infty$ for $k = \Omega(\log n)$ in $O(m^{2-\epsilon})$ time for some constant $\epsilon >0$, then HSH is false.
\end{theorem}

\begin{proof}
Let $A,B$ be an instance of the HSE-problem on sets of $n$ $d$-dimensional vectors where $d = \Theta(\log n)$ and consider $\gov = (A\cup B \cup U, E)$. As in the proof of \autoref{thm:hardlogcolordiam}, we will define a $k = d+2$ multimode graph $\Gg = (V, E_1, \ldots, E_{d+2})$ where $V = A\cup B \cup U\cup \set{x}$ where $x$ is an additional vertex not in $A,B,U$. For any $i\in [d]$ the edge set $E_i$ consists of all the edges in $E$ incident to $i\in U$. The edge set $E_{d+1}$ consists of all edges between $A$ and $U$ and the final edge set $E_{d+2}$ consists of all the edges between $A$ and $x$. 

In the YES case of the HSE problem, there exists $a\in A$ such that any $b\in B$ has $i\in [d]$ with $a[i] = b[i] = 1$. Thus, $d_\Gg(a,b) \leq d_i(a,b) = 2$. Furthermore, $d_{d+1}(a,x)\leq 2$ for any $x\in A \cup U$. Thus, $R(\Gg) \leq ecc(a) = 2$. 

In the NO case, every $v\notin A$ has infinite eccentricity, as $x$ can't reach any vertex in $B$ or $U$. Any $a\in A$ has some $b\in B$ which is orthogonal to it, and therefore $d_i(a,b) = \infty$ for any $i\in [d]$. Since $d_{d+1}(a,b) = d_{d+2}(a,b) = \infty$ we conclude $ecc(a)=\infty$ and so $R(\Gg) = \infty$.

Thus, under HSH, any algorithm that can distinguish between $k$-mode radius $2$ and $\infty$ can determine the answer to the HS-problem and therefore required $\Omega(n^{2-o(1)}) = \Omega(m^{2-o(1)})$ time.
\end{proof}

%% file: 3-Exact_Case.tex
\section{Exact $k$-mode Shortest Paths}\label{section:exactcase}
In this section we consider the problem of solving various exact shortest paths problems on multimode graphs. First we consider the $k$-APSP problem of computing all pairwise $k$-mode distances in (undirected/directed) $k$-multimode graphs.

Clearly, the $k$-mode APSP problem on $\Gg = (V, E_1, \ldots, E_k)$ can be solved by computing APSP on each $G_i$ individually and taking the minimum distance for every pair of vertices. The trivial solution takes $O(kn^3)$ time and computes the $k$-mode diameter and radius of the graph as well as the $k$-mode APSP. 

In this section we show that in general, there can be no algorithm that computes $k$-mode APSP  with polynomial improvement over the trivial solution. Under the APSP Hypothesis of Fine Grained Complexity \cite{roditty2004dynamic} (see also \cite{vsurvey}), there is no $O((kn^3)^{1-\eps})$ time algorithm for $k$-mode undirected or directed APSP for any constant $\eps>0$, in $k$-multimode graphs with arbitrary integer edge weights. The same conditional lower bound applies to $k$-mode radius in $k$-multimode graphs with arbitrary integer edge weights.

However, in directed $k$-multimode graphs with bounded weights we show that an improvement over running APSP $k$ times is in fact possible.  For directed $k$-multimode graphs with bounded integer weights in $\{-M,\ldots,M\}$, there is an $\tilde{O}(k^{1/(4-\omega)}M^{1/(4-\omega)}n^{(9-2\omega)/(4-\omega)} + kMn^\omega)$ time algorithm that computes $k$-mode APSP. Here $\omega<2.372$ is the exponent of matrix multiplication \cite{newomega2024}. For APSP in directed graphs with bounded integer weights, the fastest known algorithm by Zwick \cite{zwickbridge} runs in $\tilde{O}(M^{1/4-\omega}n^{(9-2\omega)/(4-\omega)})$ time. Since $k^{1/(4-\omega)}<<k$, this is an improvement over running Zwick's algorithm $k$ times.

Similar to Zwick's algorithm, our algorithm can be slightly improved using fast rectangular matrix multiplication. The same qualitative improvement happens in this case, so we omit the details.

In the case of diameter or radius, there is no clear relationship between the diameter/radius of $G_1, \ldots, G_k$ and the diameter/radius of $\Gg$, and so computing these values is not enough to determine that of of $\Gg$. In section \ref{sct:exactappdx} we show that in fact it is possible to compute the exact values $D(\Gg), R(\Gg)$ using an algorithm for computing diameter/radius of a standard graph as a black box, with a $k$ blowup. This is not the case for approximating these values. As our lower bounds show, in many cases the lower bound for approximating the $k$-mode diameter/radius, even for $k$ as small as 2, is greater than an existing algorithm with the same approximation factor for standard diameter/radius.

\subsection{Hardness for Multimode APSP and Radius}
The APSP hypothesis states that in the word-RAM model of computation with $O(\log n)$ bit words, any algorithm for APSP in $n$ node graphs with arbitrary integer edge weights requires $n^{3-o(1)}$ time \cite{roditty2004dynamic, vsurvey}. Vassilevska W. and Williams \cite{subcubicequiv} showed that the APSP hypothesis is equivalent to the statement that the Negative Triangle problem in graphs with $n$ nodes requires $n^{3-o(1)}$ time.

The {\em Negative Triangle} problem is as follows:
Given a tripartite graph $G=(V,E)$ with node partitions $I,J,L$ on $n$ nodes each, and integer edge weights $w:E\rightarrow \ZZ$, determine whether there are $i\in I,j\in J, \ell\in L$ such that 
$w(i,j)+w(j,\ell)+w(\ell,i)<0$.

Before presenting our reductions to Multimode APSP and Radius, we first reduce the Negative Triangle problem to a multimode shortest paths question in layered graphs.
Given $k$,  let $g=\lfloor k^{1/3}\rfloor$.
Partition the vertices of $I,J,L$ into $g$ groups each of roughly $n/g$ vertices each. Let $I_p$ be the $p$-th group of $I$, $J_r$ be the $r$-th group of $J$ and $L_s$ be the $s$-th group of $L$.

$G$ has a negative triangle if and only if for some $r,p,s\in [g]$, the subgraph induced by $I_p\cup J_r\cup L_s$ contains a negative triangle.

Fix $p,r,s\in [g]$ and consider the following graph on $4$ layers $A,B,C,D$ with $w=\lceil n/g\rceil$ nodes each.
Let $A=\{a_1,\ldots,a_{w}\}$, $B=\{b_1,\ldots,b_{w}\}$, $C=\{c_1,\ldots,c_{w}\}$, $D=\{d_1,\ldots,d_{w}\}$.
Let $I_p(i)$ be the $i$-th node of $I_p$, $J_r(j)$ be the $j$-th node of $J_r(j)$ and $L_s(\ell)$ be the $\ell$-th node of $L_s$.

For every $i,j\in [w]$, add an edge from $a_i$ to $b_j$ of weight $w(I_p(i),J_r(j))$ if the edge exists in $G$. Similarly, add an edge from $b_i$ to $c_j$ of weight $w(J_r(i),L_s(j))$ if the edge exists in $G$, and add an edge from $c_i$ to $d_j$ of weight $w(L_s(i),I_p(j))$ (if the edge exists in $G$). All these edges are directed, although we will make them undirected later.

Let $G_{p,r,s}$ denote this graph.
The Negative Triangle question now becomes: does there exist $p,r,s\in [g]$ and $i\in [w]$ so that the distance in $G_{p,r,s}$ from $a_i$ to $d_i$ is negative?

To make the graphs undirected and with positive weights we make the following changes. 
Suppose that the original edges of the Negative Triangle instance had weights in $\{-M,\ldots,M\}$. We then add $10M$ to every edge in $G_{p,r,s}$ and remove the edge directions.

With these modifications, every new edge now has weight at least $10M-M=9M$. The Negative Triangle question now becomes: are there $p,r,s\in [g],i\in [w]$ so that there is a path on 3 edges from $a_i$ to $d_i$ in $G_{p,r,s}$ of weight $<30M$. Notice that every path on at least $4$ edges has weight at least $4\cdot 9M=36M>30M$, so that the only way for the distance between $a_i$ and $d_i$ to be $<30M$ is if there exists a $3$-path of weight $<30M$ between them.

Thus, we have successfully reduced the Negative Triangle problem to the problem of determining if there are some $p,r,s\in [g],i\in [w]$ so that the distance between $a_i$ and $d_i$ in $G_{p,r,s}$ is $<30M$ and where each edge has weight in $[9M,11M]$.

Note that the graphs $G_{p,r,s}$ have the same vertex set but different edges and weights.
Given this reduction, we create a $k$-multimode graph $\Gg$ with vertex set $A\cup B\cup C\cup D$. For every $p,r,s\in [g]$ add the edges of $G_{p,r,s}$ as an edge set of $\Gg$ - $E_{p,r,s}$. Note that the number of edge sets is $g^3\leq k$.

For each $i$, the $k$-mode distance between $a_i$ and $d_i$ is exactly $\min_{p,r,s} d_{G_{p,r,s}}(a_i,d_i)$, and so $G$ has a negative triangle if and only if for some $i$, the $k$-mode distance between $a_i$ and $d_i$ is $<30M$.

Suppose that $k$-mode APSP can be solved $O((kN^3)^{1-\eps})$ time for some $\eps>0$ in $N$-node $k$-multimode graphs.
In our reduction from Negative Triangle, $N=O(n/g)=O(n/k^{1/3})$.
Thus, we can solve Negative Triangle in asymptotic time 
$$(k(n/k^{1/3})^3)^{1-\eps}\leq n^{3(1-\eps)},$$
which is truly subcubic and would refute the APSP hypothesis. 

\begin{theorem}
Under the APSP hypothesis, $k$-mode APSP cannot be solved in $O((kn^3)^{1-\eps})$ time for any $\eps>0$ in the word-RAM model of computation with $O(\log n)$ bit words.
\end{theorem}

Notice, however that in our intermediate problem we didn't need to compute all pairwise distances, but merely those between $a_i$ and $d_i$.
We can utilize this observation to reduce Negative Triangle to multimode Radius.

Consider the same $k$-multimode graph $\Gg$ with vertex set $A\cup B\cup C\cup D$ and edge sets  $E_{p,r,s}$ for all 
$p,r,s\in [g]$. 
For every $i\in [w]$ and every node $u$ different from $d_i$, add an edge $(a_i,u)$ of weight $30M-1$ to the edge set $E_{1,1,1}$ (picked arbitrarily).

Now for each $i$, $\max_{u} \min_{p,r,s} d_{G_{p,r,s}}(a_i,u)<30M$ if and only if  $\min_{p,r,s} d_{G_{p,r,s}}(a_i,d_{i})<30M$. This is because all original edge weights are large enough so the new weights cannot change the distance between $a_i$ and $d_i$.

Next, add a new node $x$ and for every $i$, add an edge $(x,a_i)$  of weight $30M-1$ to $E_{1,1,1}$.
Note that for every $z\in B\cup C\cup D$, the $k$-mode distance between $x$ and $z$ is at least $30M-1+9M>30M$ as to get from $x$ to $z$, one needs to take an edge to $A$ and then an edge between $A$ and $B$ (this can only happen in $G_{1,1,1}$). 

Thus, if the $k$-mode radius is $<30M$, then the $k$-mode center must be in $A$.
In this case, there must be some $a_i$ such that in particular the multimode distance to $d_i$ is $<30M$, but this is only if there is a negative triangle in the original graph.

Conversely, if $G$ had a negative triangle $a,b,c$, then these nodes are contained in some $G_{p,r,s}$. If $a$ is represented by $a_i$ and $d_i$ in $G_{p,r,s}$, then the distance in the Radius instance that we created between $a_i$ and $d_i$ must be $<30M$. Since all other distances between $a_i$ and other vertices are $<30M$ by construction, the $k$-mode radius of our construction is $<30M$.

Thus the $k$-mode radius of $\Gg$ is $<30M$ if and only if $G$ had a negative triangle.

\begin{theorem}
Under the APSP hypothesis, $k$-mode Radius cannot be solved in $O((kn^3)^{1-\eps})$ time for any $\eps>0$ in the word-RAM model of computation with $O(\log n)$ bit words.
\end{theorem}

\subsection{Improved Algorithm for Bounded Integer Weights}

Next we show a case in which it is possible to compute $k$-mode APSP faster than computing $APSP$ $k$-times. We prove the following theorem.

\begin{theorem}
 For directed $k$-multimode graphs with bounded integer weights in $\{-M,\ldots,M\}$ where $kM\leq n^{3-\omega}$, there is an $\tilde{O}((kM)^{1/(4-\omega)}n^{(9-2\omega)/(4-\omega)})$ time algorithm that computes $k$-mode APSP.
 \end{theorem}

The bounds in the above theorem can be improved using fast rectangular matrix multiplication \cite{legallurr} if $\omega>2$, giving the same running time as Zwick's algorithm \cite{zwickbridge}, except that $M$ is replaced by $kM$ in the running time expression. 
For simplicity, we will not use rectangular matrix multiplication, but will rather focus on running times involving $\omega$.

Our algorithm mimics Zwick's algorithm \cite{zwickbridge} with a few changes. We also use the distance oracle by Yuster and Zwick \cite{yz05}. We present a randomized algorithm. However, since both Zwick's algorithm and Yuster and Zwick's distance oracle can be derandomized, our algorithm can also be derandomized.

We begin with a standard hitting set lemma whose proof can be found e.g. in \cite{zwickbridge}.

\begin{lemma}
Let $S$ be a random set comprising $\Theta(n/L \log n)$ vertices from a given $n$-node graph $G$ (for a sufficiently large constant in the $\Theta$). With probability $1-1/\poly(n)$, for any pair of vertices $u,v$ in $G$ that have a shortest path between them of at least $L$ vertices (i.e. $L$ hops), $S$ contains a vertex that lies on at least one such shortest path between $u$ and $v$.
\end{lemma}

Now let $\mathcal{G} = (V, E_1, \ldots, E_k)$ be a $k$-multimode graph on $n$ vertices with edge weights in $\{-M,\ldots,M\}$.
Similar to Zwick's algorithm, for every $i=1,\ldots, \log_{3/2}(n)$, we consider shortest paths on a number of vertices in the range $[(3/2)^i,(3/2)^{i+1})$. 

By the above lemma, if we take a random subset of $\Theta(n/(3/2)^i \log n)$ vertices, then with high probability, for every $u,v$ with a shortest path $P_{u,v}$ on a number of nodes in $[(3/2)^i,(3/2)^{i+1})$, the set $S$ hits $P_{u,v}$. In fact, with high probability $S$ hits $P_{u,v}$ in the middle third. 

Consequently, for every $u$ and $v$, $S$ contains a node $s\in P_{u,v}$ so that the number of nodes in $P_{u,v}$ from $u$ to $s$ and from $s$ to $v$ is in $[(3/2)^{i-1},(3/2)^{i})$.
This also means that the weights of both these subpaths are in the range $[-M(3/2)^{i},M(3/2)^{i}]$.

We are now ready to describe the algorithm.
As in  Yuster and Zwick's algorithm \cite{yz05}, we define a sequence of random subsets of $V$ as follows.
Let $S_0=V$. For $i>0$, $S_i$ is a random subset of $S_{i-1}$ of size $\min\{|S_i|,Cn\log n/(3/2)^i\}$ (any constant $C\geq 9$ works). 

We compute for every mode $j = 1,\ldots, k$ and every $i$, the distance in $G_j=(V,E_j)$ from every node $u\in V$ to every node $s\in S_i$ only using paths on at most $(3/2)^i$ hops. Symmetrically, we will also compute for every mode $j$ and every $i$, the distance in $G_j=(V,E_j)$ from every node $s\in S_i$ to every node $u\in V$ only using paths on at most $(3/2)^i$ hops. 

Let $d^i_j(u,v)$ be the distance between $u$ and $v$ in $G_j$ using only paths on at most $(3/2)^i$ hops. In particular, if there exists a path from $u$ to $v$ in $G_j$ with at most $(3/2)^i$ hops, then $d^i_j(u,v)=d_j(u,v)$.

To compute these distances, we employ the distance oracle construction by Yuster and Zwick \cite{yz05} for each graph $G_j$ independently. Yuster and Zwick demonstrate\footnote{The main idea is that if $A$ is an $n\times |S_{i-1}|$  matrix with $A[u,s]=d^{i-1}(u,s)$ and if $B$ is an $|S_{i-1}|\times |S_i|$ matrix with $B[s,s']=d^{i-1}(s,s')$, then one can compute $d^i(u,s)$ for all $s\in S_i$ by computing the Min-Plus product of the matrices $A$ and $B$.} that in a total of $\tilde{O}(Mn^\omega)$ time, one can compute $d^i_j(u,s)$ for all $i$, all $u\in V$ and all $s\in S_i$.

Consequently, with an overhead of $\tilde{O}(kMn^\omega)$ time, we can assume that for every $i$ and $j\in [k]$ and all $u\in V,s\in S_i$ we have computed $d^i_j(u,s)$ and $d^i_j(s,u)$.

Define an $n\times k|S_i|$ matrix $A_i$ such that for every $u\in V, s\in S_i$ and $j\in [k]$, the entry $A_i[u,(s,j)]=d^i_j(u,s)$.
Similarly, define an $k|S_i|\times n$ matrix $B_i$ 
 such that for every $u\in V, s\in S_i$ and $j\in [k]$, $B_i[(s,j),u]=d^i_j(s,u)$.

 Then, the Min-Plus product of $A_i$ and $B_i$ in the entry $(u,v)$ is given by
 \[(A_i\star B_i)[u,v]=\min_{(s,j)}A_i[u,(s,j)]+B_i[(s,j),v]=\min_{j\in [k]} d^i_j(u,s)+d^i_j(s,v).\]

If there exists a shortest path from $u$ to $v$ in $G_j$ with a number of hops within $[(3/2)^i,(3/2)^{i+1})$, then with high probability, $S_i$ contains a node $s$ on such a shortest path such that both the $u$-to-$s$ portion and the $s$-to-$v$ portion have a number of nodes in $[(3/2)^{i-1},(3/2)^{i})$.
Then w.h.p \[\min_{j\in [k]} \min_{s\in S_i} d^i_j(u,s)+d^i_j(s,v)=\min_{j\in [k]} \min_{s\in S_i} d_j(u,s)+d_j(s,v)=d_j(u,v).\]

Conversely, if for all shortest paths from $u$ to $v$ in $G_j$, the number of hops falls outside of $[(3/2)^i,(3/2)^{i+1})$, then $d^i_j(u,s)+d^i_j(s,v)\geq d_j(u,v)$. Therefore, $(A_i\star B_i)[u,v]\geq \min_{j\in [k]} d_j(u,v)$, with equality holding w.h.p when the number of hops on a $u$-to-$v$ path in the minimizing $G_j$ lies within $[(3/2)^i,(3/2)^{i+1})$.

Next, we analyze the cost of computing $A_i\star B_i$. Notably, any entry of $A_i$ and $B_i$ that falls outside the range $[-M(3/2)^{i},M(3/2)^{i}]$ can be set to $\infty$. This is because the subpaths we are interested in have at most $(3/2)^i$ nodes, and the edge weights lie in the range $[-M,M]$.
Thus, we are computing the Min-Plus product of an $n\times \tilde{O}(kn/((3/2)^i))$ matrix by an $\tilde{O}(kn/((3/2)^i)) \times n$ matrix where both matrices have integer entries in $\{\infty\}\cup [-M(3/2)^{i},M(3/2)^{i}]$.

\begin{lemma}[\cite{AlonGM97}]
The Min-Plus product of two $N\times N$ matrices with integer entries in $[-Q,Q]\cup\{\infty\}$ can be computed in $\tilde{O}(QN^\omega)$ time.
\end{lemma}

We now have two cases. First, if $k\geq (3/2)^i$,
we can compute the Min-Plus product by partitioning the computation into $\tilde{O}(k/(3/2)^i)$ products of $n\times n$ matrices with entries in  $\{\infty\}\cup [-M(3/2)^{i},M(3/2)^{i}]$. The running time becomes, within polylog factors,
\[\frac{k}{(3/2)^i} M(3/2)^i n^\omega=kMn^\omega.\]

In the second case, if $k<(3/2)^i$,
then we can partition the computation into $\tilde{O}(((3/2)^i/k)^2)$ Min-Plus products of two $kn/((3/2)^i) \times kn/((3/2)^i)$ matrices with entries in $\{\infty\}\cup [-M(3/2)^{i},M(3/2)^{i}]$.
The running time becomes, within polylog factors,
\[\frac{(3/2)^{2i}}{k^2} \cdot M(3/2)^{i} \cdot\left(\frac{kn}{(3/2)^i}\right)^\omega=Mk^{\omega-2}((3/2)^i)^{3-\omega}n^\omega.\]

Now, if $(3/2)^i\leq D$ for a parameter $D$ to be determined later, the above running time is $\tilde{O}(Mk^{\omega-2}D^{3-\omega}n^\omega)$.

Consider the case $(3/2)^i\geq D$.
Here, for every $j\in [k]$, every $u,v\in V, s\in S_i$, we compute $\min_s A^i[u,(s,j)]+B^i[(s,j),v]$ in a brute force manner. As $|S_i|\leq \tilde{O}(n/(3/2)^i)\leq \tilde{O}(n/D)$, the running time in this case is
\[\tilde{O}(kn^3/D).\]

The overall running time, up to polylogarithmic factors, is given by
\[kMn^\omega + Mk^{\omega-2}D^{3-\omega}n^\omega+ kn^3/D.\]

%Mk^{\omega-2}D^{3-\omega}n^\omega= kn^3/D
%D^{4-\omega}= k^{3-\omega}n^{3-\omega}/M

To minimize the running time we set $D=(kn)^{(3-\omega)/(4-\omega)}/M^{1/(4-\omega)}$.
The running time becomes $$\tilde{O}(k^{1/(4-\omega)}M^{1/(4-\omega)}n^{(9-2\omega)/(4-\omega)} + kMn^\omega).$$

The running time is truly subcubic as long as $kM\leq O(n^{3-\omega-\eps})$ for $\eps>0$, in which case the first term dominates and the running time is $\tilde{O}(k^{1/(4-\omega)}M^{1/(4-\omega)}n^{(9-2\omega)/(4-\omega)}).$

\subsection{Computing $k$-mode Diameter and Radius}\label{sct:exactappdx}
In this section we consider the problem of computing the diameter and radius of a $k$-multimode graph $\Gg = (V, E_1, \ldots, E_k)$. As there is no relationship between the diameter/radius of $G_1, \ldots, G_k$ and $D(\Gg), R(\Gg)$, it is not clear if algorithms for computing the diameter and radius of a standard graph could be applicable to the multimode setting. However, in this section we show that in fact we can convert any algorithm for diameter or radius in standard graphs with integer edge weights to an algorithm for $k$-mode diameter or radius, while incurring a blowup of $k$ in both the vertices and the edges.

\begin{theorem}\label{thm:equiv_diam}
If there exists a $T(n,m)$ time algorithm for computing the diameter of a weighted, undirected (standard) graph with $n$ nodes and $m$ edges, then there exists an algorithm for computing the $k$-mode diameter of a $k$-multimode undirected graph in time $T(O(kn), O(m + kn))$.
\end{theorem}
\begin{proof}
Let $\mathcal{G} = (V, E_1, \ldots, E_k)$ be a $k$-multimode graph. Construct a graph $G$ consisting of a disjoint union of $G_1, \ldots, G_k$ and an additional copy of the vertex set $V$. In other words the vertices of $G$ consist of $k+1$ copies of $V$ -  $V, V_1, V_2, \ldots, V_k$ and an edge exists between $u,v\in V_i$ if $(u,v)\in E_i$. Denote each copy of $v\in V_i$ as $v_i$.  Set $W$ to be an even value greater than the length of a simple path in any $G_i$, $W > n\cdot \max_{e\in \cup E_i}w(e)$. For every $v\in V$, add edges of weight $W$  between $v$ and $v_i$ for any $i\in [k]$. Finally, add an additional vertex $x$ and connect it with an edge of weight $\frac{W}{2}$ to every node in $V_1, \ldots, V_k$.
% The construction is illustrated in Figure \ref{fig:undir_diam_equiv}.

% \begin{figure}[ht]
%     \centering
%     \includegraphics[width=0.5\textwidth]{Figures/Undirected_diamter_equivalence.png}
%     \caption{Reducing $k$-mode undirected diameter to undirected diameter.}
%     \label{fig:undir_diam_equiv}
% \end{figure}
The resulting graph $G$ has $O(k n)$ vertices and $O(m + k n)$ edges. We claim that the diameter of $G$ is equal to $2W + D(\Gg)$. If $D(\Gg)=\infty$, the diameter of $G$ is equal to $3W$.

Indeed, any path between two vertices $u,v\in V$ will have to enter at least one $G_i$ since there are no edges within $V$. If the path enters two distinct $G_i, G_j$ it will be of length $\geq 3W$, therefore the shortest path will have to be of the form $u\to u_i \rightsquigarrow v_i \to v$, where $u_i \rightsquigarrow v_i$ is a shortest path between $u$ and $v$ in $G_i$. We conclude that $d_G(u,v) = 2W + \min _i d_{G_i}(u,v) = 2W + d_{\mathcal{G}}(u,v)$.

For any pair of points $u\in V, v_i \in V_i$, there exists a path between them of length $2W$, $u \to u_i \to x \to v_i$. Similarly, for any pair of points $u_i\in V_i, v_j\in V_j$  we have  $d_G(u_i, v_j)\leq d_G(u_i, x) + d_G(x, v_j) = W$. Finally, $x$ is within distance $\frac{W}{2}$ of any vertex in $V_1, \ldots, V_k$ and within distance $\frac{3W}{2}$ of any vertex in $V$.

We conclude that the diameter of $G$ is $\max_{u,v} (2W + d_{\Gg}(u,v)) = 2W + D(\Gg)$. Therefore, by computing the diameter of $G$ and subtracting $2W$ we can compute the $k$-mode diameter of $\mathcal{G}$ in $T(O(kn), O(m + kn))$ time.
\end{proof}

The above result holds for directed $k$-multimode graphs as well. Using the same constructions, with the edges between the sets $V, V_1, \ldots, V_k, \set{x}$ taken in both directions, we arrive at the same conclusion. 

Next, consider computing the $k$-mode radius of a weighted $k$-multimode graph. The following theorem shows that we can convert any algorithm for computing the radius of a standard weighted graph into an algorithm for computing the $k$-mode radius. We will prove this claim for undirected graphs, from which the same statement for directed graphs follows.

\begin{theorem}
If there exists a $T(n,m)$ time algorithm for computing the radius of a weighted, undirected (standard) graph with $n$ nodes and $m$ edges, then there exists an algorithm for computing the $k$-mode radius of a $k$-multimode undirected graph in time $T(O(kn), O(m + kn))$.
\end{theorem}
\begin{proof}
Recall the construction of graph $G$ in the proof of \autoref{thm:equiv_diam}. We define our graph $G'$ to be $G$ with an additional vertex $y$, connected to all nodes $v\in V$ with an edge of weight $2W$.

As we showed in the proof of \autoref{thm:equiv_diam}, for any $u,v\in V$ and $v_i\in V_i$, $d_{G'}(u,v) = 2W + d_{\mathcal{G}}(u,v)$ and $d_{G'}(u,v_i)  \leq 2W$. Hence, the eccentricity of $u$ is $ecc_{G'}(u) = 2W + \max_v d_\mathcal{G}(u,v) = 2W + ecc_{\mathcal{G}}(u)$. For any node in $G'$ not in the set $V$, its eccentricity will be at least $3W$ since $d_{G'}(y, v_i) = 3W$ for any $v_i \in V_i$ and $d_{G'}(x,y) = \frac{7W}{2}$. 

Therefore, the radius of $G'$ is equal to $R(G') = \min_{u\in V} ecc_{G'}(u) = \min _{u\in V} (2W + ecc_{\mathcal{G}}(u)) = 2W + R(\mathcal{G})$. And so, by computing the radius of $G'$ and subtracting $2W$, we can compute the $k$-mode radius of $\mathcal{G}$ in $T(O(kn), O(m + kn))$ time.
\end{proof}

%% file: 4a-Undir_diam_appendix.tex
\section{Approximating $2$-mode Undirected Diameter} \label{section:4a_more_undir_diam}
In section \ref{section:alg_undir} we showed a $2$ and $2.5$-approximation algorithm for undirected, unweighted $2$-multimode graphs (theorems \ref{thm:2approx}, \ref{thm:2.5approx}). In the following section we complete their proof and remove the simplifying assumptions we had previously made. First, if $\Gg$ is a weighted graph we preform Dijkstra's algorithm in place of every BFS in the original algorithms, which only adds a logarithmic factor to the runtime.

In our $2$ and $2.5$ approximation algorithms we used the following lemma:
\begin{lemma}[Lemma \ref{lm:point2smallnbhds}]
    Let $\Gg$ be a $2$-multimode graph  with $2$-mode diameter $\geq D$. Given  $\frac{1}{3}< \alpha < \frac{1}{2}$ and a point $z$ with $|B_1(z, \frac{3\alpha - 1}{2} D)|\leq n^\delta$ and $|B_2(z, \frac{3\alpha - 1}{2} D)| \leq n^\delta$, we can find a pair of points $a,b$ with $d_{\Gg}(a,b) \geq \alpha D$ in time $O(n^\delta m) + M(n^\delta, n, n^\delta)$.
\end{lemma}

To prove this lemma, we assumed that $\frac{1 - \alpha}{2}D\in \NN$ and that $\Gg$ was unweighted. Under these assumptions, given a point $z$ and diameter endpoint $s$ such that $d_1(z,s)\leq \alpha D$, there exists a point $x$ on the shortest path between $z$ and $s$ such that $d_1(x,s)=\frac{1-\alpha}{2}D$, $d_1(z,x)\leq \frac{3\alpha - 1}{2}D$. 

If $\frac{1-\alpha}{2}D\notin \NN$, in an unweighted graph we can find a point $x$ on the shortest path from $z$ to $s$ such that $\frac{1-\alpha}{2}D -\frac{1}{2}\leq d_1(x,s)\leq \frac{1-\alpha}{2}D + \frac{1}{2}$. Now, $d_1(x,z)\leq \frac{3\alpha - 1}{2}D + \frac{1}{2}$. 

Furthermore, if we have two points $x,y$ such that $d_1(x,s)\leq \frac{1-\alpha}{2}D + \frac{1}{2}$ and $d_2(y,t)\leq \frac{1-\alpha}{2}D + \frac{1}{2}$ we can perform the same algorithm as we showed in section \ref{section:alg_undir} and find a pair of points $a,b$ with $d_\Gg(a,b)\geq \alpha D - 1$.

In a weighted graph with edge weights bounded by $M$ we can find a point $x$ such that $\frac{1-\alpha}{2}D - \frac{M}{2} \leq d_1(x,s)\leq \frac{1-\alpha}{2}D + \frac{M}{2}$  and $d_1(x,z)\leq \frac{3\alpha - 1}{2}D + \frac{M}{2}$. Using this, we can complete the proof of the following lemma the same way we proved lemma \ref{lm:point2smallnbhds}.

\begin{lemma}[Lemma \ref{lm:point2smallnbhds}, removing assumptions]
    Let $\Gg$ be a $2$-multimode graph with non-negative edge weights bounded by $M$ and $2$-mode diameter $\geq D$. Given  $\frac{1}{3}< \alpha < \frac{1}{2}$ and a point $z$ with $|B_1(z, \frac{3\alpha - 1}{2} D + \frac{M}{2})|\leq n^\delta$ and $|B_2(z, \frac{3\alpha - 1}{2} D) + \frac{M}{2}| \leq n^\delta$, we can find a pair of points $a,b$ with $d_{\Gg}(a,b) \geq \alpha D - M$ in time $O(n^\delta m) + M(n^\delta, n, n^\delta)$.
\end{lemma}

\subsection{$2$-Approximation for $2$-mode Diameter}
To obtain a $2$-approximation algorithm for $2$-mode diameter we used lemma \ref{lm:point2smallnbhds} with $\alpha = \frac{1}{2}$. In theorem \ref{thm:2approx}, we showed that if we sample a point of distance $<\frac{D}{4}$ from some diameter endpoint, we can use it to find two points of distance $\geq \frac{D}{2}$. In this case we didn't use any additional assumptions about the input. 

If instead we sample a point $z$ of red distance $<\frac{D}{4} + \frac{M}{2}$ from some diameter endpoint $s$, a similar algorithm gives us a pair of points with $d_\Gg(a,b)\geq \frac{D}{2} - M$. Take $A = B_1(z,\frac{D}{4} + \frac{M}{2}), B = V \setminus B_1(z, \frac{3D}{4} - \frac{M}{2})$. Our assumption says $s\in A$ and so we must have $t\in B$. All points $a\in A, b\in B$ have $d_1(a,b)\geq \frac{D}{2}- M$ while the blue $ST$-diameter of $A,B$ is $D$. Therefore, a $2$-approximation to the $ST$-diameter of $A,B$ in the blue graph will produce a pair of points $a,b$ with $d_2(a,b)\geq \frac{D}{2}$ and so $d_\Gg(a,b)\geq \frac{D}{2}-M$.

If one of the diameter endpoints had a large $\frac{D}{4} + \frac{M}{2}$ neighborhood in either color, we would be able to sample a point in that neighborhood and finish. Otherwise, we find a point with small $\frac{D}{4} + \frac{M}{2}$ neighborhood in both colors and apply lemma \ref{lm:point2smallnbhds}. 

Therefore, after removing the simplifying assumptions we remain with the following theorem.

\begin{theorem} [Theorem \ref{thm:2approx}, removing assumptions]
Given an undirected $2$-multimode graph $\Gg$ with non-negative edge weights bounded by $M$ and an integer $D$, if $\omega = 2$, there exists an $\Tilde{O}(m\cdot n^{3/4})$ time algorithm that does one of the following with high probability:
\begin{enumerate}
    \item Finds a pair of points $a,b$ with $d_\Gg(a,b) \geq \frac{D}{2}-M$.
    \item Determines that the 2-mode diameter of $G$ is $< D$.
\end{enumerate}
If $\omega > 2$ the runtime of the algorithm is $\Tilde{O}\pars{m\cdot \pars{\frac{n^{1.5}}{(m\sqrt{n})^{1/\omega}} + \pars{\frac{m}{n}}^{1/(\omega - 2)}}}$.
\end{theorem}

By binary searching over $D$, the above algorithm allows us to find a value $\Tilde{D}$ such that $D(\Gg)\geq \Tilde{D}\geq \frac{D(\Gg)}{2} - M$. This means $D(\Gg)\leq \Tilde{D}\leq 2\cdot D(\Gg) + 2M$, or that $\Tilde{G}$ is a $(2,2M)$ approximation to the $2$-mode diameter of $\Gg$.

\subsection{$2.5$-Approximation for $2$-mode Diameter}
To obtain a $2.5$-approximation algorithm for $2$-mode diameter we used lemma \ref{lm:point2smallnbhds} with $\alpha = \frac{2}{5}$. Denote a pair of diameter endpoints by $s,t$. In theorem \ref{thm:2.5approx}, we showed that if we sample a point $x$ of distance $<\frac{D}{10}$ from  $s$ (w.l.o.g $d_1(x,s)<\frac{D}{10}$), we can then take a point $y$ of distance $d_1(x,y)=\frac{D}{2}$  and guarantee that running BFS from $y$ will find a point $z$ with $d_\Gg(y,z)\geq \frac{2}{5}$. 

If instead we only assume $d_1(x,s)< \frac{D}{10} + \frac{M}{2}$ we can find a point $y$ such that $\frac{2D}{5} - \frac{M}{2} \leq d_1(x,y)\leq \frac{2D}{5} + \frac{M}{2}$. Now we have either $d_1(s,y)\geq \frac{2D}{5} - M$ or $d_1(y,t)\geq \frac{2D}{5} - M$, using the same triangle inequalities as in the proof of theorem \ref{thm:2.5approx}. Therefore, running Dijkstra's algorithm from $y$ will find a point of distance $\geq \frac{2D}{5}-M$ from $y$.

We conclude that if either $s$ or $t$ had a large $\frac{D}{10} + \frac{M}{2}$ neighborhood in either color, we would be able to sample a point in that neighborhood and finish. Otherwise, we find a point with small $\frac{D}{10} + \frac{M}{2}$ neighborhood in both colors and apply lemma \ref{lm:point2smallnbhds}. 

After removing the simplifying assumptions we remain with the following theorem.

\begin{theorem} [Theorem \ref{thm:2.5approx}, removing assumptions]
Given an undirected $2$-multimode graph $\Gg$  with non-negative edge weights bounded by $M$ and integer $D$, if $\omega = 2$ there exists an $\Tilde{O}(m\cdot \sqrt{n})$ time algorithm that does one of the following:
\begin{enumerate}
    \item Finds a pair of points $a,b$ with $d_\Gg(a,b) \geq \frac{2D}{5} - M$.
    \item Determines that the 2-mode diameter of $G$ is $< D$.
\end{enumerate}
If $\omega > 2$ the runtime of the algorithm is $\Tilde{O}\pars{m\cdot \pars{\frac{n}{m^{1/\omega}} + \pars{\frac{m}{n}}^{1/(\omega - 2)}}}$.
\end{theorem}

By binary searching over $D$, the above algorithm allows us to find a value $\Tilde{D}$ such that $D(\Gg)\geq \Tilde{D}\geq \frac{2D(\Gg)}{5} - M$. This means $D(\Gg)\leq \Tilde{D}\leq 2.5\cdot D(\Gg) + 2.5M$, or that $\Tilde{D}$ is a $(2.5,2.5M)$ approximation to the $2$-mode diameter of $\Gg$.